%% file: paper.tex
\documentclass{sig-alternate-hotpets15}
\usepackage{color}
\usepackage[hyphens]{url}
\usepackage{longtable}
\usepackage{graphicx}
\usepackage{enumitem}
\usepackage{pdfpages}

\def\etal{{\it et al.~}}

\newenvironment{packed_item}{
\begin{itemize}
  \setlength{\itemsep}{1pt}
  \setlength{\parskip}{0pt}
  \setlength{\parsep}{0pt}
}{\end{itemize}}

\begin{document}
\input{tex-inputs/title.tex}
\input{tex-inputs/abstract.tex}
\input{tex-inputs/intro.tex}
\input{tex-inputs/methodology.tex}

\input{tex-inputs/results-intro.tex}
\input{tex-inputs/results-type.tex}

\input{tex-inputs/results-recipient.tex}
\input{tex-inputs/results-devices.tex}
\input{tex-inputs/results-techrank.tex}
\input{tex-inputs/results-openended.tex}
\input{tex-inputs/results-demographics.tex}
\input{tex-inputs/results-regression.tex}

\input{tex-inputs/discussion.tex} 
\input{tex-inputs/related_work.tex} 
\input{tex-inputs/conclusion.tex} 

\bibliographystyle{abbrv}
\bibliography{bibliography.bib} 
\input{tex-inputs/appendixes.tex}

\input{tex-inputs/collapsed.tex}
\input{tex-inputs/riskben.tex}
\input{tex-inputs/VURcombined.tex}

\end{document}

%% file: tex-inputs/title.tex

\title{Risk Perceptions for Wearable Devices}
\numberofauthors{1}
\author{
 \alignauthor Linda N. Lee\textsuperscript{1}, Serge Egelman\textsuperscript{1,2}, Joong Hwa Lee\textsuperscript{1}, David Wagner\textsuperscript{1}\\
   \vspace{0.5em}
   \affaddr{\textsuperscript{1}University of California, Berkeley, \{lnl,egelman,daw\}@cs.berkeley.edu}, dlwndghk94@berkeley.edu\\
   \affaddr{\textsuperscript{2}International Computer Science Institute, egelman@icsi.berkeley.edu}\\
}

\maketitle

%% file: tex-inputs/abstract.tex

\begin{abstract}
\indent\indent Wearable devices, or ``wearables,'' bring great benefits but also potential risks that could expose users' activities without their awareness or consent. In this paper, we report findings from the first large-scale survey conducted to investigate user security and privacy concerns regarding wearables. We surveyed 1,782 Internet users in order to identify risks that are particularly concerning to them; these risks are inspired by the sensor inputs and applications of popular wearable technologies. During this experiment, our questions controlled for the effects of what data was being accessed and with whom it was being shared. We also investigated how these emergent threats compared to existent mobile threats, how upcoming capabilities and artifacts compared to existing technologies, and how users ranked technical and non-technical concerns to sketch a concrete and broad view of the wearable device landscape. We hope that this work will inform the design of future user notification, permission management, and access control schemes for wearables.
\end{abstract}



\keywords{Privacy, Security, User Studies, Risk Perception, Ubiquitous Computing, Wearable Devices} 

%% file: tex-inputs/intro.tex

\section{Introduction}

Wearables are a \$700 million, growing industry~\cite{cmo}. With 20\% of the general population owning at least one wearable and 10\% using it daily~\cite{WearableStatNews}, wearables are bringing ubiquitous computing to everyday life. This trend will likely continue, as 52\% of technology consumers are aware of wearables and 33\% are likely to buy one~\cite{NPD}.  

Wearable devices enable many benefits, ranging from interaction with virtual objects in an augmented reality world to healthier, fitness-data inspired lifestyles. However, wearable devices also bring new potential privacy and security risks that could expose users' activities without their awareness or consent. Although wearable devices are still in their infancy, we have already seen manifestations of these risks. Fitbit's default privacy settings inadvertently exposed information about some of their users' sexual activity~\cite{Fitbit}. Public discomfort toward facial recognition caused Google to prohibit Google Glass applications from using facial recognition~\cite{GlassDetection}, but still resulted in tech hate crimes against its users ~\cite{1_russell_2014, 16_gross_2014}. Google Glass has since been discontinued.

For smartphones, security and privacy risks are generally addressed by communicating data capture to users. However, many users are habituated to these notifications, because they see them frequently, often for things that they do not care about~\cite{felt2012android}. Once habituated to seemingly benign privacy and security warnings, users tend to ignore more sensitive warnings that are similarly designed~\cite{Egelman08}. 

Wearables' sensor capabilities, continuous access, and ubiquitous presence will result in a firehose of familiar and unfamiliar types of data, at a rate which will likely dwarf the amount of data currently captured by smartphones. Bystanders of wearable devices have already expressed interest in such communication, desiring notification before data about them is captured~\cite{denning2014situ}. However, subjecting people to increased notifications is not a sound option, as it has shown to lead to negative effects, such as frustration and habituation~\cite{bohme2011security}. An understanding of user concerns may allow for targeted and effective communication with the user, inform design of future permission systems, or provide insight for access control mechanisms. 

The goal of this work is to shape the still-malleable future of wearable platforms and interaction models, with research on user-centric concerns. To our knowledge, this is the first large-scale study to investigate user security and privacy concerns for wearable devices. Our survey of 1,782 Internet users contributes the following: 

\begin{itemize} \itemsep1pt \parskip0pt \parsep0pt
\item We report how 72 types of data likely to be captured by wearable devices were perceived by our participants and rank them by relevance. 
\item We repeat this across 4 types of recipients to also illustrate the contribution of the data recipient to the overall perceived risk. 
\item We sketch a landscape of users' self-reported concerns regarding wearable devices, spanning concerns outside of security and privacy. 
\item We compare emergent risks with existing risks and find that participants perceive risks similarly to physical risks---for instance, facial detection was perceived as risky as using a lawnmower.
\end{itemize}

%% file: tex-inputs/methodology.tex

\section{Methodology}
To obtain a comprehensive list of possible risks that wearable devices might present in the future, we examined the sensors, capabilities, permissions, and applications of the most popular wearable devices on the market. At the time of this study (August 2014) the most popular wearable devices included the Fitbit fitness tracker, which continuously monitors heartbeat, steps taken, and sleep patterns;
the Pebble smartwatch, which can take pictures, send texts, show notifications from online, and push notifications to services; 
and Google Glass, which can take pictures, record video, and perform a subset Internet-based tasks such as search, reading emails, etc. 
These devices' capabilities and requested permissions were our inspiration to develop a list of possible security and privacy risks that users will encounter.

We designed a survey to gauge the relevancy of all these possible risks.
Our survey contained two main sections. A set of questions presented participants with several scenarios---something undesirable that might happen with their wearable device---and asked them to rate their level of concern if each scenario were to happen. This was intended to elicit their perception of the severity and impact of the risk.
The format of this section was based on Felt {\it et al.}'s study of user perceptions of security and privacy risks with mobile devices~\cite{Felt}. Another set asked participants to compare the risks and benefits of wearable technologies to those of better-understood technologies, following the same methodology from Fischhoff {\it et al.}'s seminal study in risk perception~\cite{Fischhoff}.

\subsection{Related Work}
In this section, we describe the two prior works on which we based our survey format: Felt {\it et al.}'s survey of smartphone-based risks~\cite{Felt} and Fischhoff {\it et al.}'s survey of a wide range of general technology-based risk perceptions~\cite{Fischhoff}.

\subsubsection{Smartphone Risk Scenarios}
Felt \etal previously studied the security concerns of smartphone users by conducting a large-scale online survey~\cite{Felt}. Their survey asked 3,115 smartphone users about 99 risk scenarios. Participants were asked how upset they would be if a certain action occurred without their permission. Participants rated each situation on a Likert scale ranging from ``indifferent (1)'' to ``very upset (5).''
Our methodology closely follows that study, but with scenarios chosen to shed light on the security and privacy risks of wearable devices.

\subsubsection{Technology Risk Perceptions}
Fischhoff \etal performed a seminal study of the perceived risks and benefits surrounding 30 widely used technologies~\cite{Fischhoff}. Participants were asked to separately think of the risks then benefits, considering all people affected, long-term effects, and short term effects. Then, the participants respectively rated these technologies on a numerical scale, being instructed to rate the least risky or least beneficial technology a 10 and scaling the ratings linearly (e.g., a technology with a risk rating of 20 is considered twice as risky as compared to a technology with a risk rating of 10). We apply their methodology to evaluate perceived risks and benefits of technologies related to wearable computing with respect to more familiar technologies.

\subsection{Survey Questions}
\noindent Each participant answered 27 questions across 5 sections:   \\[-.5cm]

\begin{itemize} \itemsep1pt \parskip0pt \parsep0pt
\item 2 reading comprehension questions
\item 6 questions about wearable computing scenarios 
\item 2 questions about smartphone scenarios 
\item 2 Fischhoff-style risk/benefit questions 
\item 15 demographic questions 
\end{itemize}

We randomized the order participants saw sections of the survey (with the exception of the comprehension and demographic questions, which were always first and last, respectively), as well as the order of questions in each section.

\subsubsection{Comprehension Questions}
Because participants might be biased to specific companies (e.g., visceral reactions to Google Glass based on popular media stories), we based our questions on a fictitious wearable. Thus, the beginning of the survey introduced participants to the ``Cubetastic3000,'' which was the basis for all questions on wearables risks. We highlighted the capabilities of this device and described use cases:

\begin{quotation}
{\it Imagine that you are the proud owner of the Cubetastic3000, a new, high-tech computing device designed to be worn on your head. Imagine also that you wear this device all the time, because it is very lightweight, durable, and convenient.

The Cubetastic3000 has the capability to capture video, photos, audio, and biometrics (biological data about you, such as heart rate). Just like other devices today, you can install third-party applications from an app store, and these applications can use the information that the Cubetastic3000 captures.

With a wide range of applications and capabilities, your device can do all sorts of things, such as:\\

\noindent---measuring heart rate, breathing, and other things to keep track of your fitness level and overall health\\

\noindent---look at what you see to provide information about what's around you\\

\noindent---allow you to take notes just by telling the device what you need to remember\\

\noindent---take videos of you or what you see to share with others\\

\noindent---automatically take photos or video so that you can replay events that previously happened\\

\noindent---play music that you like for you when it detects that no one is around\\

\noindent---infer information about you so you don't need to log in or search for the same thing over and over\\

\noindent ...and much more!}
\end{quotation}

Since this device is the basis for many of our questions, we ensured that participants had understood its capabilities by asking them two multiple-choice comprehension questions. We filtered out responses from participants who could not answer these questions.

\subsubsection{Wearable Scenarios}
We presented scenarios involving data captured by the Cubetastic3000 and asked participants to rate how upset they would be if a particular type of data (e.g., video, audio, name, etc.) was shared without permission with a particular data recipient (see Figure \ref{fig:prompt}). Responses were reported on a 5-point Likert scale (from ``indifferent'' to ``very upset''), following Felt \etal\cite{Felt}. Questions were of the form: 

\begin{quotation}
\noindent
\textit{``How would you feel if an app on your Cubetastic3000 learned <data> and shared it with <recipient>, without asking you first?''}
\end{quotation}

We combined 72 data types (<data>) with 4 data recipients (<recipient>) to form an initial pool of 288 questions: 

\begin{figure}[t]
	\centering
	\includegraphics[width=\columnwidth]{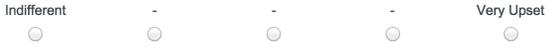}
	\caption{An example of a wearable scenario question participants saw while taking the survey.}
	\label{fig:prompt}
\end{figure}

\begin{packed_item}
\item Your work contacts
\item Your friends
\item The public
\item The app's server (but didn't share it with anyone else) 
\end{packed_item}

The purpose for using this question format was to determine how upset participants would be if data were inappropriately shared, and the extent to which their reactions were based on the data type and recipient. Each participant answered 6 questions that were randomly drawn from a pool of 293: the 288 described here, plus 5 that we describe in the next section (Section \ref{sec:smartphones}).

\subsubsection{Smartphone Scenarios}
\label{sec:smartphones}
We presented participants with a second set of scenarios to control for the type of device being used. Rather than using the previous pool of 288 <data> and <recipient> combinations, we selected 5 scenarios that Felt \etal found least and most concerning to their participants~\cite{Felt}:

\begin{enumerate}[topsep=0pt,itemsep=-1ex,partopsep=1ex,parsep=1ex]
\item {\it How would you feel if an app on your <device> vibrated your phone without asking you first?}
\item {\it How would you feel if an app on your <device> connected to a Bluetooth device (like a headset) without asking you first?}
\item {\it How would you feel if an app on your <device> un-muted a phone call without asking you first?}
\item {\it How would you feel if an app on your <device> took screenshots when you were using other apps, without asking you first?}
\item {\it How would you feel if an app on your <device> sent premium (they cost money) calls or text messages, without asking you first?}
\end{enumerate}

In the previously described section of our survey, <device> was set to ``Cubetastic3000'' and not every participant received one of these questions (i.e., these 5 questions were among the pool of 293 questions from which participants were randomly assigned 6). In the separate smartphone section of the survey, every participant received exactly two of these questions, where <device> was set to ``smartphone.'' This allowed us to perform controlled comparisons based on whether the same misbehavior was occurring on a smartphone (i.e., a better understood device) or the Cubetastic3000 (i.e., a fictitious wearable device).

\subsubsection{Risk and Benefit Assessment}
In addition to investigating reactions to particular scenarios, we examined broad perceptions of new technologies and how those compared to perceptions of other understood technologies. We modeled this section after a seminal risk perception study by Fischhoff \etal\cite{Fischhoff}, in which participants ranked technologies by their relative risk and benefit to society. We ask participants to perform this exercise for 4 technologies previously examined by Fischhoff {\it et al.}: handguns, motorcycles, lawnmowers, and electricity.  These technologies were chosen to span varying levels of risks and benefits.

Our goal was to ask about familiar technologies such as the Internet, general and specific wearable artifacts, and a range of new capabilities. To do this, we asked participants to evaluate one of 20 technologies relevant to wearables along 4 previously studied technologies. The 20 technologies were: Internet, email, laptops, smartphones, smart watches, fitness trackers, Google Glass, Cubetastic3000, discrete camera, discrete microphone, facial recognition, facial detection, voice recognition, voice-based emotion detection, location tracking, speech-to-text, language detection, heart rate detection, age detection, and gender detection.

To parallel Fischhoff {\it et al.}'s risk perception study, we gave our participants a similar prompt to numerically express the perceived gross risk/gross benefit over a long period of time for all parties involved. We randomized whether they performed the ranking for risks or benefits first. The prompt is listed in Appendix \ref{sec:prompt}. The question format was as follows:

\begin{quotation}
{\it \noindent Fill in your <risk/benefit> numbers for the following:\\

\noindent Handguns: \_\_\_\_\_\_\_ \\
Motorcycles: \_\_\_\_\_\_\_\\
Lawnmowers: \_\_\_\_\_\_\_\\
<Wearable Technology>: \_\_\_\_\_\_\_\\
Electricity: \_\_\_\_\_\_\_\\ }
\end{quotation} 

\subsubsection{Additional Questions}
The exit portion of the survey contained demographic questions asking for age, gender, and education. We then asked about wearables ownership so we could control for prior exposure. An open-ended question on what would be the most likely risks associated with wearable devices was included to catch user concerns more broadly. To avoid biasing the open-ended question, we asked before concluding with the 10-question Internet Users' Information Privacy Concerns (IUIPC) index~\cite{malhotra2004internet}, which was included so we could control for participants' general privacy attitudes.

\subsection{Focus Group}
We conducted a one-hour focus group to validate our design, gauge comprehension, and measure fatigue. The focus group began with participants taking the survey then giving feedback on the format and the content, noting any instructions or questions that were unclear. The focus group concluded with a discussion of possible benefits and risks of wearable devices, in order to brainstorm any additional scenarios to include. Finally, we compensated participants with \$30 in cash. We recruited all of our focus group participants from Craigslist. Of the 13 participants, 54\% were female, and ages ranged from 18 to 64 ($\mu = 36.1$, $\sigma = 15.3$).  Education backgrounds ranged from high school to doctorate degrees, and professions included student, artist, marketer, and court psychologist.

\subsection{Recruitment and Analysis Method}
We recruited 2,250 participants over August 7--13, 2014 via Amazon's Mechanical Turk. We restricted participants to those over 18, living in the United States, and having a successful HIT completion rate of 95\% or above. We compensated each participant with \$1.75 upon successfully completing the survey. Based on incorrect responses to either of the two comprehension questions, we filtered out 366 (16\% of 2,250) participants. We filtered out an additional 99 participants (4\% of 2,250) due to incomplete responses, and three participants for being under 18, leaving us with a total sample size of 1,782. Of these, 57.9\% were male (1,031), 41.0\% were female (731), and 20 participants declined to state their genders. Ages ranged from 18 to 73, with a mean of 32.1 ($\sigma$ = 10.37). Almost half of our participants had completed a college degree or more (49.2\% of 1,782), which includes the 219 (12.3\% of 1,782) who reported graduate degrees. While our sample was younger and more educated than the U.S. population as a whole, we believe it is still consistent with the U.S. Internet-using population.

In performing our analysis in the next section, we chose to focus on the very upset rate (VUR) of each scenario.  The VUR is defined as the percentage of participants who reported a `5' on the Likert scales. 
We use the VURs rather than the average of all Likert scores for the same reasons as Felt {\it et al.}: the VUR does not presume that the ratings, ranging from ``indifferent'' to ``very upset,'' are linearly spaced. Additionally, most people are likely to be upset, at least a little, in all scenarios, because a device is taking action without permission (rating distribution: ``1''= 759, ``2'' = 918, ``3'' = 1,452, ``4''' = 2,421, ``5'' = 8,344). Thus, the distinguishing factor is whether a participant was maximally upset, rather than if they were upset. A limitation of this approach is that it only allows us to make {\it relative} comparisons between scenarios, rather than being able to definitively state how upset people might be if a single scenario were to occur.

We followed Fischhoff {\it et al.}'s methodology and did not normalize the numerical responses. Rather, we report medians and quartiles, which are not impacted by outliers. For the open-ended question at the end (i.e., additional privacy concerns), two researchers independently coded 1,782 responses, with an initial agreement rate of 89.7\%. The researchers discussed and resolved any disagreements so that the final codings reflect unanimous agreement.

%% file: tex-inputs/results-intro.tex
\section{Results}
In this section, we present participants' responses to the various data-sharing scenarios, and discuss how and which various factors contributed to their risk perceptions. We also discuss participants' risk/benefit assessment of various new technologies relative to well-established technologies. We conclude the section with participants' self-reported concerns about the biggest risks in owning wearable devices.

\subsection{Concern Factors}
Many factors can potentially impact participants' concern levels:  the data being shared, with whom the data is shared, and device in question. We analyze each factor individually, as well as present a statistical model of participants' concerns as a function of the above factors and additional demographic traits.

%% file: tex-inputs/results-type.tex

\subsubsection{Data Type}
\label{sec:datatypes}

Based on our statistical models (later reported in Section \ref{sec:regression}), we observe that the largest effect on participants' VURs stemmed from the data being shared; with whom the data is shared and which device the user is interacting with had weaker impacts on overall VURs. The most and least concerning data types are listed in Table \ref{top10-table}, and the full list can be seen in Table \ref{full-vur-table} in Appendix \ref{sec:concerns-appendix}. 

Participants were most concerned about photos and videos, especially if they contained embarrassing content, nudity, or financial information. As seen in Table \ref{top10-table}, photos and videos accounted for five of the top ten concerns, and are almost unanimously considered to be concerning. Information that could be used to impersonate someone (e.g., usernames/passwords for websites), or photos of someone at home, were also among the most concerning data types. 

Participants were least concerned about data that could be collected through observations of public behavior, such as demographics (e.g., age, gender, language) or information available to advertisers (e.g., TV shows watched, music on device). As seen in Table \ref{top10-table}, participants' responses had a greater amount of variance.  This greater variance and overall decreased concern may be because of uncertainty with how the data would be used, or because the financial, social, or physical consequences would be less immediate.

Although certain data is considered unanimously upsetting to have shared, it is interesting to note that no data was considered unanimously non-upsetting to have shared, nor were there any data types that evoked strong disagreement between participants (i.e., bimodal). Generally, the average concern magnitude was inversely correlated with the standard deviation, which suggests the presence of ceiling effects for the most concerning data types. For the complete ranked list of data types in this study, see Appendix \ref{sec:concerns-appendix}.

\begin{table*}[t]
\begin{center}
\small
\begin{tabular}{| r | l | r | r |c |}
\hline
Rank & Data & VUR & $\sigma$ & Distribution \\
\hline
1 & video of you unclothed & 95.97\% & 0.31 & \includegraphics[width = 2cm, height = 0.5cm]{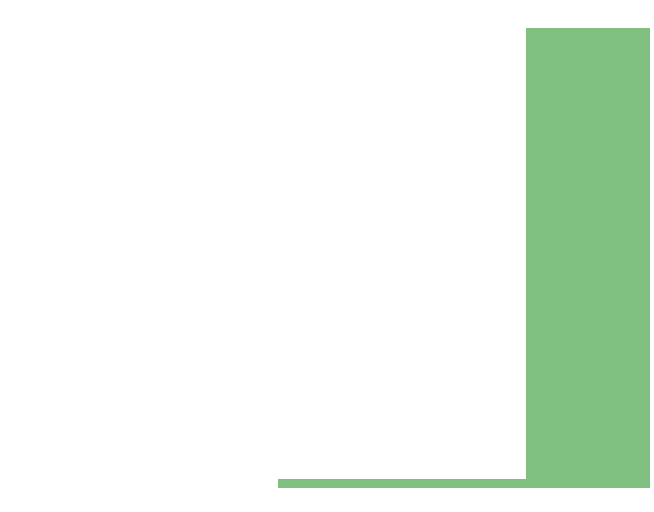} \\
2 & bank account information & 95.91\% & 0.35 & \includegraphics[width = 2cm, height = 0.5cm]{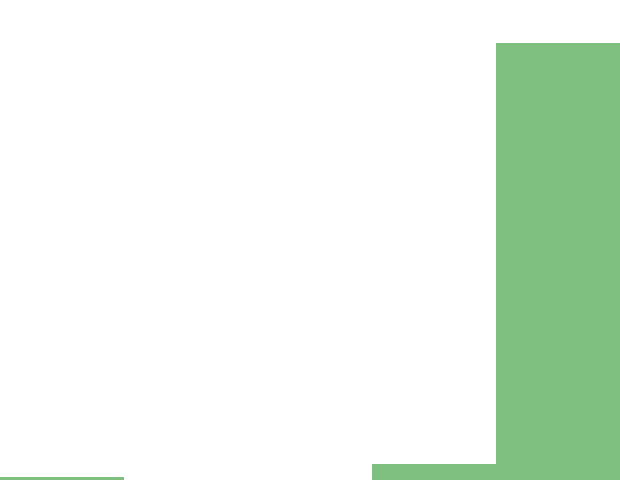}  \\
3 & social security number & 94.84\% & 0.26 & \includegraphics[width = 2cm, height = 0.5cm]{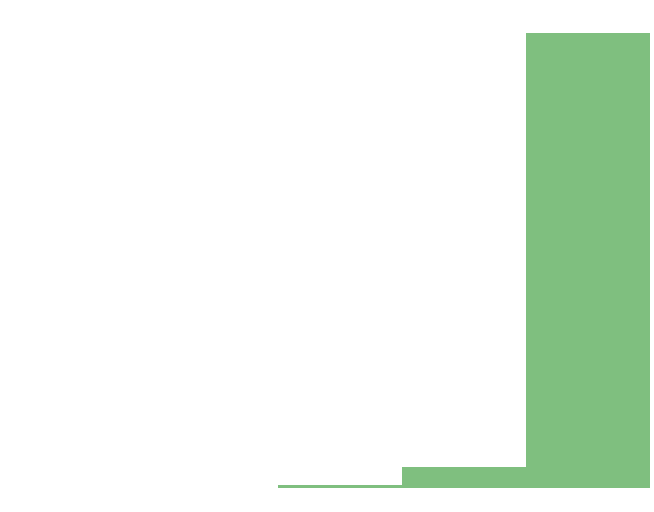}\\
4 & video entering in a PIN at an ATM & 92.67\% & 0.47 & \includegraphics[width = 2cm, height = 0.5cm]{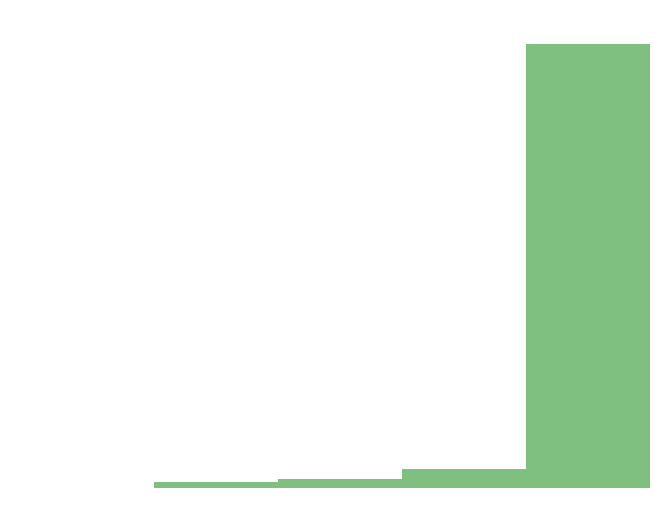}\\
5 & photo of you unclothed & 92.59\% & 0.46 & \includegraphics[width = 2cm, height = 0.5cm]{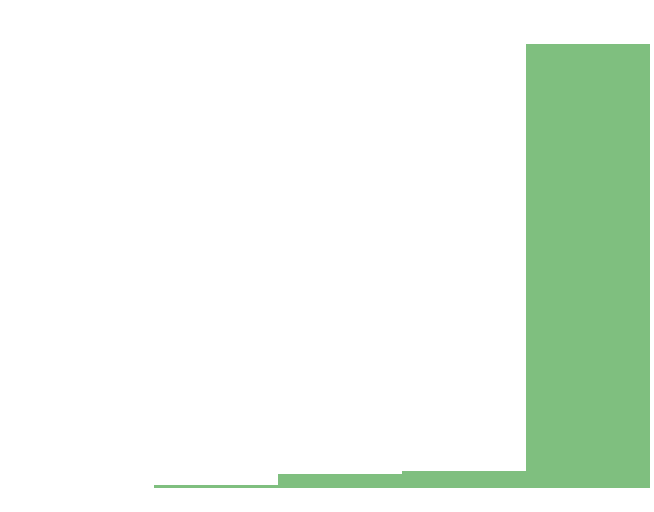}\\
6 & photo of you that is very embarrassing & 91.39\% & 0.55 & \includegraphics[width = 2cm, height = 0.5cm]{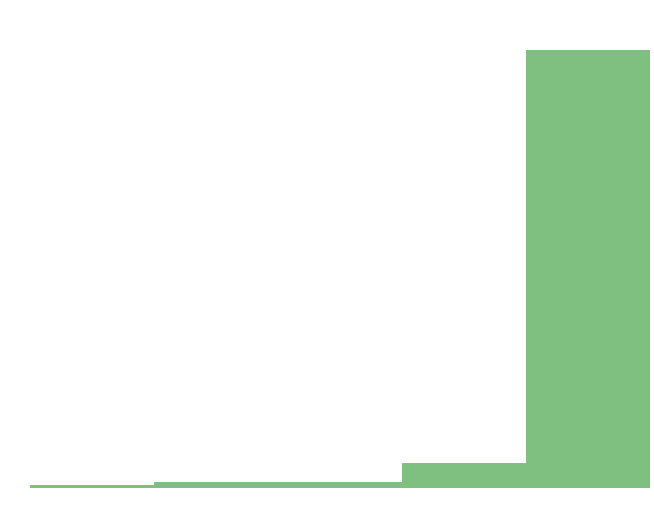}\\
7 & username and password for websites & 89.55\% & 0.62 & \includegraphics[width = 2cm, height = 0.5cm]{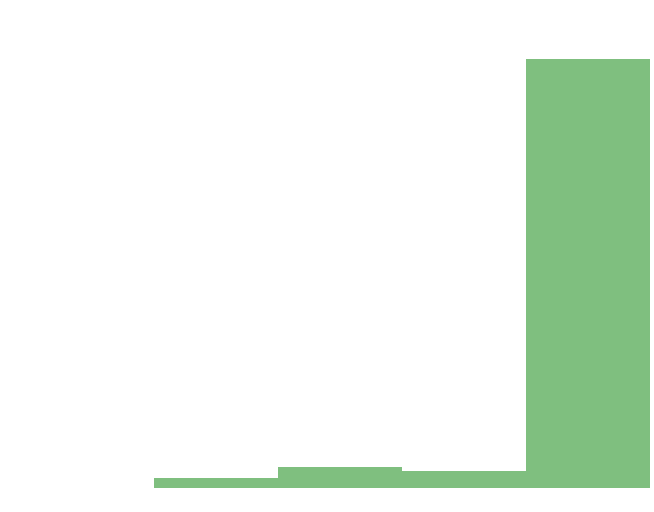}\\
8 & credit card information & 88.98\% & 0.56 & \includegraphics[width = 2cm, height = 0.5cm]{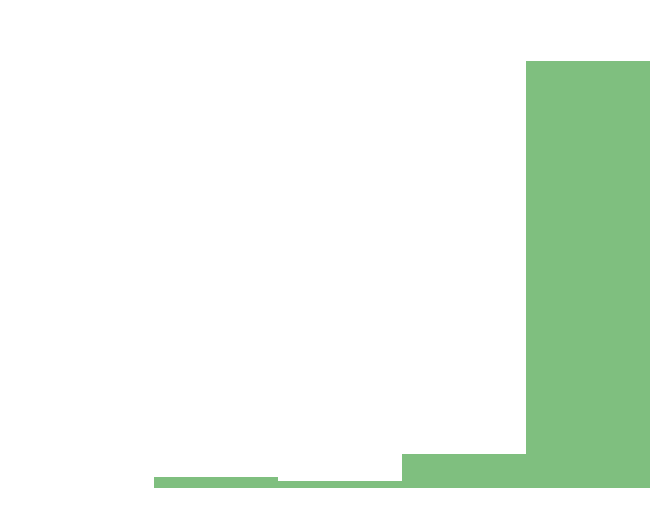}\\
9 & video of you that is very embarrassing & 88.41\% & 0.53 & \includegraphics[width = 2cm, height = 0.5cm]{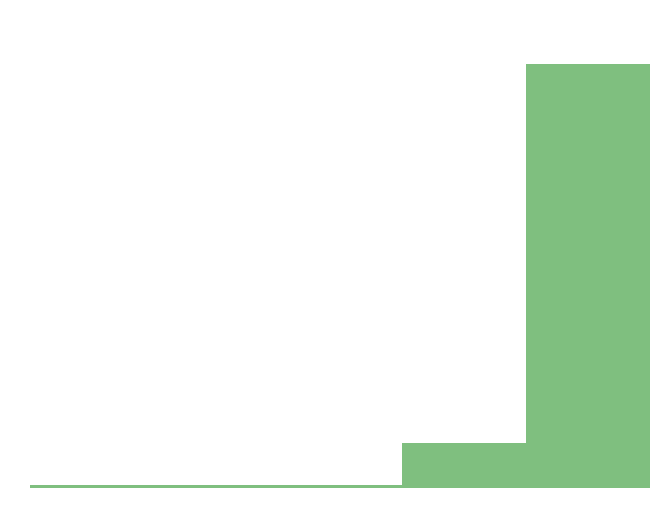}\\
10 & photo of you at home & 87.50\% & 0.60 & \includegraphics[width = 2cm, height = 0.5cm]{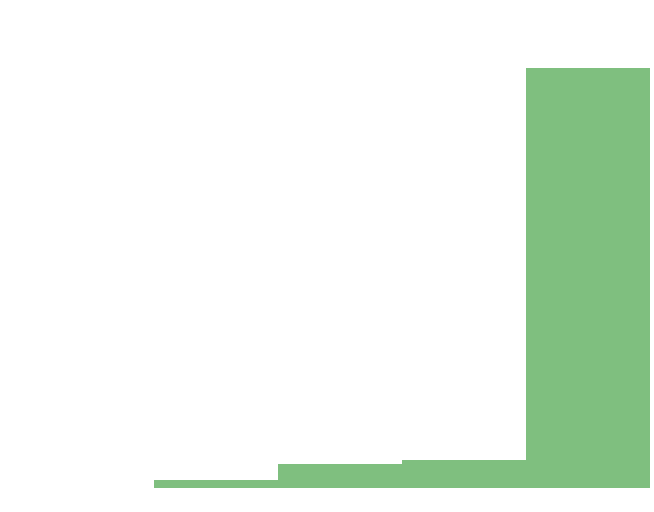}\\
 & \vdots & & & \\
64 & eye patterns (for eye tracking) & 40.51\%& 1.27 & \includegraphics[width = 2cm, height = 0.5cm]{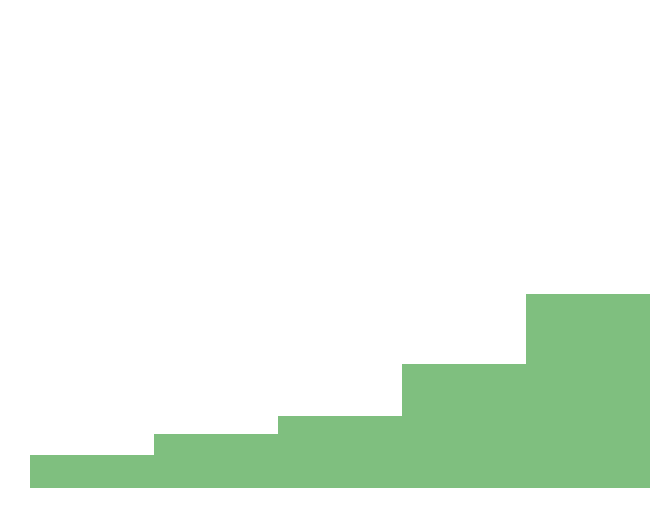} \\
65 & exercise patterns  & 38.66\% & 1.26 & \includegraphics[width = 2cm, height = 0.5cm]{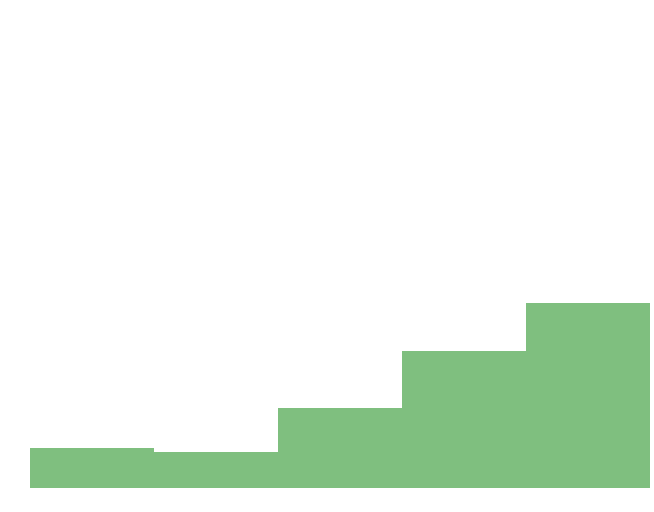}\\
66 & when you are happy or having fun  & 34.75\% & 1.27 & \includegraphics[width = 2cm, height = 0.5cm]{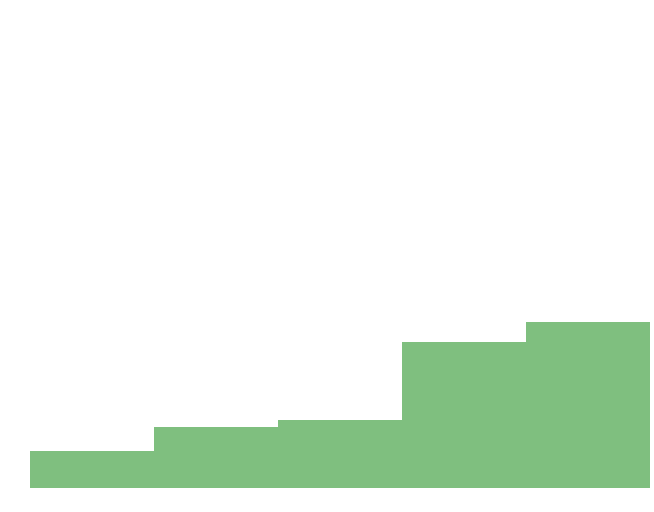}\\
67 & television shows watched & 30.20\% & 1.40 & \includegraphics[width = 2cm, height = 0.5cm]{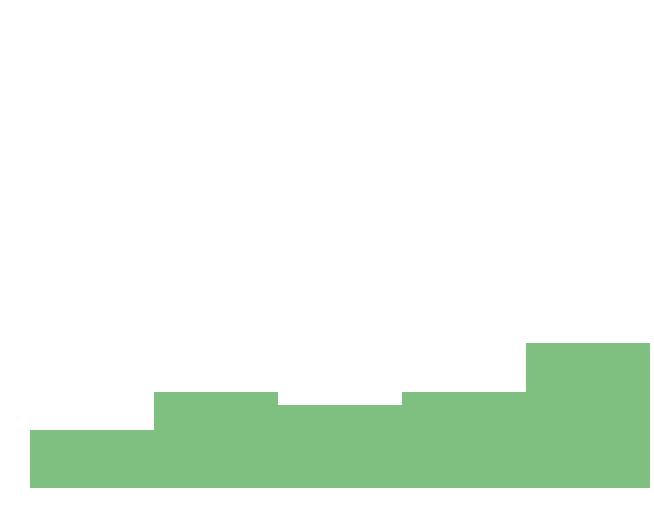}\\
68 & when you are busy or interruptible  & 29.50\% & 1.26 & \includegraphics[width = 2cm, height = 0.5cm]{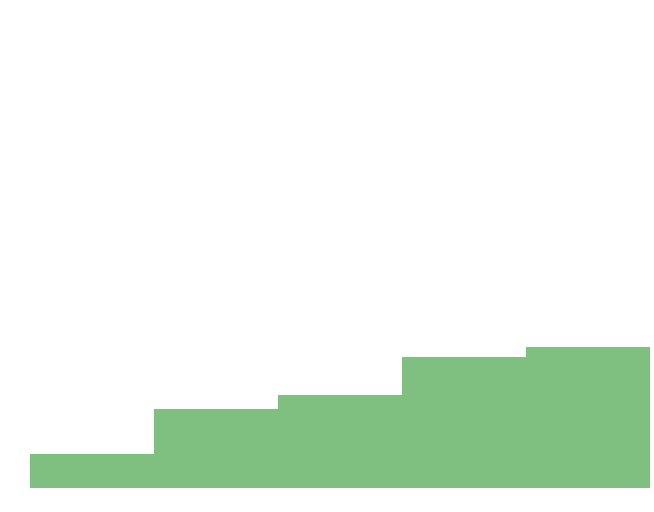}\\
69 & music on device  & 28.06\% & 1.43 & \includegraphics[width = 2cm, height = 0.5cm]{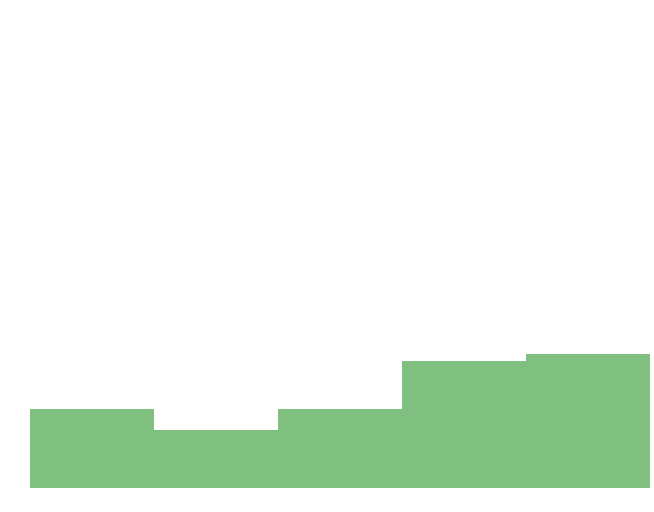}\\
70 & your heart rate & 27.50\% & 1.40 & \includegraphics[width = 2cm, height = 0.5cm]{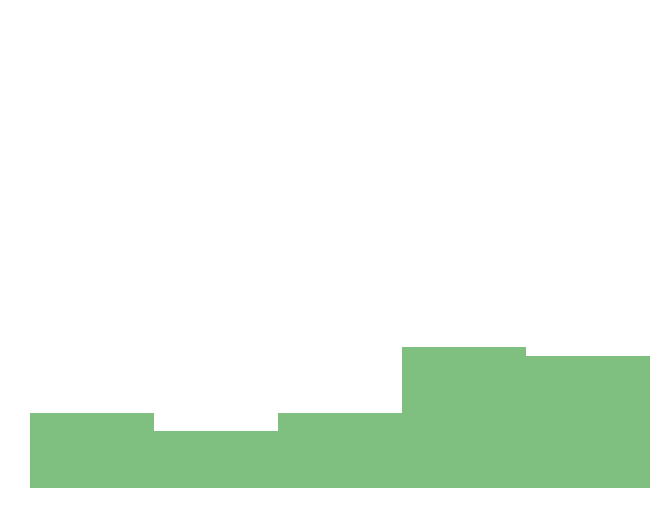} \\
71 & age & 24.29\% & 1.43 & \includegraphics[width = 2cm, height = 0.5cm]{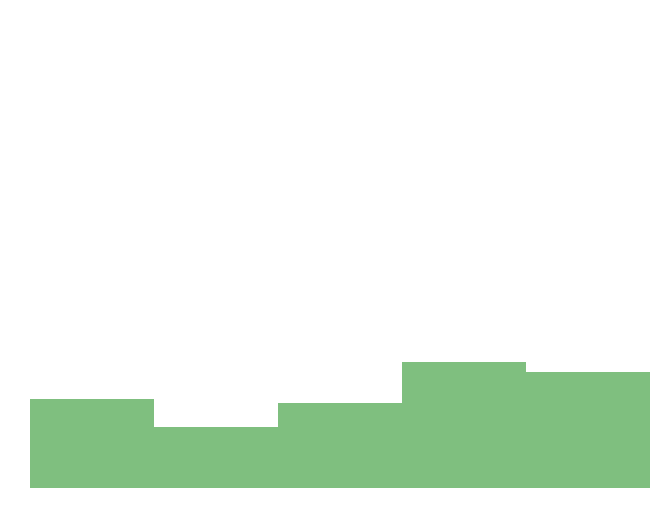}\\
72 & language spoken & 15.86\% & 1.49 & \includegraphics[width = 2cm, height = 0.5cm]{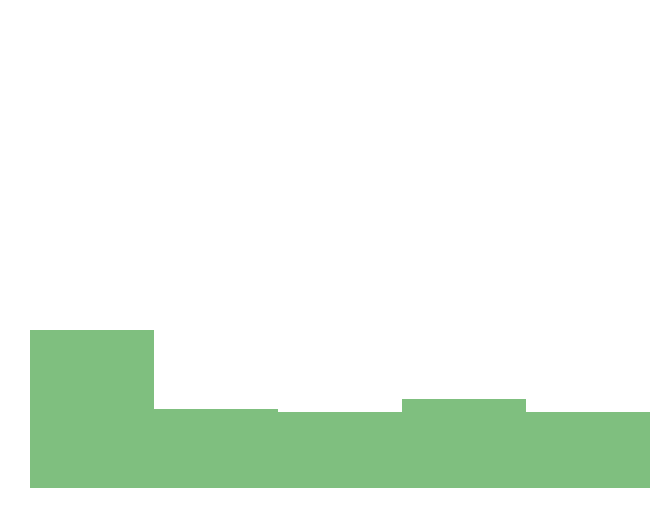}\\
73 & gender & 15.00\% & 1.45 & \includegraphics[width = 2cm, height = 0.5cm]{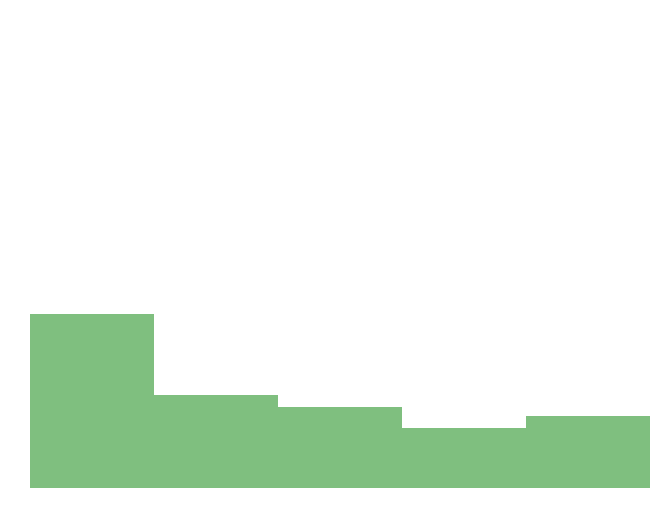}\\ 
\hline
\end{tabular}
\caption{The 10 most and least upsetting data types, across all recipients. For the complete list of all data types across all recipients, see Appendix \ref{sec:concerns-appendix}.}
\label{top10-table}
\end{center}
\end{table*}


%% file: tex-inputs/results-recipient.tex

\subsubsection{Data Recipient}
A statistically significant difference in VUR exists between data shared with an application versus data shared with human recipients. On average, 42\% of participants stated that they would be ``very upset'' if their data was shared with only an application's servers, whereas the VURs for friends (70\%), work contacts (75\%), and the public (72\%) were almost double (Table \ref{recipient-vur}). A chi-square test indicated that these differences were statistically significant (Table \ref{recipient-chi}). However, these effect sizes were small: the largest effect was between work contacts and an app's server ($\phi=0.11$); while the VUR for sharing with work contacts was significantly higher than sharing with friends, the effect size was negligible ($\phi=0.004$). 

We note that this chi-square test violates the assumption of independent observations, since participants responded to multiple scenarios with multiple recipients. But based on the randomization of treatments and large sample size, we do not believe that this significantly impacted our results. Similarly, we are unaware of a more appropriate test, given our data format. Cochran's Q requires binary outcomes (i.e., participants would have had to answer only one question for each data recipient, preventing us from adequately controlling for data type) and a repeated measures ANOVA requires normality (our data was not normally distributed). Nonetheless, we repeated our analysis using only one randomly-selected data point per participant and found that our selected test was robust to this violation. Therefore, we conclude that participants were significantly more concerned about having their data seen by a human versus an application, though differences between specific human groups such as the public, friends, and work contacts were not as significant. Our results motivate mechanisms for data taint tracking and accidental data sharing prevention.

However, we do not claim that there are no distinctions between the friends, public, and work contact recipients. People are more comfortable sharing certain data types with certain human recipients. For instance, participants were significantly uncomfortable sharing if they were lying, nervous, or stressed to work contacts compared to the rest of the data recipients. Table \ref{all-vur} Appendix \ref{sec:concerns-appendix} shows the  complete VUR and rankings of all data types by recipient.


\begin{table}[t]
\begin{center}
\small
\begin{tabular}{| r | l | r | l |c |}
\hline
Rank & Recipient & VUR & sigma & Distribution \\
\hline
1 & Work Contacts & 75.16\% & 0.94 & \includegraphics[width = 2cm, height = 0.5cm]{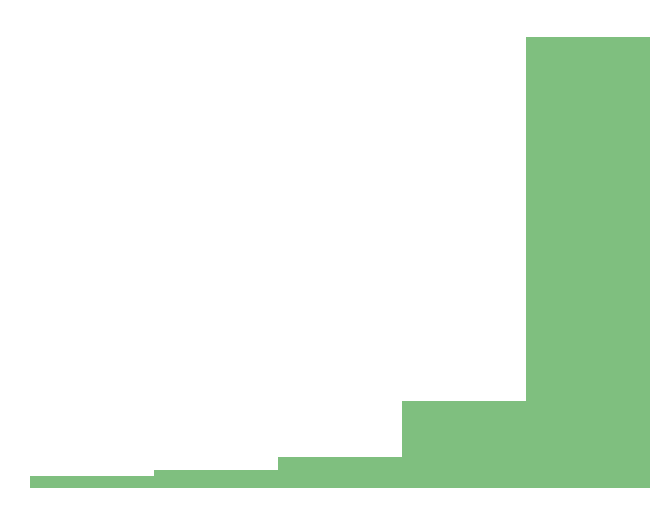} \\
2 & Public & 72.41\% & 0.98 & \includegraphics[width = 2cm, height = 0.5cm]{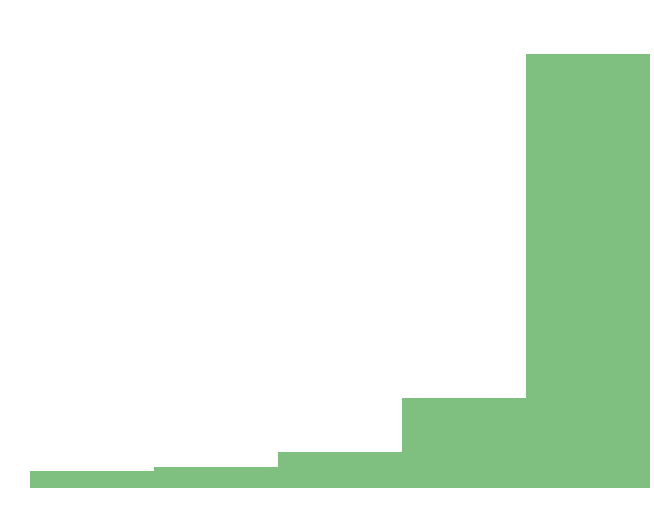}  \\
3 & Friends & 69.47\% & 1.02 & \includegraphics[width = 2cm, height = 0.5cm]{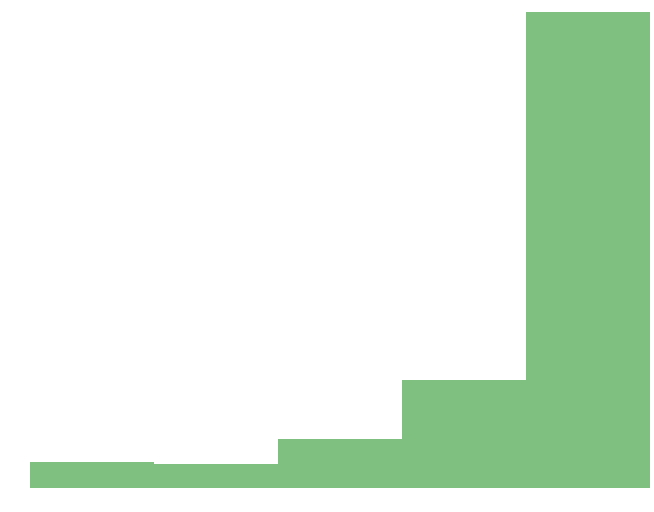}\\
4 & App's Server & 42.28\% & 1.15 & \includegraphics[width = 2cm, height = 0.5cm]{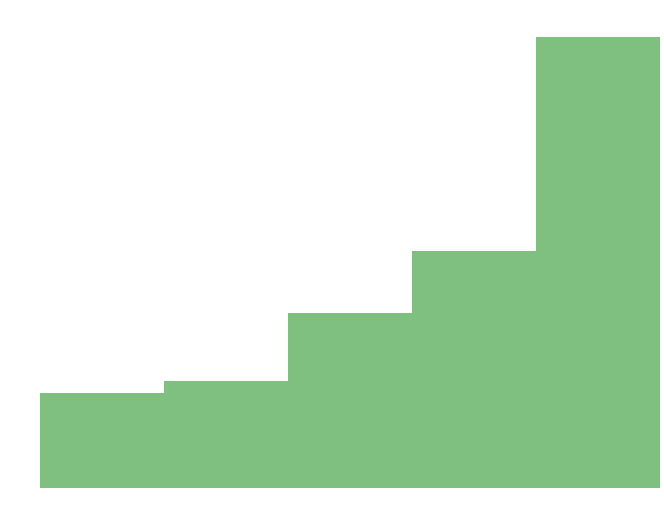}\\
\hline
\end{tabular}
\caption{The overall upset rate for all recipients.}
\label{recipient-vur}
\end{center}
\end{table}

\begin{table}[t]
\small
\begin{center}
\begin{tabular}{|l|r|r|r|r|}
\hline
Recipients	& $\chi^2$ & p-value 	& n & $\phi$ \\
\hline
Work-App	& 565.910 & <0.0001 & 5,083 & 0.111\\
Public-App	& 481.776 & <0.0001 & 5,1988& 0.093\\
Friends-App & 381.653 & <0.0001 & 5,096 & 0.075\\
Friends-Work & 20.39 & <0.0001 & 5,037 & 0.004\\
Friends-Public & 5.41 & <0.0200 & 5,142 & 0.001\\
Work-Public&  5.00 & <0.0253 & 5,129	& 0.001\\
\hline
\end{tabular}
\caption{Results of a chi-square test to examine VUR based on data recipient, across all data points.}
\label{recipient-chi}
\end{center}
\end{table}

%% file: tex-inputs/results-devices.tex

\begin{table}[t]
\small
\begin{center}
\begin{tabular}{| c | r | r |}
\hline
 Misbehavior &  Cubetastic3000 & Smartphone \\
 \hline
 \hline
 All & 58.79\% & 46.67\%\\
 \hline
Vibration & 14.81\%  &  6.14\%\\
Bluetooth & 44.12\%  &  19.86\%\\
Unmuted Call & 87.10\%  &  58.44\%\\
Screenshot & 52.78\%  & 55.74\%\\
Premium Calls/Texts & 86.49\%  &  91.94\%\\ 
\hline
\end{tabular}
\caption{VURs for the five questions about device misbehaviors described in Section \ref{sec:smartphones}, contrasting smartphones with the Cubetastic3000.}
\label{deviceVUR}
\end{center}
\end{table}

\subsubsection{Device}
Each participant answered two questions drawn from a set of five smartphone misbehaviors. To compare these misbehaviors with misbehaviors on the Cubetastic3000, we included these same five questions amongst the pool of 293 Cubetastic3000 scenarios, modifying the device in question. In this manner, all 1,782 participants received two smartphone questions, while 159 participants received at least one of these five Cubetastic3000 misbehavior questions. Across all participants, the VUR was 46.7\% (of 1,782) for smartphones, whereas the VUR increased to 58.8\% (of 159) for same misbehaviors on the Cubetastic3000. The VURs for both devices for these questions are in Table \ref{deviceVUR}.

To ensure independence of observations, we performed a Mann-Whitney U test to compare participants' average VURs for the Cuebtastic3000 scenarios (i.e., 159 participants) to the remaining participants' average VURs for the smartphone scenarios (i.e., 1,623 participants). This difference is statistically significant ($U=108,664.0$ with $p<0.0005$) but the effect size is very small ($r=0.08$). Because of this, we did not further reduce our statistical power by separately comparing the five misbehaviors. We conclude that users are likely more wary of misbehaviors on wearables than smartphones, but this difference is likely negligible. The entire effect may be due to participants' familiarity with smartphones, and therefore this effect may disappear as they increasingly encounter more wearables.




%% file: tex-inputs/results-techrank.tex

\subsection{Risk and Benefit Rankings} 
\begin{figure}[t]
	\centering
	\includegraphics[width=\columnwidth]{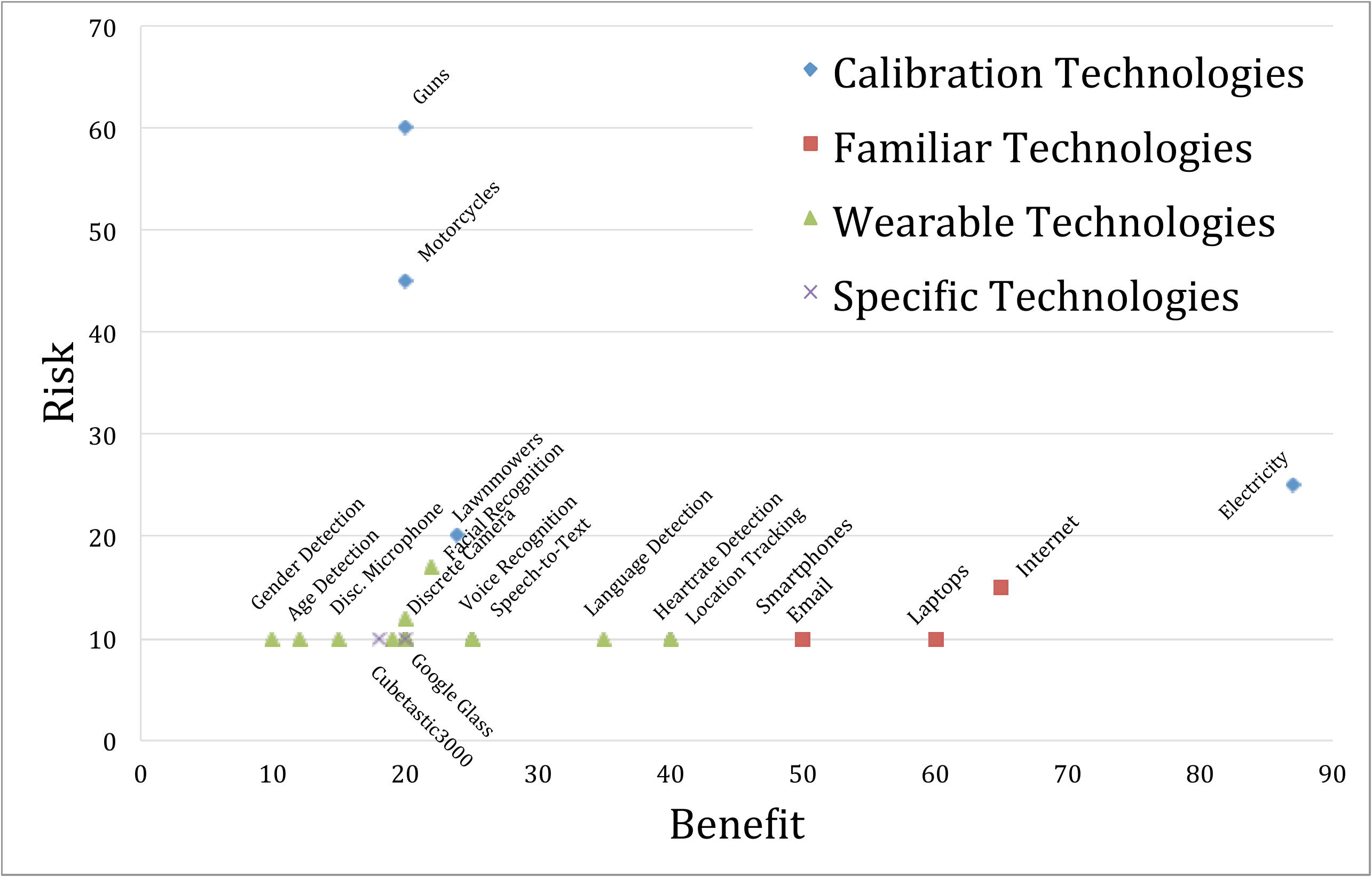}
	\caption{Participants' median risk-benefit ratings of technologies examined by Fischhoff \etal\cite{Fischhoff}, which we used for calibration, alongside familiar technologies (e.g., laptops, the Internet, etc.), wearable technologies, as well as two specific wearable devices (Google Glass and the Cubetastic3000).}
	\label{fig:techplot}
\end{figure}

We ask participants to rate new capabilities related to wearable technologies (e.g., facial recognition) in terms of their  benefits and risk. We also ask them about technologies with which they were likely to be more familiar (e.g., smartphones and laptops) and two specific wearable devices, Google Glass and the fictitious Cubetastic3000. To calibrate our results, we  ask about four well-established technologies studied by Fischhoff \etal\cite{Fischhoff}. We find that participants generally rated technologies related to wearables as being low-risk comparatively to other technologies (Figure \ref{fig:techplot}). Tables \ref{risk} and \ref{benefit} in Appendix ~\ref{sec:concerns-appendix} shows participants' median, quartiles, and distributions of risks and benefit ratings for all technologies. 

Participants were prompted to rate technologies with respect to all considerations (see Appendix~\ref{sec:prompt}), including risk of physical harm to bystanders, financial cost, distress, misuse, or impact on public, personal, and private life. Participants may have still evaluated the risks with an emphasis toward physical risk and without an emphasis on privacy risk. Among the five presented options, the wearable-related one is the only one without some physical risk scenario, and physical risk is a clear, tangible risk. 

We examined participants' perceptions, and therefore responses may not be reflective of actual risks or benefits. However, they also reflect the general public's exposure to these technologies and show that people perceive specific risks and benefits. We suspect that the similarity in assessments between the various wearable technologies are because most people are not consciously aware of the possibilities and that performing this experiment longitudinally may yield more interesting results, as these technologies become pervasive (and more familiar to participants).

As a group, participants rated more familiar technologies as more beneficial. We believe this is the result of exposure to these technologies---most people use these technologies daily and therefore see what the benefits of these technologies are. It is true that people perceive unfamiliar technologies as less beneficial at the moment, but this will change as the use of these technologies evolve and adoption increases. Most calibration technologies, with the exception of electricity, were seen as lower benefit than the others. However, Google glass and Cubetastic3000 were about equally beneficial, and gender and age recognition were less beneficial. 

We find that the more familiar a participant was with a technology, the more likely they were to rate it as risky.  Again, we believe that this is due to the exposure to these technologies. Of the wearable technologies, the riskiest technologies included facial recognition, the Internet, and discrete cameras, whereas the remainder of the technologies were seen as having minimal, equivalent risk levels (i.e., a median of ``10''). We did not test the differences in risk between the different wearable-related technologies for statistical significance, but given their minimal spread compared to the calibration options, the differences appears to be negligible. Interestingly, privacy risks were perceived as being comparable to physical risks; for instance, the capacity for facial detection on a wearable device was perceived as being almost as risky as interacting with a lawnmower. 

%% file: tex-inputs/results-openended.tex
\subsection{Open-Ended Concerns for Wearables}
We captured participants' reactions to wearable devices as a whole by asking the following open-ended question:

\begin{quotation}
\noindent
\textit{What do you think are the most likely risks associated with wearable devices?}
\end{quotation}

This question was asked with demographics questions, but before any IUIPC questions (which asked a lot of direct privacy-related questions, to avoid biasing the recipients). The participants were presented with a blank box to write in, with no character limit to their open-ended responses. 

Table \ref{openresponses} shows common user concerns related to wearable devices. Appendix \ref{sec:coding} details the responses categorized in each coding label. Most are related to privacy and security, but this open-ended data gives a sense of what broad categories of concerns are most relevant to users. This can be used to guide research in unexplored use cases. 

In addition to privacy and security in the general sense, significant concerns included being unaware of what the device is collecting, doing, or which information it is using (Being Unaware). Other orthogonal concerns included long-term health effects caused from wearing the device such as cancer from EMF waves (Health) and safety hazards from wearing the device, such as distractions that cause car accidents (Safety). Interestingly, participants were also concerned with resulting changes in social behaviors, such as dependence on devices or spending less time with loved ones (Social Impact). 

\begin{table}[t]
\begin{center}
\begin{tabular}{|l|r|r|}
\hline
Concern &  Responses &  Frequency   \\
\hline
Privacy & 452 & 25.32\% \\
Being Unaware & 275 & 15.40\% \\
Health Risk & 191 & 10.70\%\\
Safety & 185 & 10.42\%\\
Social Impact &	157 & 8.80\%\\
Financial Cost & 151 & 8.46\%\\
Security &	144 & 8.07\%\\
Accidental Sharing &	69 & 3.87\%\\
Miscellaneous &	57 & 3.19\%\\
None	& 51 & 2.86\%\\
Social Stigma &	39 & 2.18\%\\
False Information & 33 & 1.85\%\\
Don't know & 31 & 1.74\%\\
Aesthetics 	& 19 & 1.06\%\\
Don't care 	& 11 & 0.62\%\\
\hline
\end{tabular}
\caption{The most common open-ended risks associated with owning a wearable device.}
\label{openresponses}
\end{center}
\end{table}

%% file: tex-inputs/results-demographics.tex
\subsubsection{Demographic Factors}

We see that privacy is a main concern for wearables users. Additionally, we show that privacy preferences should also be a consideration for configuring a user's device. A participant's self-reported level of privacy concerns is the biggest demographic predictor of participants' VUR rate. This is determined by the IUIPC scale~\cite{malhotra2004internet}. A Spearman correlation yielded a statistically significant effect between average IUIPC scores and VUR ($\rho=0.446$, $p<0.0005$), which suggests responses to questions were mostly based on privacy preferences. Additionally, we observed that age was another significant predictor of VUR ($\rho=0.121$, $p<0.0005$), but we suspect that this effect is due to the significant correlation between age and IUIPC scores ($\rho=0.188$, $p<0.0005$). Others have observed that older individuals tend to be more privacy protective~\cite{varian2005demographics}.

While we initially observed an effect on VURs based on whether or not participants claimed to already own wearables (57.0\% vs. 60.8\%, respectively; Mann-Whitney $U=202,896$, $p<0.032$), this difference did not remain significant upon correcting for multiple testing (Bonferroni corrected $\alpha=0.01$). The effect of a participant's gender also did not remain significant upon correcting for multiple testing. We observed no correlation between a participant's education level and VUR.

%% file: tex-inputs/results-regression.tex

\subsubsection{Regression Models} 
\label{sec:regression}
In order to examine the relative effect of each factor on participants' VURs, we constructed several statistical models to predict whether a participant would be ``very upset'' with a given scenario based on the data type, data recipient, and their demographic factors (i.e., age, education, gender, and privacy attitudes). We performed binary logistic regressions using generalized estimating equations, which account for our repeated measures experimental design (i.e., each participant contributed multiple data points).

\begin{table}[t]
\centering
\begin{tabular}{|l| r| r| r|}
\hline
Parameters & $\chi^2$ & $df$ & QIC\\
\hline
\hline
(Intercept) & 423.96 & 1 & 13,209.1\\
\hline
(Intercept) & 207.07 & 1 & 12,551.49\\
IUIPC (covariate) & 368.5 & 1 & \\
Gender (covariate) & 6.30 & 1 & \\
\hline
(Intercept) & 411.66 & 1 &12,458.86\\
Data Recipient & 599.72 & 3 & \\
\hline
(Intercept) & 418.02 & 1 & 11,382.75\\
Data Type & 1,141.40 & 71 & \\
\hline
(Intercept) & 66.18 & 1 & 9,609.65 \\
Data Recipient & 617.25 & 3 & \\
Data Type & 1,288.51 & 71 & \\
IUIPC (covariate) & 105.73 & 1 & \\
Gender (covariate) & 9.74 & 1 & \\
IUIPC $\times$ Gender & 8.33 & 1 &\\
\hline
\end{tabular}
\caption{Goodness-of-fit metrics for various binary logistic models of our data using general estimating equations to account for repeated measures. The columns represent the Wald test statistic for each parameter and the overall Quasi-Akaike Information Criterion (QIC) for each model. Each parameter listed was statistically significant at $p<0.005$.}
\label{regression}
\end{table}

We created several models using two independent variables as predictors: data and recipient. Because the device (i.e., whether they were using the Cubetastic3000 or a smartphone) is only varied for the five smartphone misbehaviors listed in Section \ref{sec:smartphones}, we removed these five  from our models, resulting in a total of 72 types of data shared with 4 possible recipients. We used the following demographic factors as covariates: age, gender, education, wearable device ownership (yes/no), and mean IUIPC score. For each model, we performed Wald's test to examine the model effects attributable to each of these parameters. The only covariates that had an observable effect on our models were participants' gender and IUIPC scores, which also exhibited an interaction effect with each other. Thus, we opted to remove the other covariates from our analysis. Table \ref{regression} shows the various models that we examined and the Quasi-Akaike Information Criterion (QIC), which is a goodness-of-fit metric for model selection that also accounts for complexity (lower relative values indicate better fit). As shown, the type of data being shared (data type) was found to be the strongest predictor of a high VUR.

While these models illustrate the relative weights that users place on information when determining a scenario as truly upsetting, one shortcoming of this approach is its generalizability: the data predictor is categorical and limited to the data that we specifically chose for this study. To make our data set more generalizable to other use cases, we coded each data type in two ways: in terms of broad descriptions of the type of data (e.g., video, audio, etc.) and the type of risk it presents. Two researchers agreed on a codebook and independently coded each of the 72 data types.

The data types fell into these six categories:

\begin{enumerate}[topsep=0pt,itemsep=-1ex,partopsep=1ex,parsep=1ex]
\item Photo
\item Video
\item Audio
\item Behavioral Information
\item Biometric Information
\item Demographic Information
\end{enumerate}

While the first three categories are self-explanatory, the latter three categories are all based on different user characteristics. We defined {\it behavioral information} as observations about the user's activities; {\it biometric information} as measurements of the user's body; and {\it demographic information} as non-biometric information about the user's traits. 

The risks for data types fell into these five categories:

\begin{enumerate}[topsep=0pt,itemsep=-1ex,partopsep=1ex,parsep=1ex]
\item {\bf Financial:} the loss of money or property.
\item {\bf Image:} the loss of control over one's self-image (e.g., publicizing something embarrassing).
\item {\bf Medical:} the disclosure of medical information.
\item {\bf Physical:} physical harm to the user.
\item {\bf Relationships:} damage to the user's inter-personal relationships.
\end{enumerate}

After independently coding, the researchers met to resolve any disagreements, such that the results reflect unanimity. There was  83\% agreement prior to resolution. Cohen's $\kappa$ was 0.81 for the data categories and 0.75 for the  risk categories, both indicating ``excellent'' agreement~\cite{Fleiss2003}. 

With regard to data types, the most concerning type of data is video (78.0\%), which was ranked similarly to photos (76.2\%). Next are audio (66.8\%) and demographic data (65.4\%), followed by behavioral (53.1\%) and biometric (46.3\%) data. We suspect that demographic data was more concerning because it included information such as a Social Security Number, bank account information, and other financial information. We chose to categorize them as such as they are non-biological descriptors of the user. We were very surprised that biometric information was seen as relatively benign compared to the other broad categories of data. One hypothesis is that since most home users do not use biometric authentication, they may have an inaccurate understanding of the types of systems that might be at risk if biometric data is stolen and abused.

With regard to the presented risks, we observed that average VURs were highest for financial information disclosure (82.0\%). Information regarding relationships (69.2\%), physical safety (66.4\%), and self-image (65.8\%) followed. VURs were lowest for medical information disclosure (47.4\%). One reason why medical risks were ranked relatively low is that this category broadly covered scenarios involving data about the user's health, but also included more basic medical information, such as age, gender, and emotional state. As mentioned in Section \ref{sec:datatypes}, participants saw these as publicly observable and unconcerning.

\begin{table}[t]
\centering
\begin{tabular}{|l| r| r| r|}
\hline
Parameters & $\chi^2$ & $df$ & QIC\\
\hline
\hline
(Intercept) & 442.66 & 1 & 12,727.42\\
Risk & 405.18 & 4 & \\
\hline
(Intercept) & 380.39 & 1 & 12,681.86\\
Data Category & 439.45 & 5 & \\
\hline
(Intercept) & 256.15 & 1 & 12,061.87\\
Risk & 157.84 & 4 & \\
Data Category & 183.90 & 5 & \\
Risk $\times$ Data Category & 259.81 & 8 & \\
\hline
(Intercept) & 62.65 & 1 & 10,406.35\\
Risk & 205.21 & 4 & \\
Data Category & 250.35 & 5 & \\
Recipient & 546.89 & 3 & \\
IUIPC (covariate) & 103.94 & 1 & \\
Gender (covariate) & 9.80 & 1 & \\
IUIPC $\times$ Gender & 8.21 & 1 & \\
Risk $\times$ Data Category & 303.44 & 8 & \\
Recipient $\times$ Risk & 39.14 & 12 & \\
\hline
\end{tabular}
\caption{Metrics for additional binary logistic models of our data using general estimating equations to account for repeated measures. The columns represent the Wald test statistic for each parameter and the overall Quasi-Akaike Information Criterion (QIC) for each model. Each parameter listed was statistically significant at $p<0.005$.}
\label{regression2}
\end{table}

Using these two new variables as additional independent variables (and removing the previous data type variable), we created a second set of models. Because these risk categories and mediums are less likely to change over time, models that take these into account are likely to be more useful and less likely to be overfit. What these models show us is that both risk and medium are relatively strong predictors by themselves, and have an even stronger interaction effect. When the data recipient and covariates are added to the model, the resulting goodness-of-fit is not much worse than that of the model using the actual data type. 

%% file: tex-inputs/discussion.tex

\section{Discussion}

\subsection{Limitations}
One of the limitations of our experiment is that our participants might not have knowledge or interest in wearables and their capabilities; 83\% of our participants reported that they do not own a wearable. Because of this, our participants may be over or underestimating the risk, stemming form an unawareness of what can be inferred from the data, not having clear relations of new technology with respect to familiar ones, and a higher likelihood of being influenced by reports of recent events.\footnote{Recently, stories of exploding batteries were in the news~\cite{1_levin_2014}, which were explicitly reported as a concern in our open-ended question.}  For instance, biometrics were generally not a concern for our participants, although there are many security and privacy implications~\cite{prabhakar2003biometric}. Our participants also did not differentiate between the benefits of risks of various new capabilities.

We recruited both wearable users and non-users in order to yield a more representative sample of the general population. We could have easily recruited only wearables owners or people specifically interested in wearables. However, that would have its own biases and limitations. At the time of this writing, about 85\% of the general population do not own wearable devices~\cite{Nilsen,WearableStatNews}, indicating our study is reflective of the current population. 

Because of the privacy paradox, participants' stated responses may differ from how they may react to these same scenarios in real life~\cite{norberg2007privacy, jensen2005privacy}. At the same time, our results do reflect actual perceptions of wearable devices and the associated privacy scenarios involving them. This is an unavoidable, yet important distinction to make with of studies of this nature: our primary goal was to examine perceptions and preferences, so that future systems can be designed with these in mind. We do not expect that such systems will satisfy users in all situations, however, we believe that user-centered design will still be a vast improvement over post hoc approaches (or ignoring user concerns altogether).

\subsection{Future Research Directions}
We find that although people have opinions on applications which are familiar, users do not know the risk associated with new data or unfamiliar applications. We hope our work both informs the direction for future research to secure video, audio, and other currently considered sensitive sensor input channels, but also encourage work for contextual and user-input- independent permission models and access control schemes.

Further work can be done to expand various aspects of this study. Investigating more fine-grained data types (e.g., investigating specific instances of location data, versus location data in general) would be a useful endeavor to gain further insight into user perceptions. Adding additional recipients, such as ``advertisers'' or ``acquaintances'' may lead to more nuanced results. Additionally, the open-ended concerns illuminate areas of possible future research, such as the design of a distraction-free interface to prevent safety issues, and how to minimize negative social impact. 

%% file: tex-inputs/related_work.tex

\section{Additional Related Work}

\subsection{Wearables and Privacy}
We are rapidly moving towards a world of ubiquitous sensing and data capture, with ensuing privacy challenges~\cite{abowd2000charting,palen2003unpacking,camp2000internet}. Roesner {\it et al.}\ urge the community to address potential concerns for wearable devices before the technologies become widespread~\cite{roesner2014security} and explore the unique and difficult problems these devices present in terms for law and policy~\cite{roesner2014augmented}.

Many researchers have worked to study how privacy can be preserved in the presence of ubiquitous devices. Examples of such efforts include frameworks to design for privacy~\cite{bellotti1993design,camp2003designing,langheinrich2001privacy}, protocols for anonymous communication~\cite{cornelius2008anonysense}, evaluation metrics for privacy~\cite{scholtz2004toward}, and privacy models~\cite{hong2004privacy, jiang2002approximate}. Our work aims to augment works like these with an understanding of what privacy means to the end user. 

\subsection{Lessons form Smartphones}
Research shows that perceptions of risk change based on the particular device being used. Chin {\it et al.}\ found that users' attitudes toward security and privacy for smarphones significantly differed from attitudes towards traditional computing systems, stemming from differences in how people used these systems ~\cite{chin2012measuring}. Undoubtedly, this will hold true for wearables as well as wearables have a sparse, world-driven interaction model. 





Tsai {\it et al.}\ found that when mobile users get feedback about releasing data, such as who has viewed location information, it greatly impacts future behaviors~\cite{Tsai2009}. Although this type of feedback is not provided to smartphone users, there is potential for impacting wearables users and shaping their behaviors so that they keep users safe.

\subsection{User Perceptions}
While risk communication for the physical world has been examined for several decades (e.g.,~\cite{Fischhoff,Morgan2001}), research into effectively communicating computer-based risks has only recently been researched. For example, both Garg {\it et al.}\ and Blythe {\it et al.}\ show that due to varying perceptions and abilities that correlate with demographic factors, computer-based risk communication should employ some degree of demographic targeting~\cite{Garg2012,Blythe2011}. While this work is likely applicable to wearable computing risk communication, we believe that a better understanding of users' risk perceptions in this domain is warranted, prior to examining risk communication.

One limitation of user perceptions is that people do not always have enough information to make privacy-sensitive decisions. Even if users did have this information, it has been shown that users often trade off long-term privacy for short-term benefits~\cite{acquisti2005privacy}. Furthermore, actual behavior may deviate from stated privacy preferences~\cite{spiekermann2001privacy}. However, understanding user concerns is a necessary first step not only for risk communication, but preventative measures in general against breaches of privacy and security in a new threat landscape. 

%% file: tex-inputs/conclusion.tex

\section{Conclusion}

 Our survey of 1,784 Internet users, is the first large-scale study to investigate user-centric security and privacy concerns for wearable devices. We contribute a comprehensive ranking of possible risks associated with wearable devices, across various recipients. We calibrate our results with mobile devices and existing technologies; additionally, we verify with an open-ended question and find that privacy and security are at the top of user's overall concerns. Wearables are still in their infancy. Perceptions of situations and capabilities will change rapidly with advancements and increased exposure. However, there is not much work done to determine which threats in the emerging threat landscape are pertinent to focus on. Inspection of possible data concerns agree with previous studies of smartphone studies to find that video capture and financial data to be most sensitive. Various systems which detect and take actions for sensitive objects in photos and videos will be critical as wearables and other devices become more ubiquitous. We also find that users' self-reported privacy preferences are correlated with how participants may react, even with respect to situations that they are unfamiliar with. Permissions and access control mechanisms which do not depend on user inputs can still benefit from being informed by user preferences. We hope that this work has given a comprehensive overview of user concerns and inform designs of future privacy and security work for wearable devices. 

%% file: tex-inputs/appendixes.tex
\appendix

\section{Fischhoff Prompts}
\label{sec:prompt} 

\textit{We would like to ask you to rate the <risks/benefits> associated with each of the following technologies.}

{\bf Risks:} \textit{Consider all types of risks: the risk of physical harm or death, the risk to others or bystanders, the financial cost of the technology, any distress caused by the technology, what the consequences would be if the technology was misused, any impact on the public, work, or personal life, and other considerations. (e.g. for electricity, consider the risk of electrocution, the pollution caused by coal, the risk that miners need to take to mine the coal, the cost to build the infrastructure to deliver electricity, etc.) Give a global estimate over a long period of time (say, a year) of both intangible and tangible risks.} 

\textit{Do not consider the costs or risks associated with these items. It is true, for example, that sometimes swimmers can drown. But evaluating such risks is not your present job. Your job is to assess the gross benefits, not the net benefits which remain after the costs and risks are subtracted out.}

\textit{Please rate the following technologies below with a number. We know that this might be a bit hard to do, but please try to be as accurate as possible, adjusting the numbers until they feel they are right for you. Start with the least risky technology at 10 and assign higher numbers for the more risky technologies. (For instance, a technology rated 14 is half as risky as a technology rated 28.)}

{\bf Benefits:} \textit{Consider all types of benefits: how many jobs are created, how much money is generated directly or indirectly, how much enjoyment is brought to people, how much a contribution is made to the people's health and welfare, what this technology promotes, and so on. (e.g. for swimming, consider the manufacture and sale of swimsuits, the time spent exercising, the social interactions during swimming, and the sport created around the activity.) Give a global estimate over a long period of time (say, a year) of both intangible and tangible benefits.}

\textit{Do not consider the costs or benefits associated with these items. It is true, for example, that electricity also creates a market for home appliances. But evaluating such benefits is not your present job. Your job is to assess the gross risks, not the net risks which remain after the costs and risks are subtracted out.} 

\textit{Please rate the following technologies below with a number. We know that this might be a bit hard to do, but please try to be as accurate as possible, adjusting the numbers until they feel they are right for you. Start with the least beneficial technology at 10 and assign higher numbers for the more beneficial technologies. (For instance, a technology rated 34 is twice as beneficial as a technology rated 17.)}

\section{Coding Label Definitions}
\label{sec:coding}
Researchers coded the self reported answers as follows:\\
{\bf Privacy}: ``privacy,'' mention of personal details, spying. \\
{\bf Security}:  ``security,'' mention of malware, hacking. \\
{\bf GPS tracking}: ``location,'' ``GPS,'' mention of monitoring. \\
{\bf Being Unaware}: mention of using, collecting, and disclosing data without permission. \\
{\bf False information}: inaccurate or maliciously false data.\\
{\bf Health Risk}: mention of radiation, cancer, or other effects.\\
{\bf Safety}: mention of distractions causing car crashes and injuries, violence due to the device, injuries from malfunctions.\\
{\bf Discomfort}: mention of eye strain, headache, irritation. \\
{\bf Financial cost}: cost of buying or using the device. \\
{\bf Theft}: mention of device theft. \\
{\bf Social Impact}: mention of dependency, distance from people, changes in decision making, etc. \\
{\bf Social Stigma}: mention of judgment, hate, or bystanders.\\ 
{\bf Aesthetics}: mention of fashion or looking dorky. \\
{\bf Miscellaneous}: odd comments, uncommon concerns. \\
{\bf None}: ``None,'' mention of no threat, or no real concerns \\
{\bf Don't know}: ``do not know,'' general confusion \\
{\bf Don't care}: `` do not care,'' nonchalant answers

%% file: tex-inputs/collapsed.tex

\section{Concern Factors}
\label{sec:concerns-appendix} 
We show the full, fine-grained results of our survey in this appendix. This includes how participants had ranked each technology in response to the Fischhoff-style questions, the VUR rates for all seventy-two data types, across all recipients and by recipient, and the details of the full regression models used in our analyses. 

\begin{table*}[t]
\begin{center}
\small
\begin{tabular}{| r | l | r | r | r | r |}
\hline
Rank & Question & VUR & $\sigma$ & Distribution \\
\hline
1 & video of you unclothed & 95.97 & 0.31 &  \includegraphics[width = 2cm, height = 0.5cm]{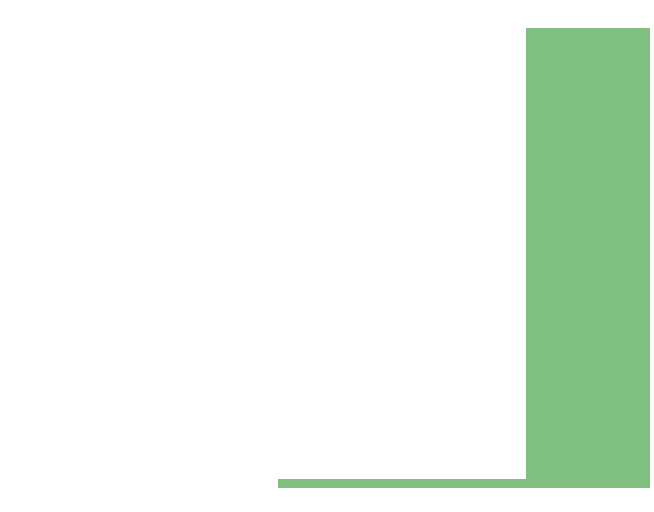} \\ 
2 & bank account information & 95.91 & 0.35 & \includegraphics[width = 2cm, height = 0.5cm]{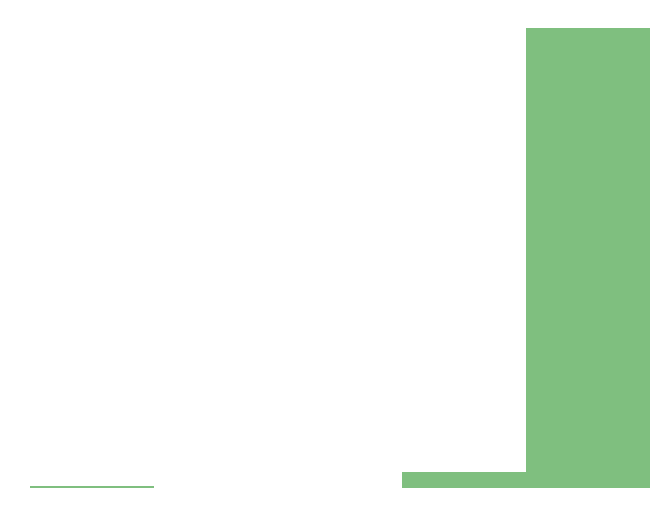} \\ 
3 & social security number & 94.84 & 0.26 & \includegraphics[width = 2cm, height = 0.5cm]{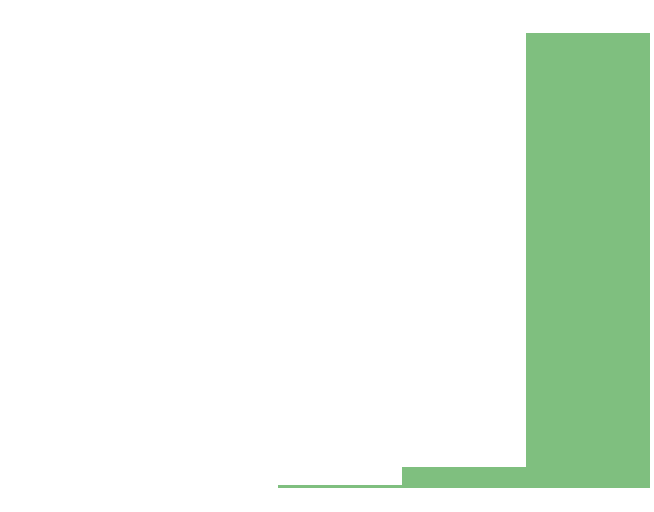} \\ 
4 & video entering in a PIN at an ATM & 92.67 & 0.48 &  \includegraphics[width = 2cm, height = 0.5cm]{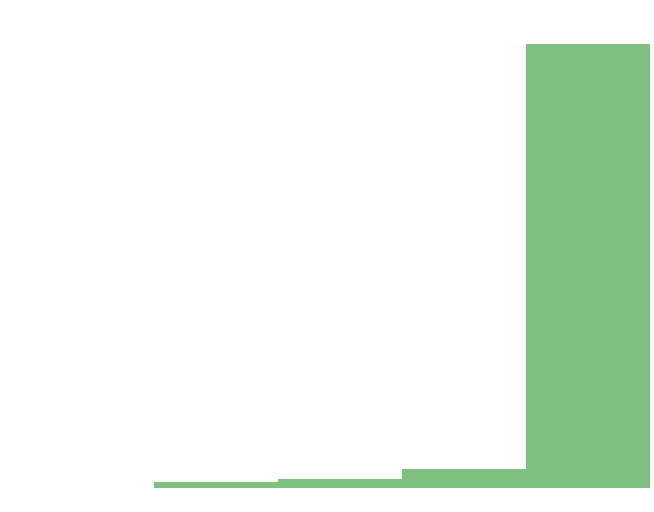} \\ 
5 & photo of you unclothed & 92.59 & 0.45 &  \includegraphics[width = 2cm, height = 0.5cm]{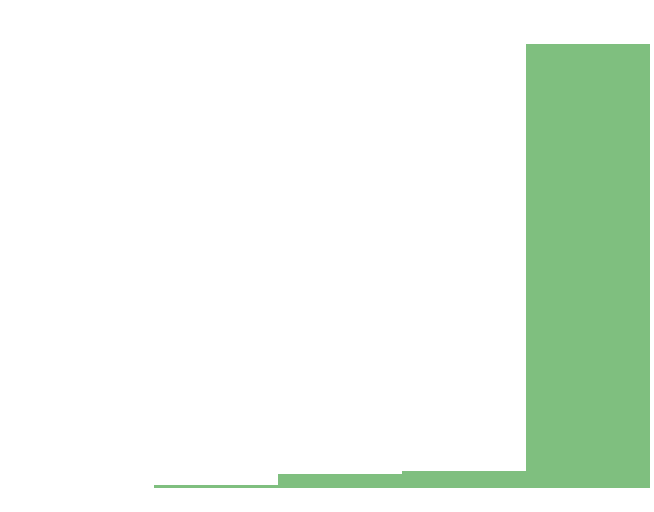} \\ 
6 & photo of you that is very embarrassing & 91.39 & 0.56 &  \includegraphics[width = 2cm, height = 0.5cm]{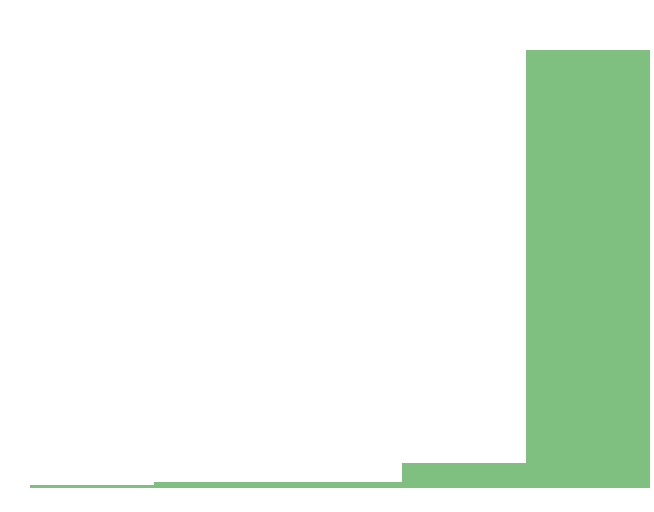} \\ 
7 & username and password for websites & 89.55 & 0.62 & \includegraphics[width = 2cm, height = 0.5cm]{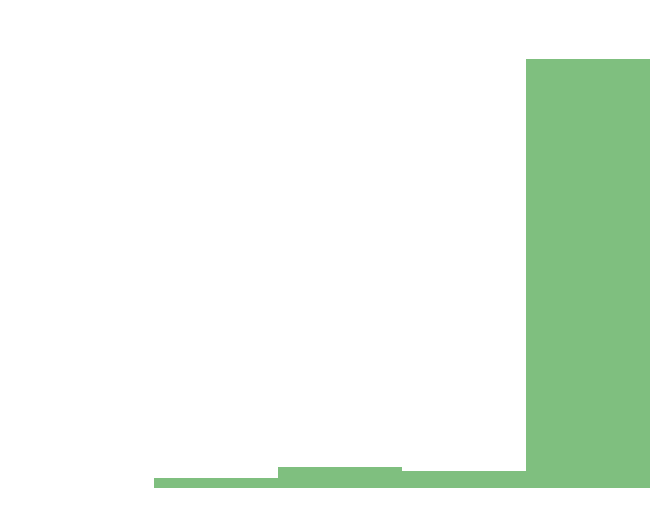} \\ 
8 & credit card information & 88.98 & 0.56 & \includegraphics[width = 2cm, height = 0.5cm]{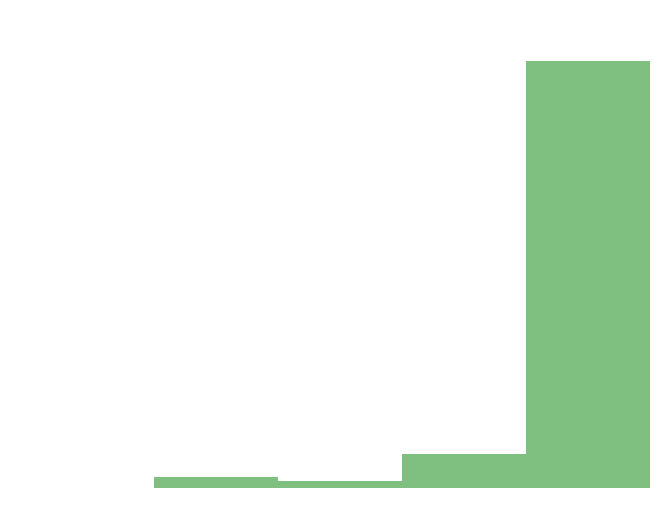} \\ 
9 & video of you that is very embarrassing & 88.41 & 0.53 & \includegraphics[width = 2cm, height = 0.5cm]{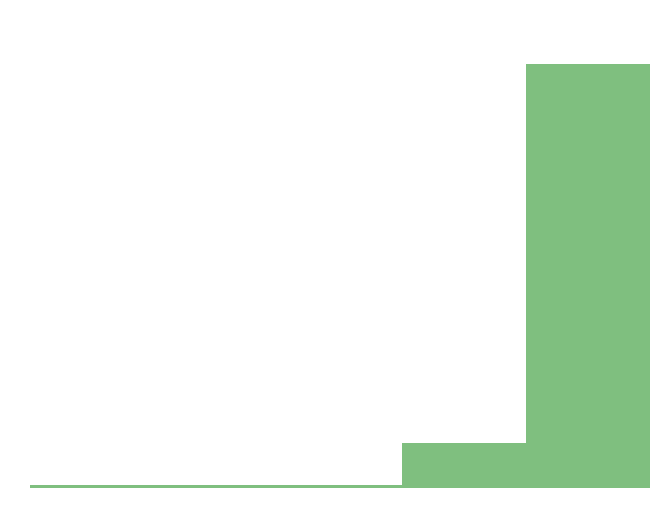} \\ 
10 & photo of you at home & 87.5 &0.60 & \includegraphics[width = 2cm, height = 0.5cm]{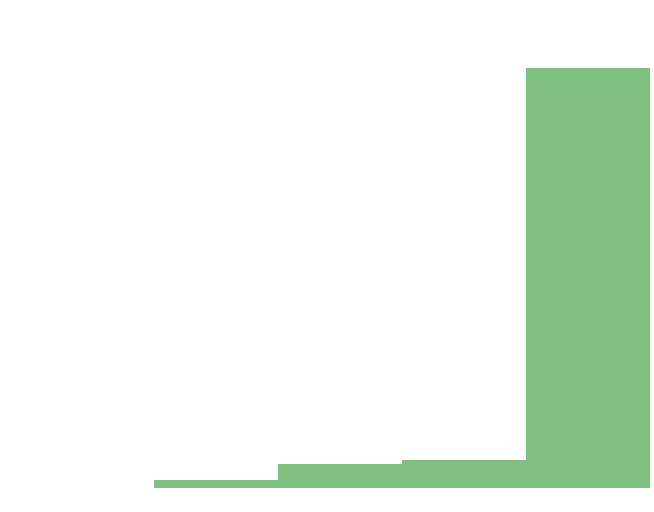} \\ 
11 & audio recording of work conversations & 86.82 & 0.76 & \includegraphics[width = 2cm, height = 0.5cm]{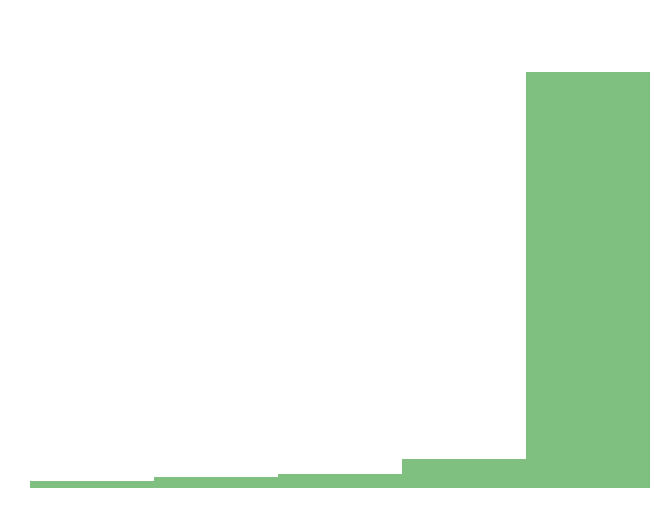} \\ 
12 & video of entering in a passcode to a door & 85.53 & 0.62 & \includegraphics[width = 2cm, height = 0.5cm]{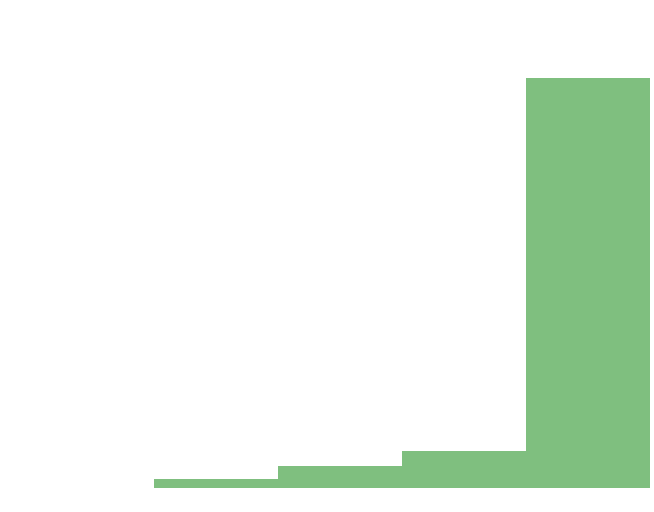} \\ 
13 & audio recording of phone conversations & 85.16 & 0.61 &  \includegraphics[width = 2cm, height = 0.5cm]{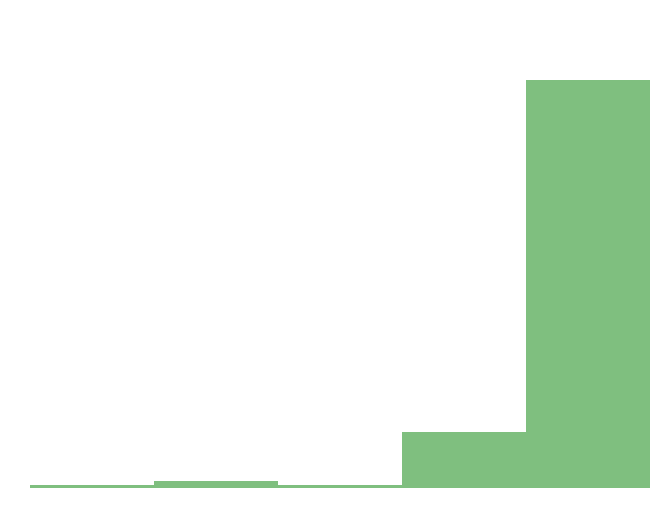} \\ 
14 & amount of money you have & 84.44 & 0.61 & \includegraphics[width = 2cm, height = 0.5cm]{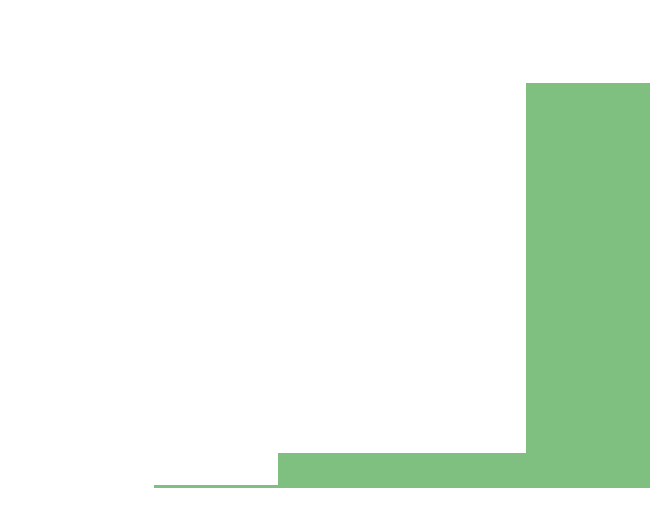} \\ 
15 & video of you intoxicated & 83.21 & 0.72 & \includegraphics[width = 2cm, height = 0.5cm]{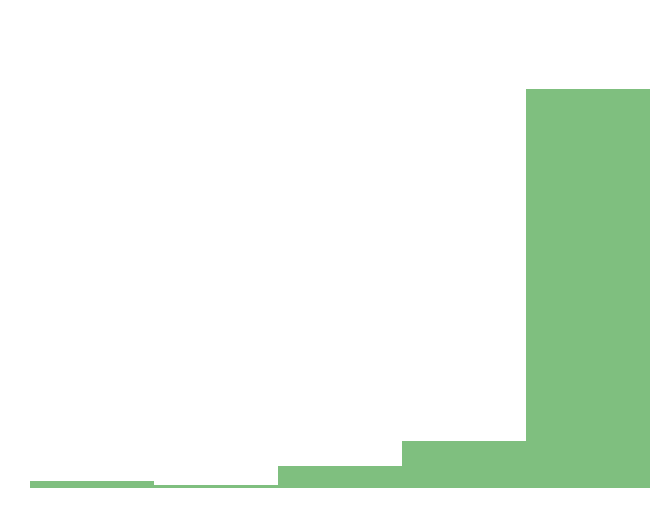} \\ 
16 & when you have sex & 81.95 & 0.82 & \includegraphics[width = 2cm, height = 0.5cm]{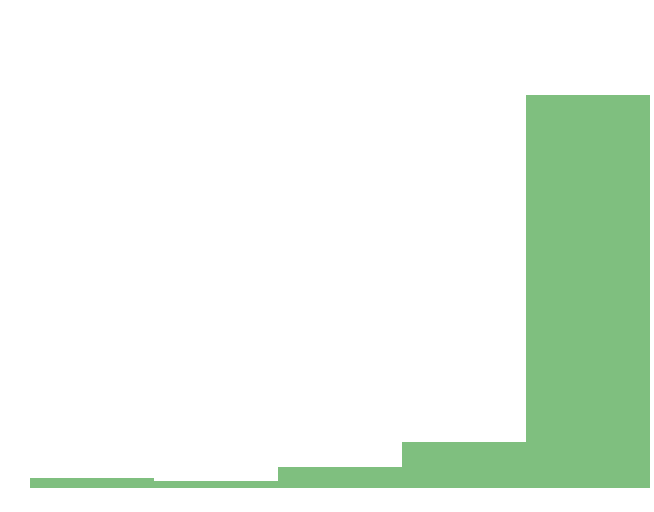} \\ 
17 & how much debt you have & 81.12  & 0.54 & \includegraphics[width = 2cm, height = 0.5cm]{tex-inputs/table-images/tookavideoofyouintoxicatedcombined} \\ 
18 & video of you at home & 81.05 & 0.60 & \includegraphics[width = 2cm, height = 0.5cm]{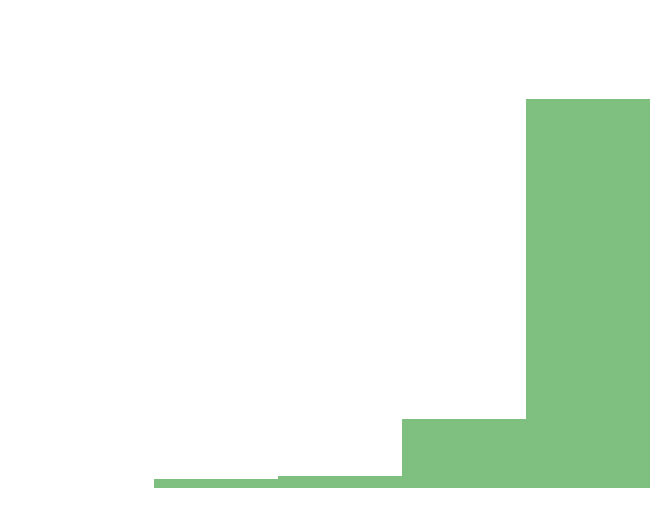} \\ 
19 & photo of you intoxicated & 78.95 & 0.82 &  \includegraphics[width = 2cm, height = 0.5cm]{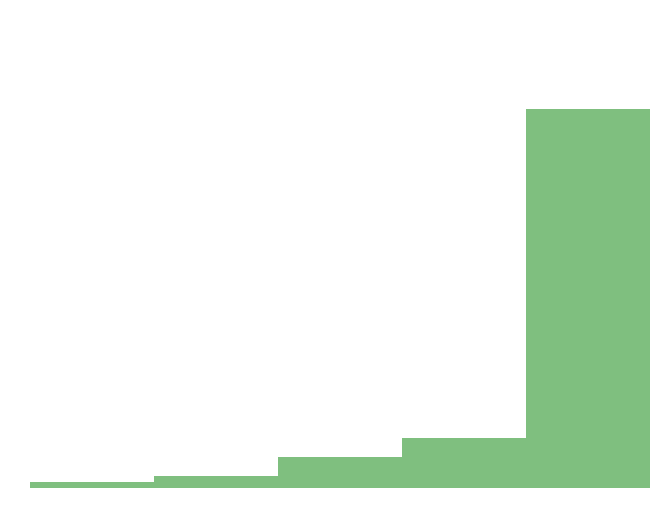} \\ 
20 & photo of you at random & 78.76 & 0.85 & \includegraphics[width = 2cm, height = 0.5cm]{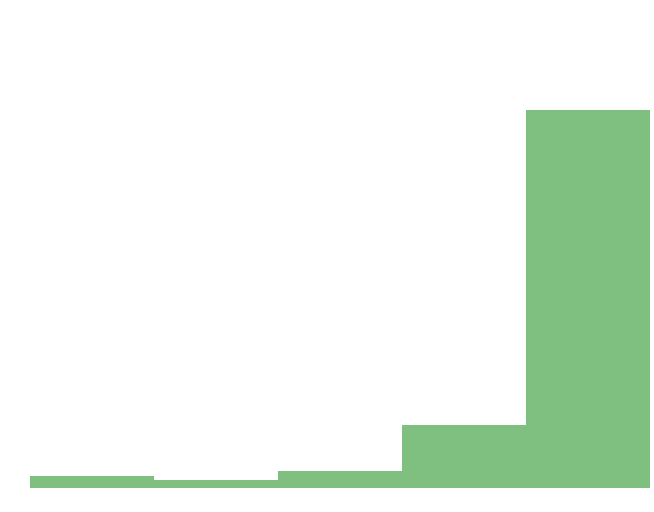} \\ 
21 & audio recording of conversations & 78.13 & 0.83 & \includegraphics[width = 2cm, height = 0.5cm]{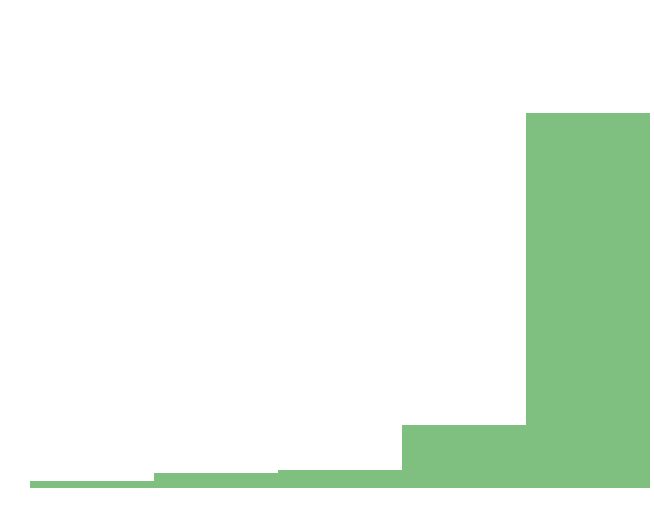} \\ 
22 & medical conditions & 77.7 & 0.86 &  \includegraphics[width = 2cm, height = 0.5cm]{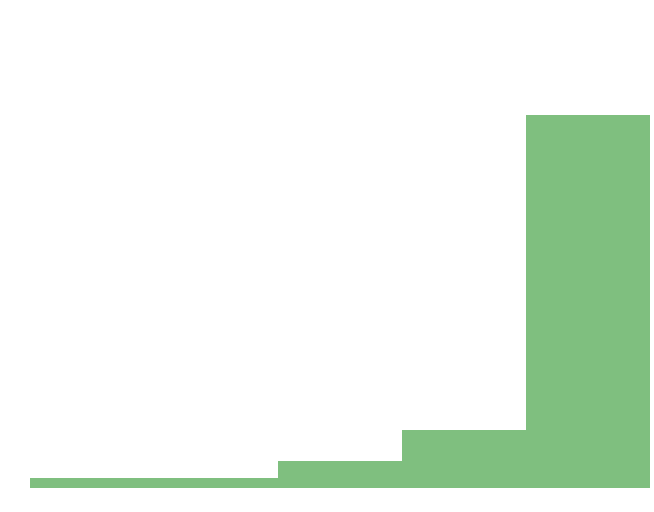} \\ 
23 & video of you at random & 76.19 & 0.59 & \includegraphics[width = 2cm, height = 0.5cm]{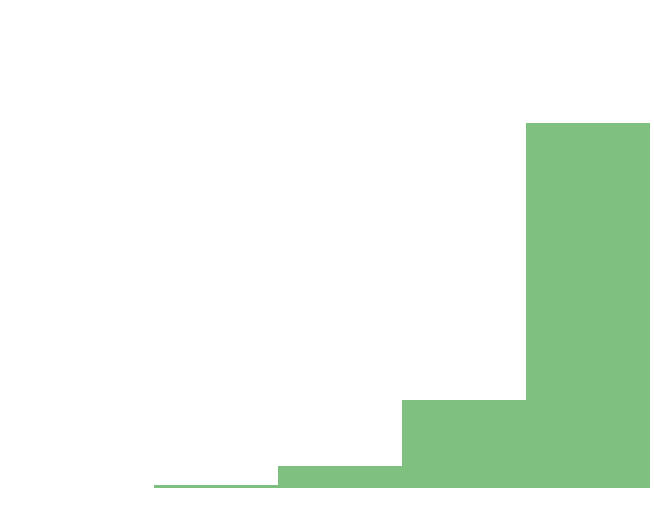} \\ 
24 & video of you off-guard & 76.0 & 0.62 & \includegraphics[width = 2cm, height = 0.5cm]{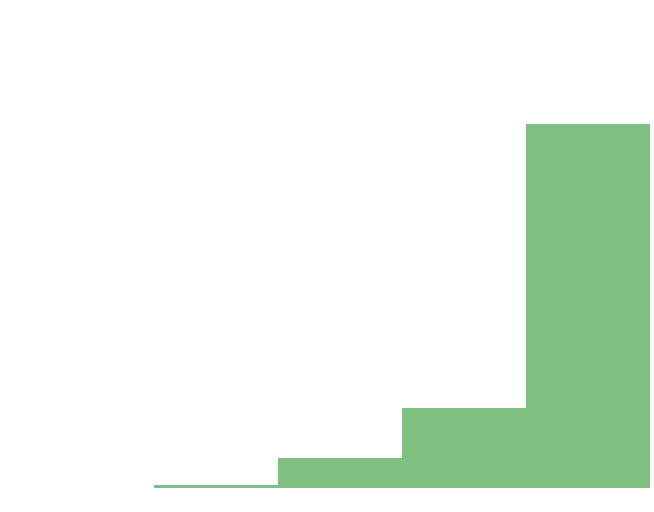} \\ 
25 & photo of your work or workplace & 74.62 & 0.90 & \includegraphics[width = 2cm, height = 0.5cm]{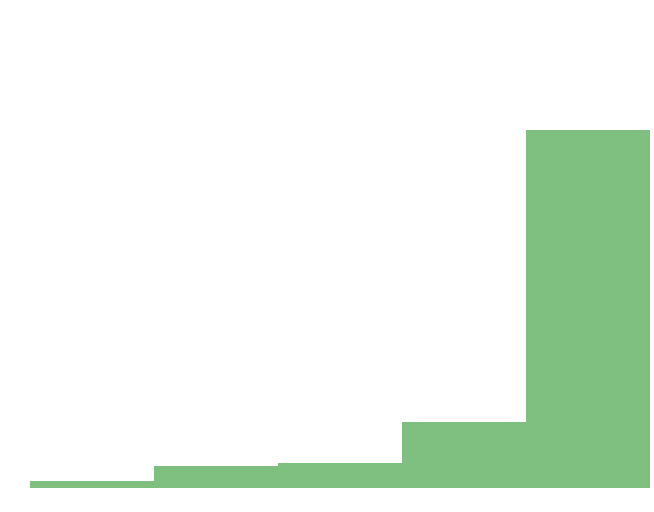} \\ 
26 & username for websites & 73.44 & 0.83 &  \includegraphics[width = 2cm, height = 0.5cm]{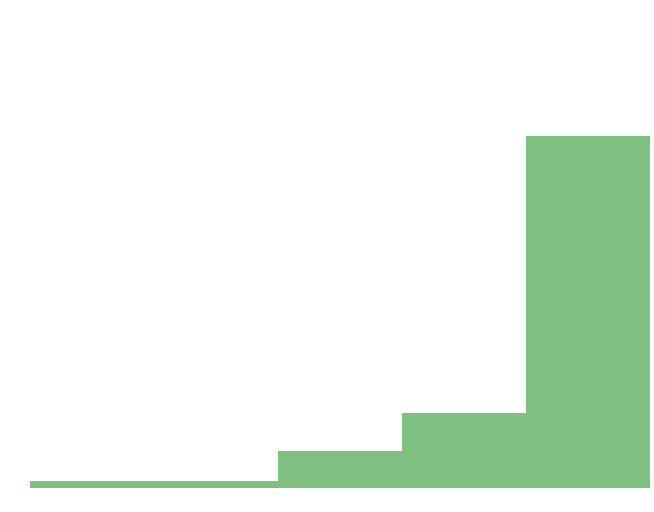} \\ 
27 & address & 72.61 & 0.86 & \includegraphics[width = 2cm, height = 0.5cm]{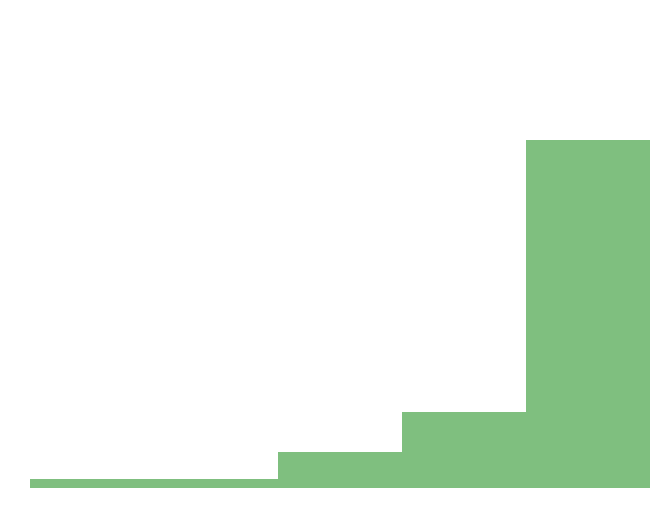} \\ 
28 & audio recording you captured & 72.55  & 0.70 & \includegraphics[width = 2cm, height = 0.5cm]{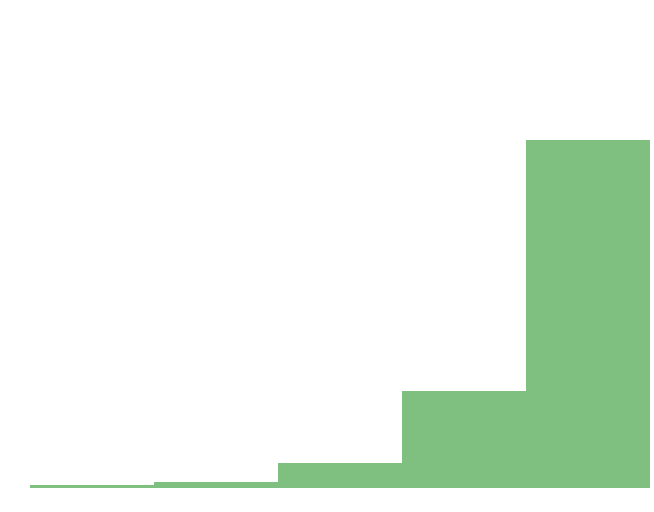} \\ 
29 & photo of you off-guard & 72.55 & 0.77 & \includegraphics[width = 2cm, height = 0.5cm]{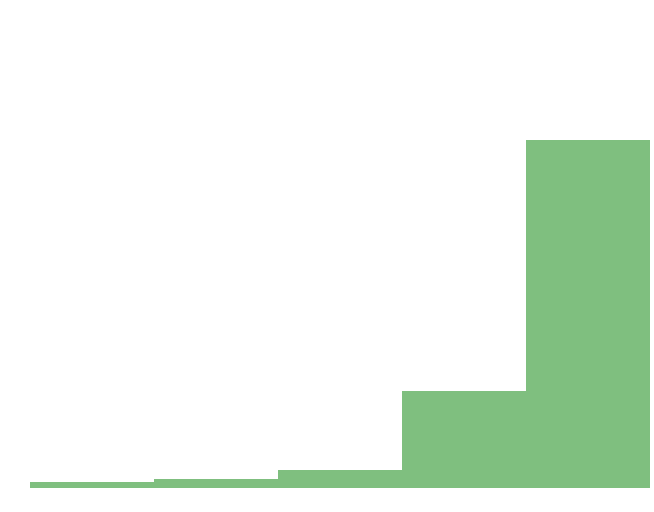} \\ 
30 & photo downloaded from internet & 71.81 & 0.90 & \includegraphics[width = 2cm, height = 0.5cm]{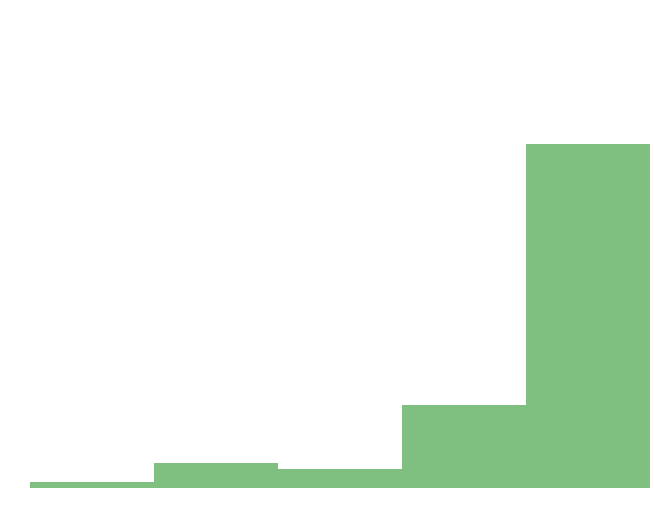} \\ 
31 & photo others sent you & 71.63 & 1.03 & \includegraphics[width = 2cm, height = 0.5cm]{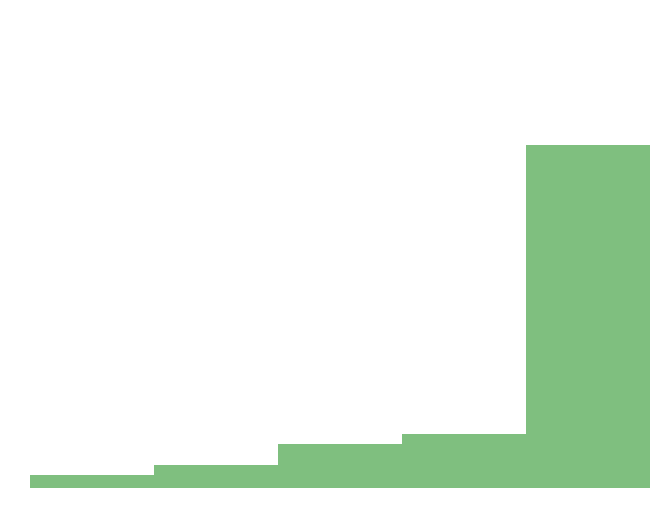} \\ 
32 & video others sent you & 70.59 & 0.81 & \includegraphics[width = 2cm, height = 0.5cm]{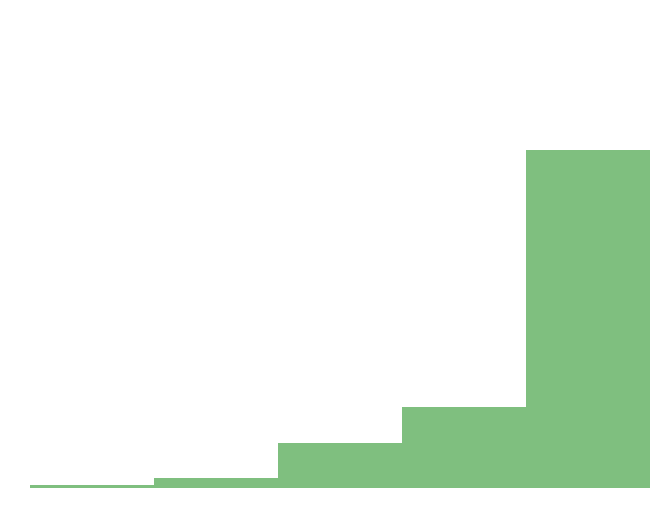} \\ 
33 & video of your work or workplace & 70.54 & 0.90 & \includegraphics[width = 2cm, height = 0.5cm]{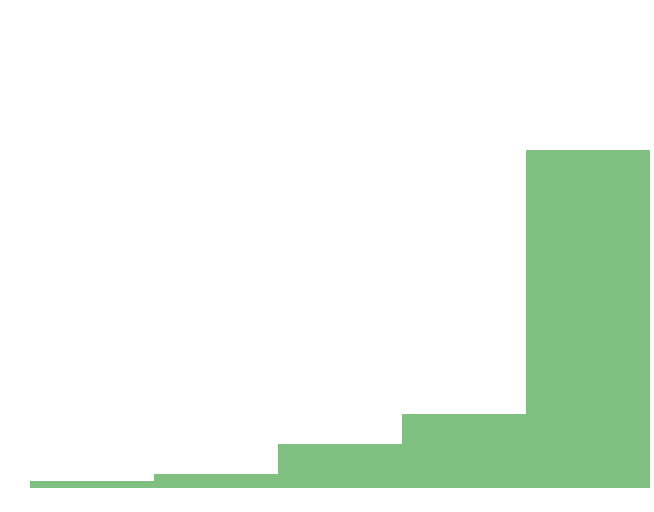} \\ 
34 & fingerprint & 70.12 & 0.86 & \includegraphics[width = 2cm, height = 0.5cm]{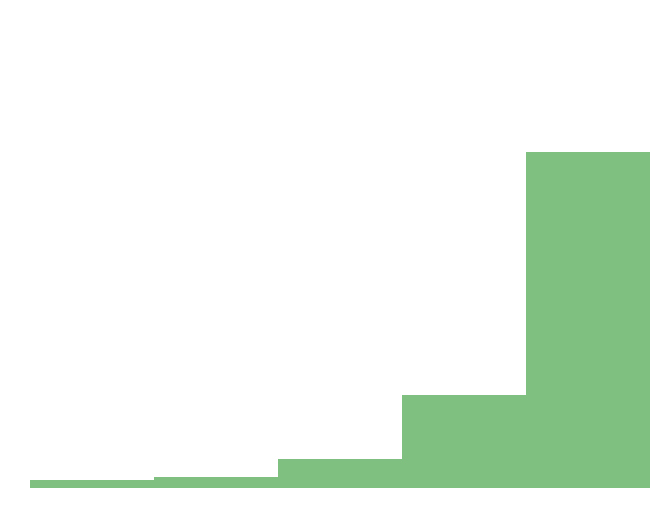} \\ 
35 & when you were lying nervous or stressed & 69.74 & 0.91 & \includegraphics[width = 2cm, height = 0.5cm]{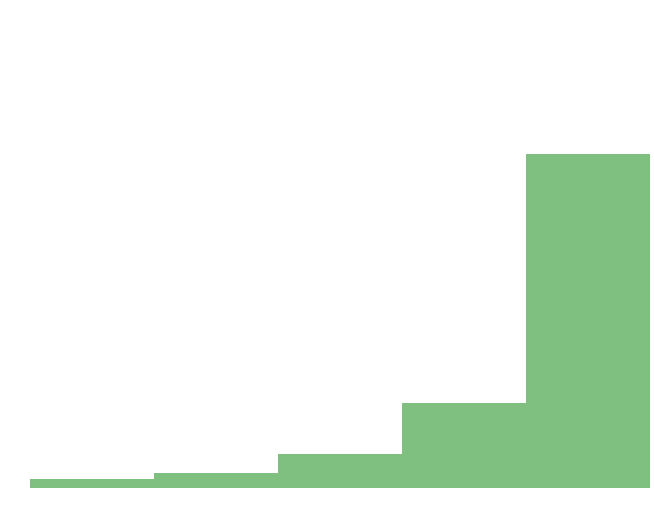} \\ 
36 & audio recording of you (voice notes) & 69.59 & 0.91 & \includegraphics[width = 2cm, height = 0.5cm]{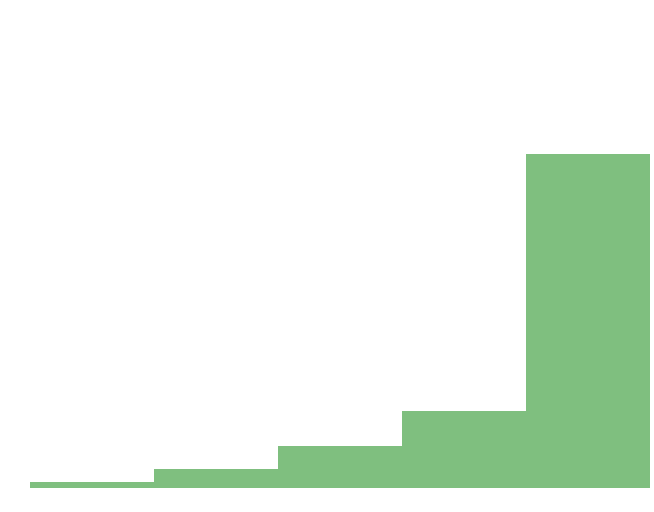} \\ 
\hline
\end{tabular}
\caption{VUR for all data types (1-36), across all recipients.}
\label{full-vur-table}
\end{center}
\end{table*}

\begin{table*}[t]
\begin{center}
\small
\begin{tabular}{| r | l | r | r | r | r |}
\hline
Rank & Question & VUR & $\sigma$ & Distribution \\
\hline
37 & medication taken & 69.49 & 1.01 & \includegraphics[width = 2cm, height = 0.5cm]{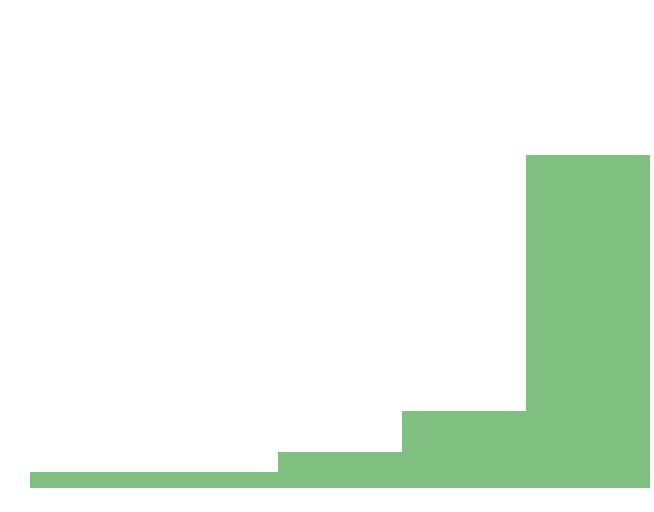} \\ 
38 & videos already on device & 68.89 & 0.88 & \includegraphics[width = 2cm, height = 0.5cm]{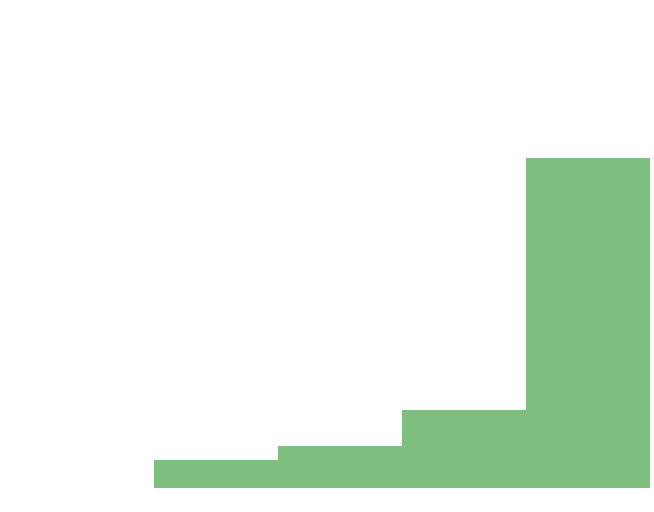} \\ 
39 & photo of your signature & 68.07 & 0.84 & \includegraphics[width = 2cm, height = 0.5cm]{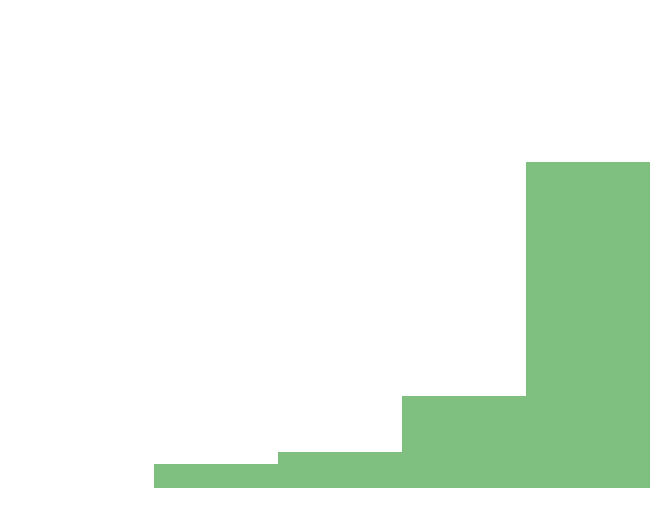} \\ 
40 & web history & 66.44 & 1.01 & \includegraphics[width = 2cm, height = 0.5cm]{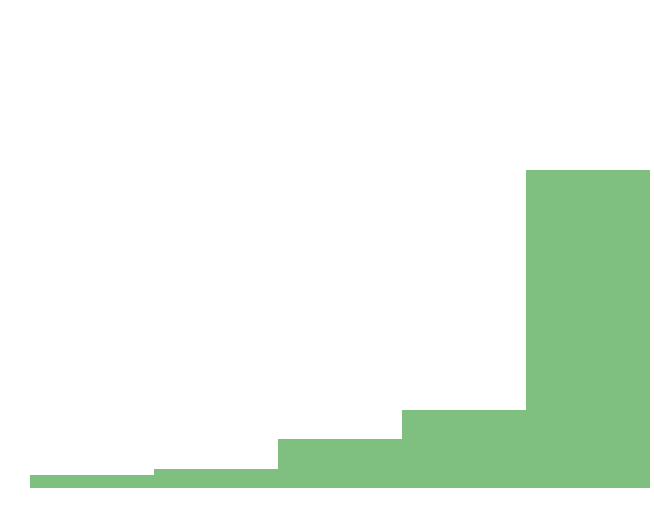} \\ 
41 & photos taken on device & 66.21 & 1.02 & \includegraphics[width = 2cm, height = 0.5cm]{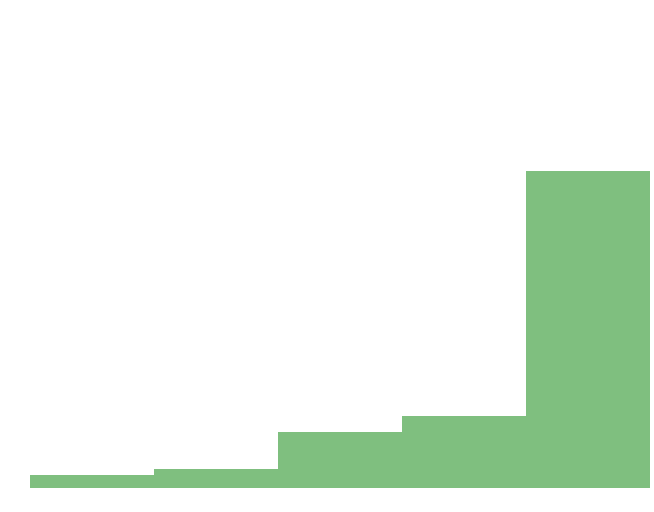} \\ 
42 & home address & 65.0 & 0.97 & \includegraphics[width = 2cm, height = 0.5cm]{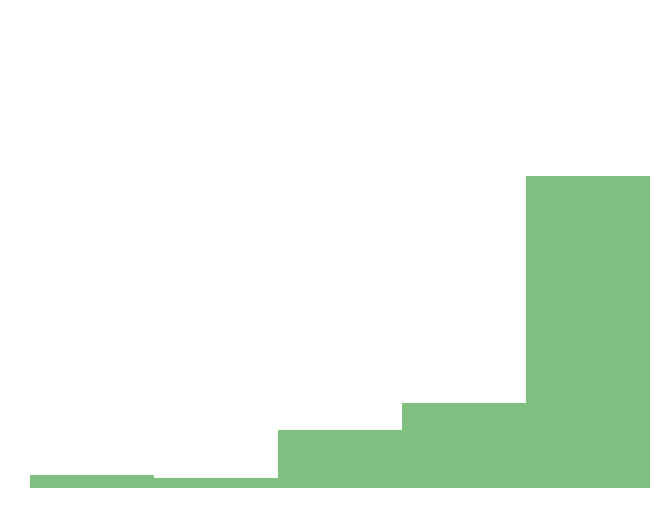} \\ 
43 & fine-grained location tracking (+/- cm) & 63.51 & 0.99 & \includegraphics[width = 2cm, height = 0.5cm]{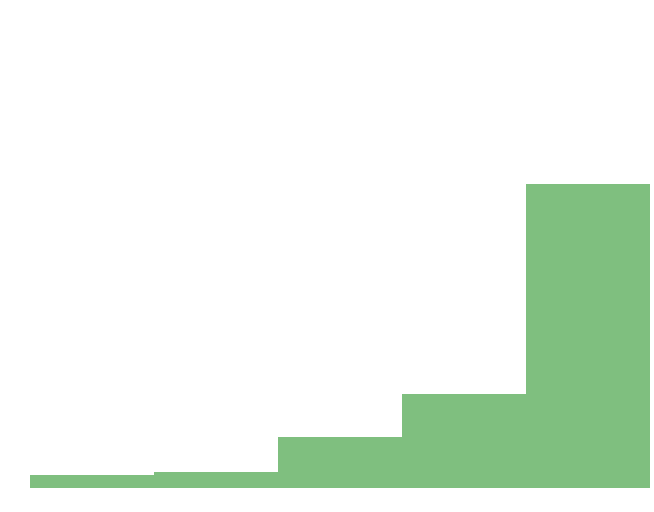} \\ 
44 & photo of people at random & 61.94 & 1.06 & \includegraphics[width = 2cm, height = 0.5cm]{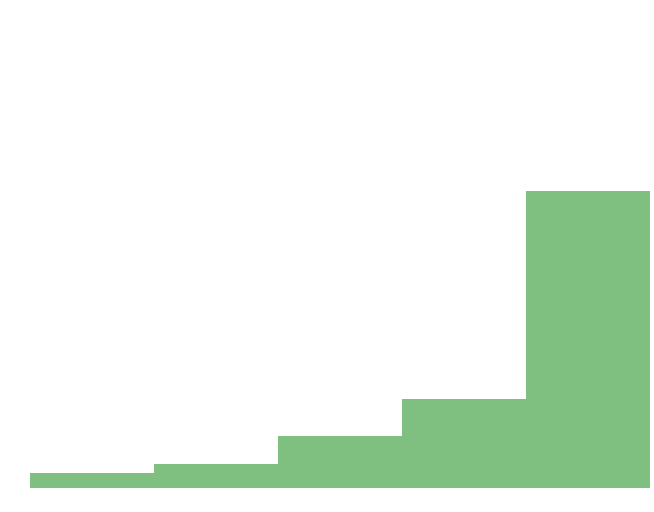} \\ 
45 & video downloaded from the internet & 61.49 & 1.00 & \includegraphics[width = 2cm, height = 0.5cm]{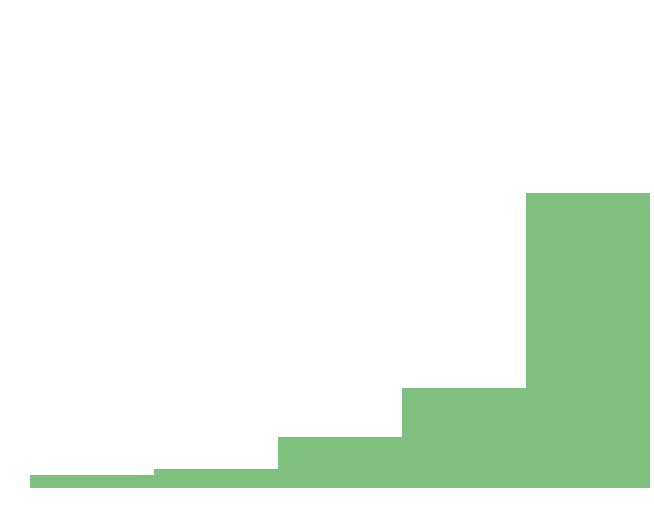} \\ 
46 & when you are alone & 61.27 & 0.99 & \includegraphics[width = 2cm, height = 0.5cm]{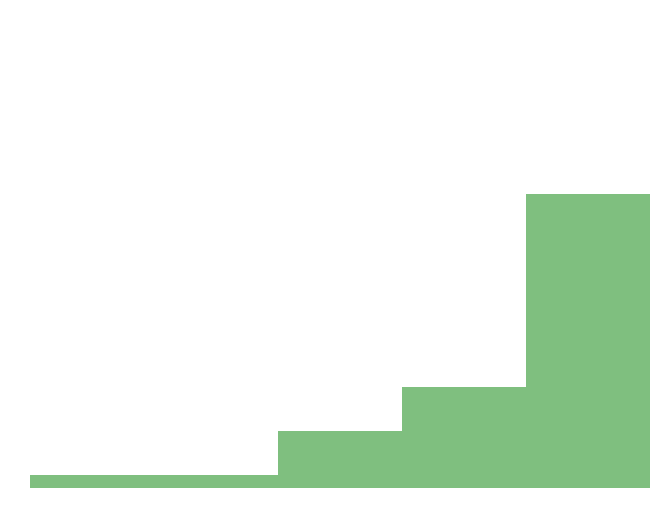} \\ 
47 & location tracking (+/- m) & 61.24 & 1.08 & \includegraphics[width = 2cm, height = 0.5cm]{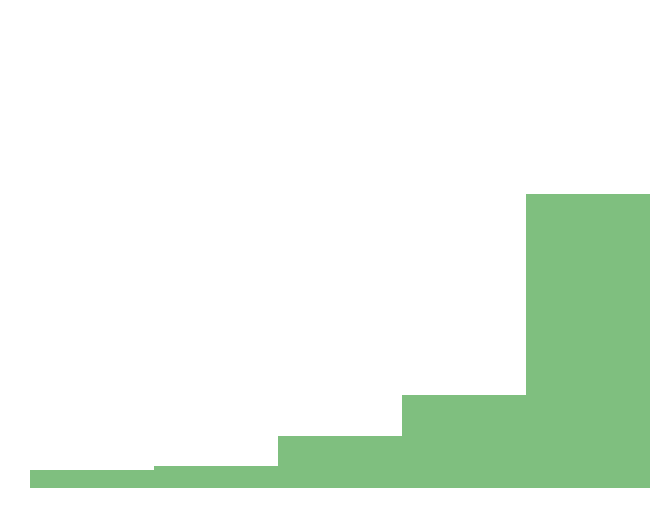} \\ 
48 & videos of people at random & 61.04 & 0.95 &  \includegraphics[width = 2cm, height = 0.5cm]{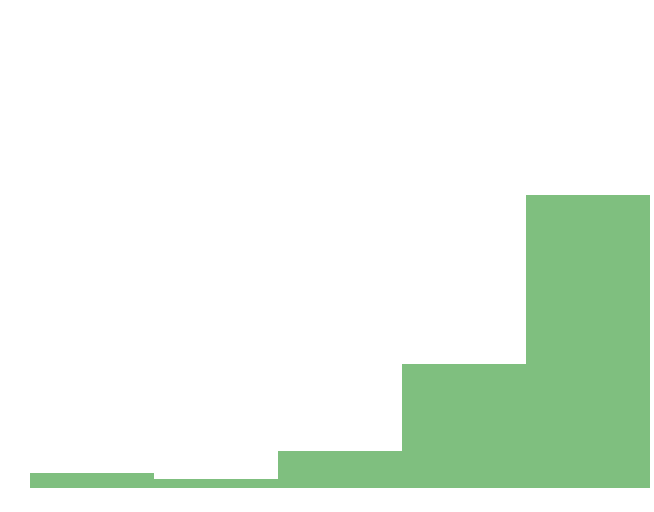} \\ 
49 & where you are currently going & 60.87 & 0.97 &  \includegraphics[width = 2cm, height = 0.5cm]{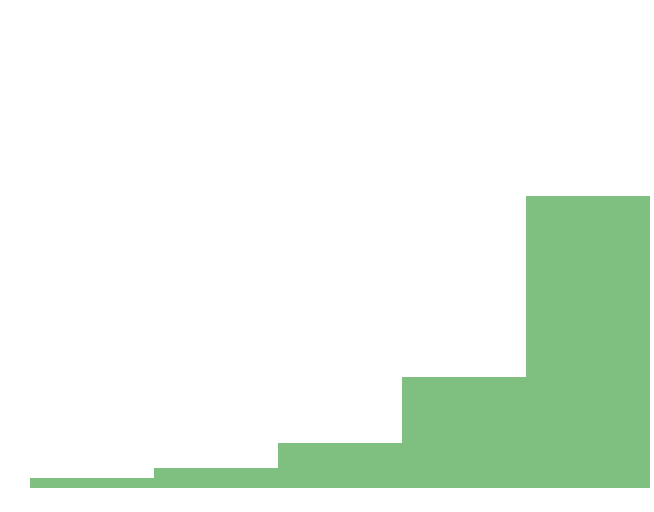} \\ 
50 & recording of sound around you & 60.45 & 0.94 & \includegraphics[width = 2cm, height = 0.5cm]{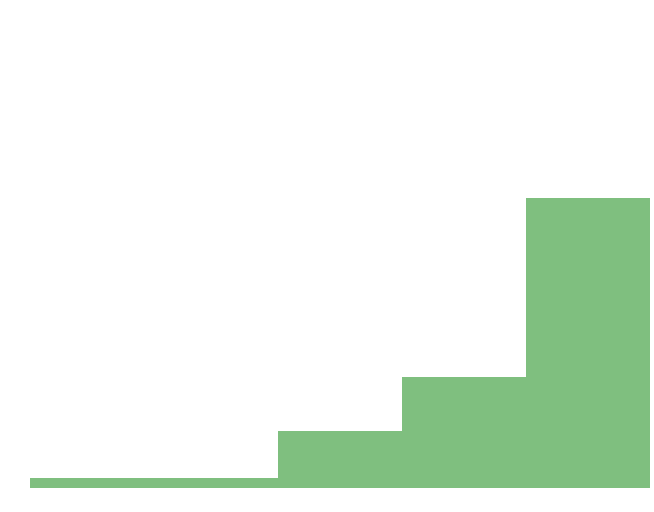} \\ 
51 & people you spend time with & 60.0 & 1.13 &  \includegraphics[width = 2cm, height = 0.5cm]{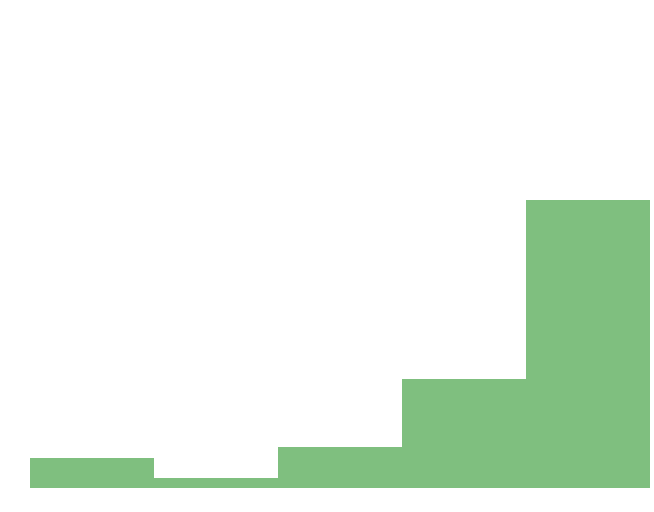} \\ 
52 & workplace address & 58.09 & 1.16 &\includegraphics[width = 2cm, height = 0.5cm]{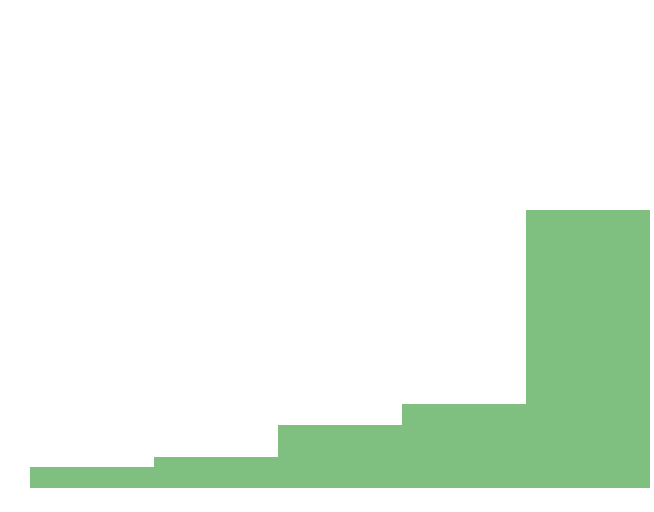} \\ 
53 & sounds on device (notifications, etc) & 54.4 & 1.29 &  \includegraphics[width = 2cm, height = 0.5cm]{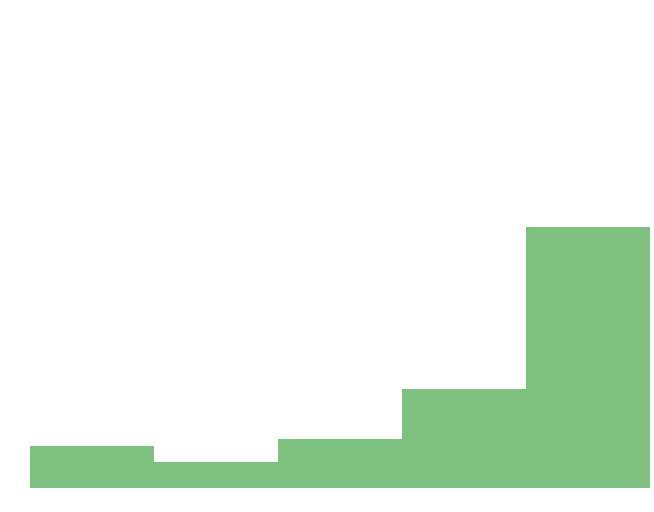} \\ 
54 & phone usage & 51.95 &1.22 &  \includegraphics[width = 2cm, height = 0.5cm]{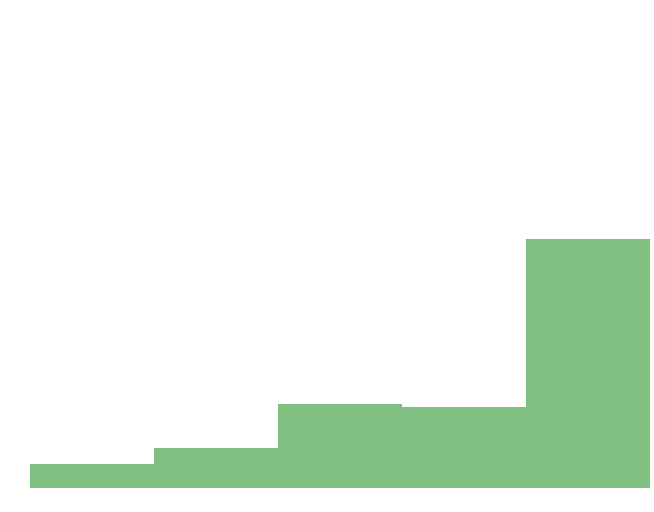} \\ 
55 & purchased products & 50.0 & 1.09 & \includegraphics[width = 2cm, height = 0.5cm]{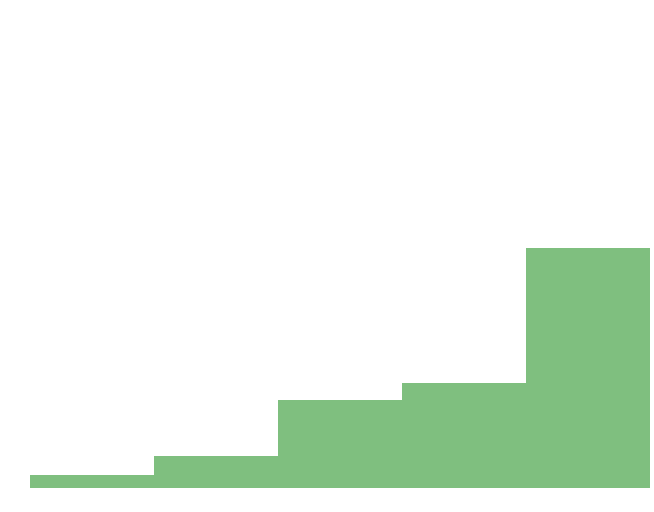} \\ 
56 & when you are sick or healthy & 48.17  & 1.27 & \includegraphics[width = 2cm, height = 0.5cm]{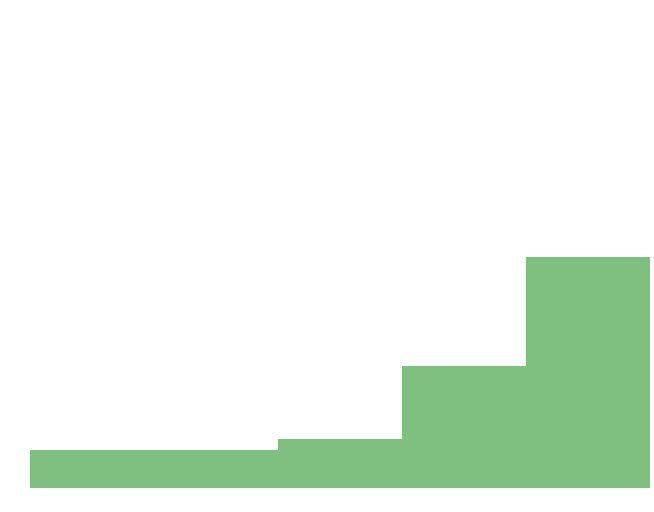} \\ 
57 & how close you are to interacting people & 46.98 & 1.12 & \includegraphics[width = 2cm, height = 0.5cm]{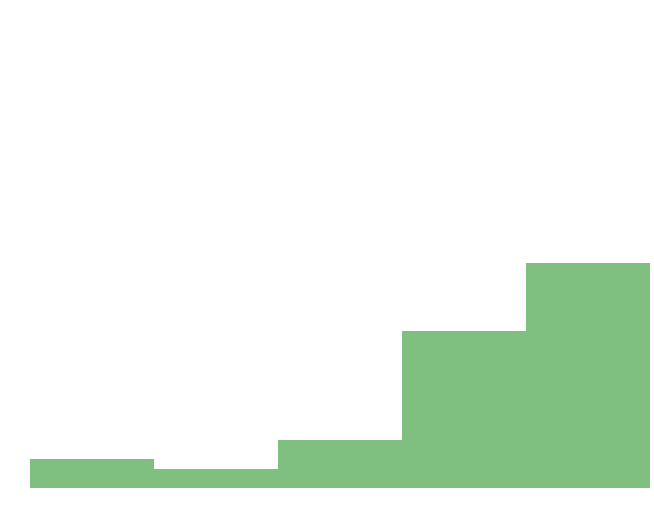} \\ 
58 & feelings (based on biometrics) & 46.81 & 1.31 & \includegraphics[width = 2cm, height = 0.5cm]{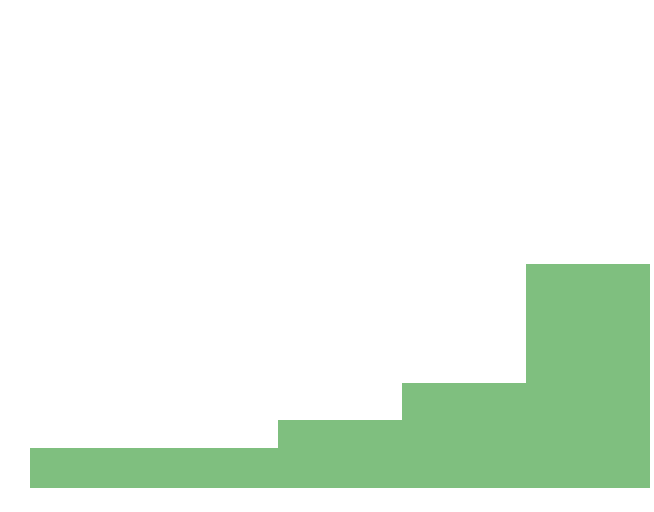} \\ 
59 & computer usage & 44.93 & 1.16 & \includegraphics[width = 2cm, height = 0.5cm]{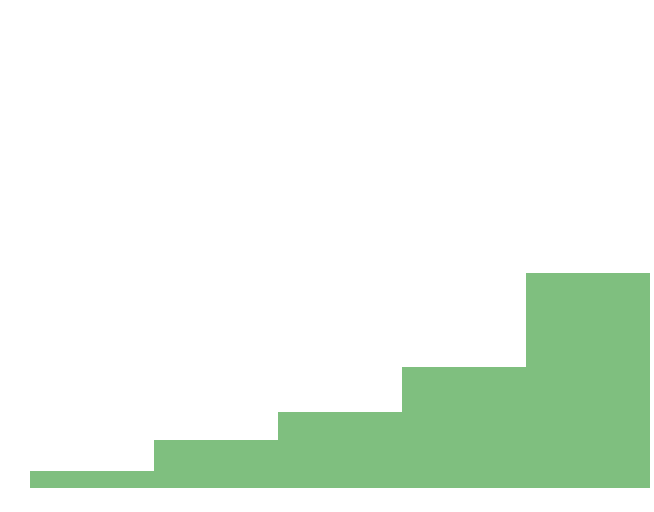} \\ 
60 & eating patterns & 42.86 & 1.27 & \includegraphics[width = 2cm, height = 0.5cm]{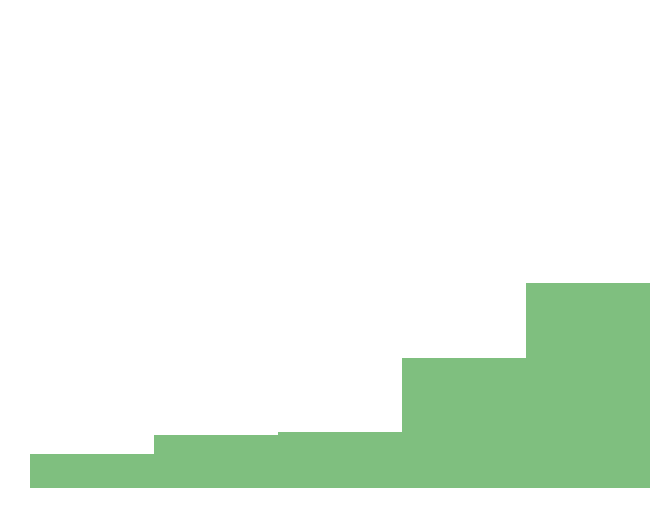} \\ 
61 & name & 42.54 & 1.40 & \includegraphics[width = 2cm, height = 0.5cm]{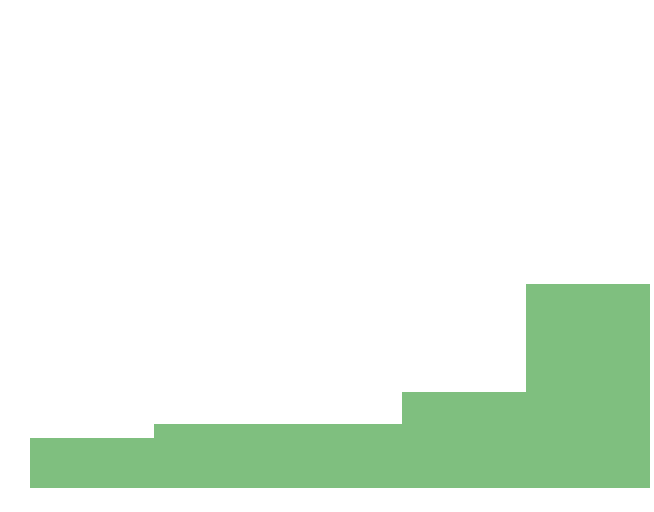} \\ 
62 & sleeping patterns & 40.56 & 1.34 & \includegraphics[width = 2cm, height = 0.5cm]{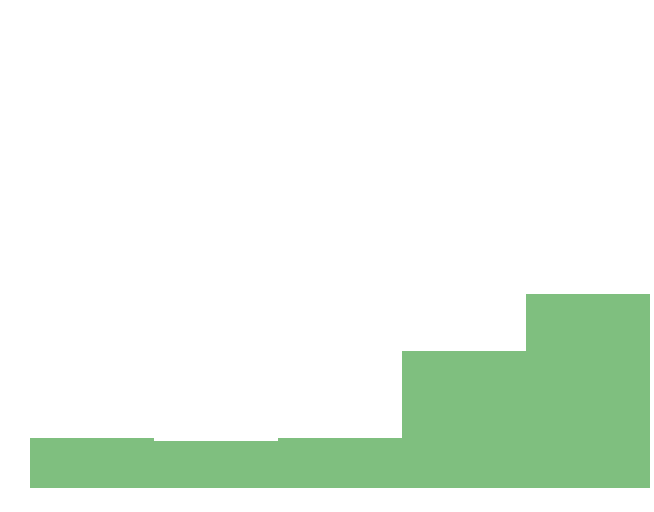} \\ 
63 & eye patterns (for eye tracking) & 40.51 & 1.27 & \includegraphics[width = 2cm, height = 0.5cm]{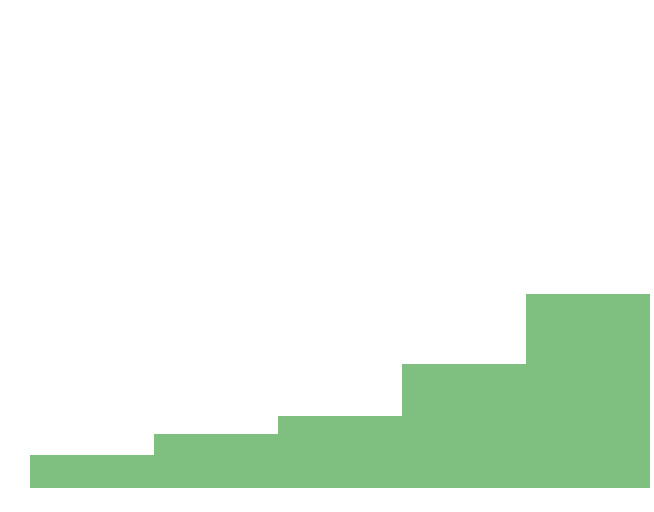} \\ 
64 & exercise patterns & 38.66 & 1.26 & \includegraphics[width = 2cm, height = 0.5cm]{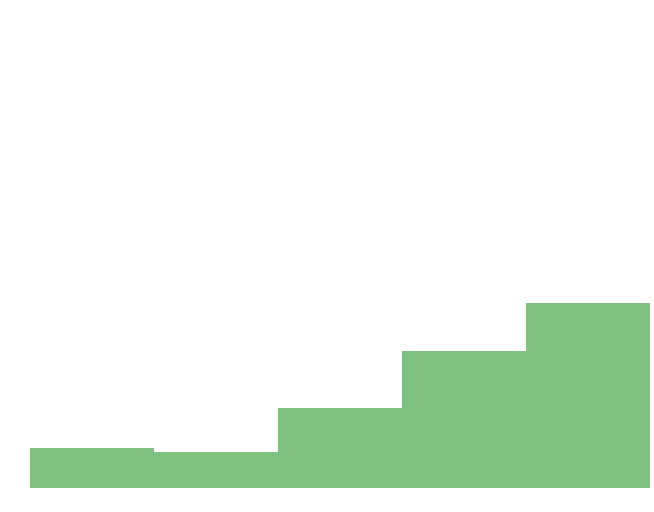} \\ 
65 & when you are happy or having fun & 34.75 & 1.27 & \includegraphics[width = 2cm, height = 0.5cm]{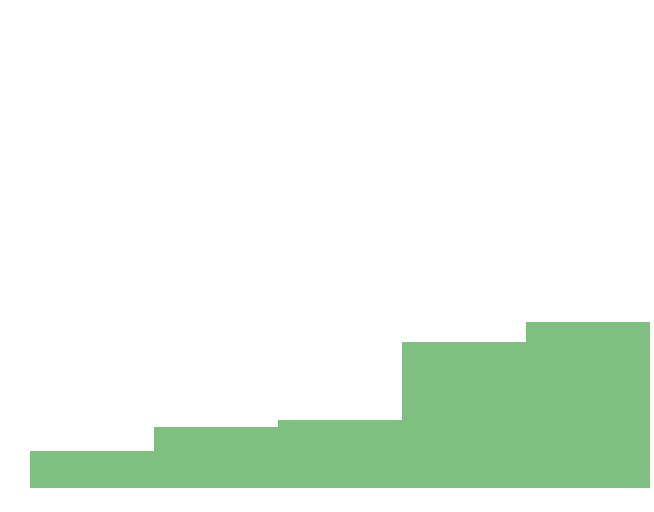} \\ 
66 & television shows watched & 30.2 & 1.40 & \includegraphics[width = 2cm, height = 0.5cm]{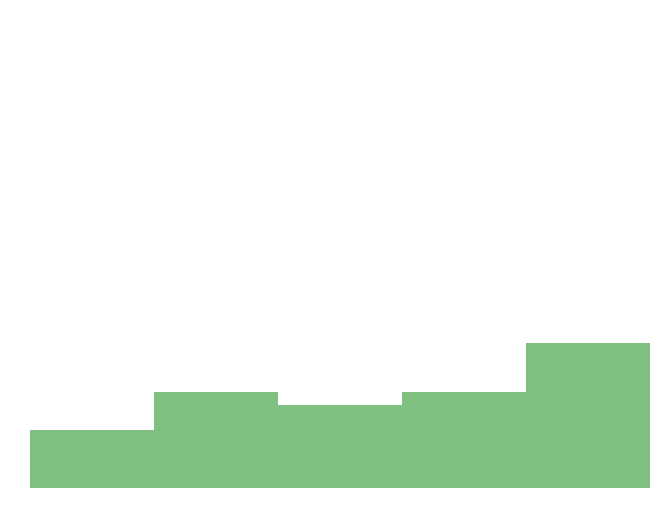} \\ 
67 & when you are busy or interruptible & 29.5 & 1.26 & \includegraphics[width = 2cm, height = 0.5cm]{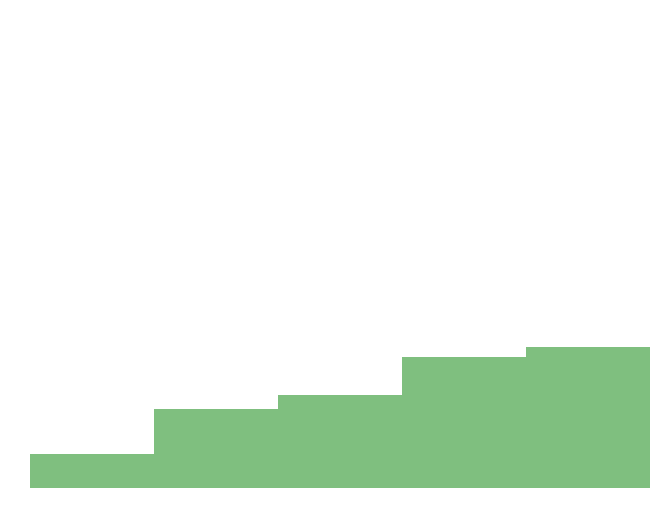} \\ 
68 & music on device & 28.06 & 1.43 & \includegraphics[width = 2cm, height = 0.5cm]{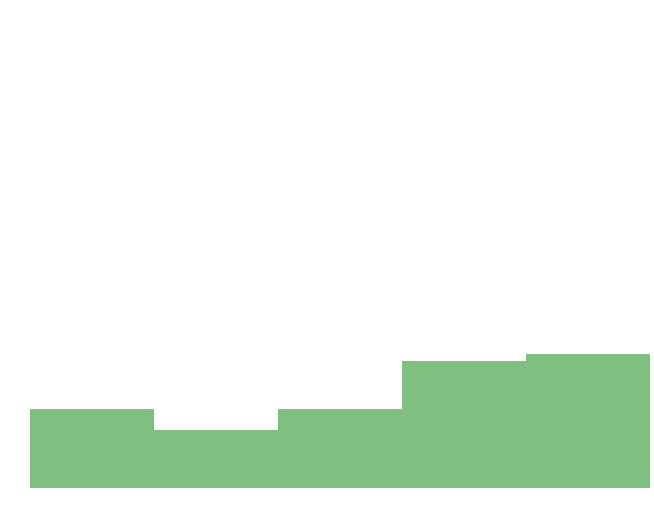} \\ 
69 & heart rate & 27.5 & 1.40 & \includegraphics[width = 2cm, height = 0.5cm]{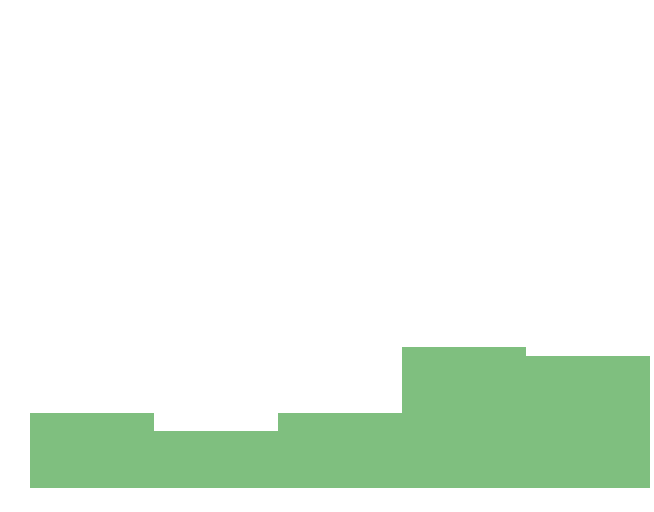} \\ 
70 & age & 24.29 & 1.43 &  \includegraphics[width = 2cm, height = 0.5cm]{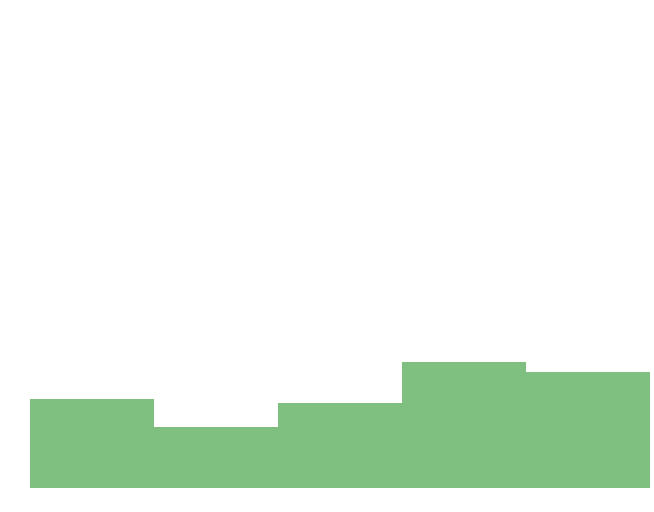} \\ 
71 & language spoken & 15.86 & 1.49 & \includegraphics[width = 2cm, height = 0.5cm]{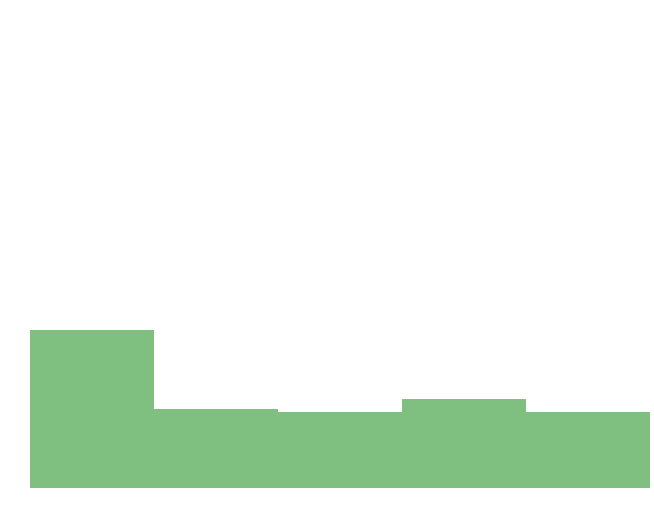} \\ 
72 & gender & 15.0 & 1.46 &  \includegraphics[width = 2cm, height = 0.5cm]{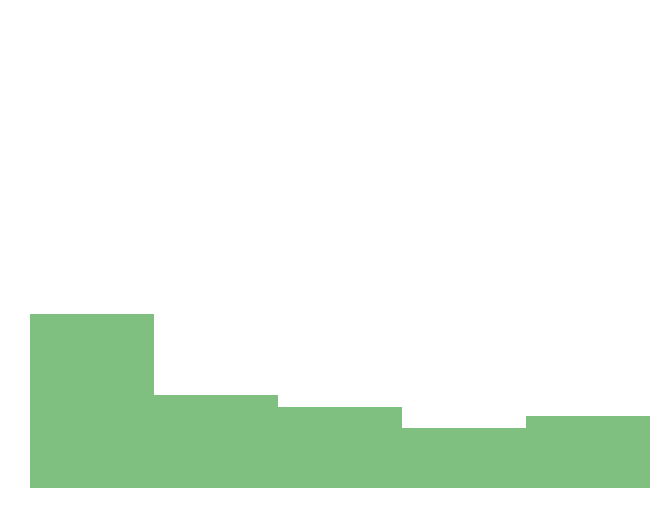} \\ 
\hline
\end{tabular}
\caption{VUR for all data types (37-72), across all recipients.}
\label{full-vur-table}
\end{center}
\end{table*}

%% file: tex-inputs/riskben.tex
\begin{table}[h]
\begin{center}
\small
\begin{tabular}{| p{2.6cm} | p{.8cm} | p{.8cm} | p{.8cm} | c |}
\hline
Technology & Q1 &  Median & Q3 & Distribution  \\ 
\hline
Location Tracking & 10.0 & 10.0 & 20.0 & \includegraphics[width = 2cm, height = 0.5cm]{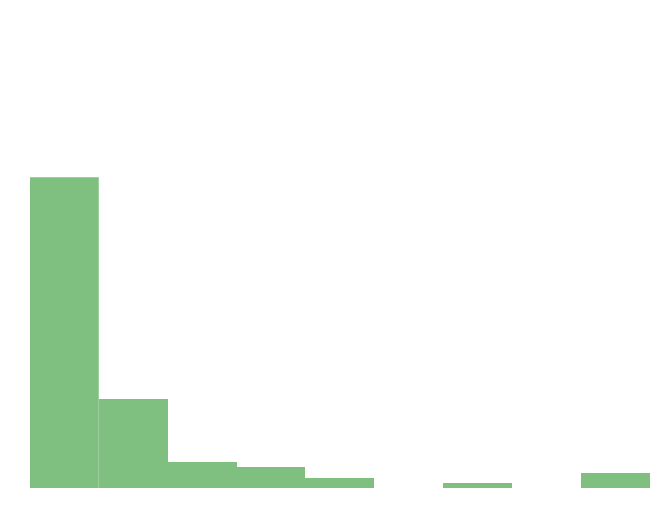} \\ 
Speech To Text & 10.0 & 10.0 & 10.0 & \includegraphics[width = 2cm, height = 0.5cm]{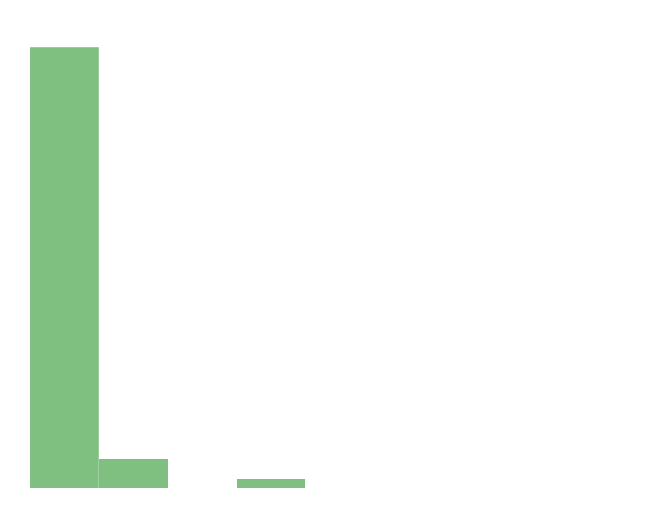} \\ 
Discreet Microphone & 10.0 & 10.0 & 20.0 & \includegraphics[width = 2cm, height = 0.5cm]{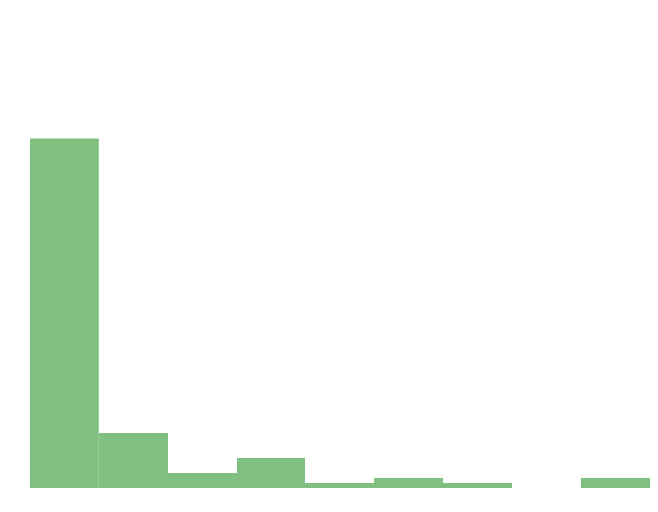} \\ 
Smartwatches & 10.0 & 10.0 & 10.0 & \includegraphics[width = 2cm, height = 0.5cm]{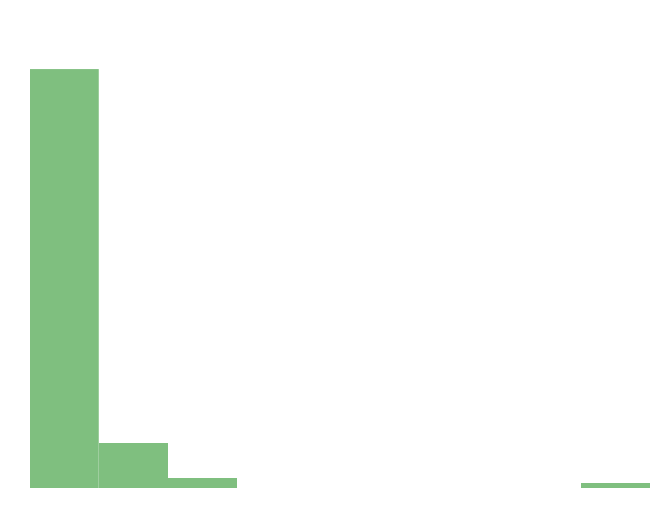} \\ 
Language Detection & 10.0 & 10.0 & 10.0 & \includegraphics[width = 2cm, height = 0.5cm]{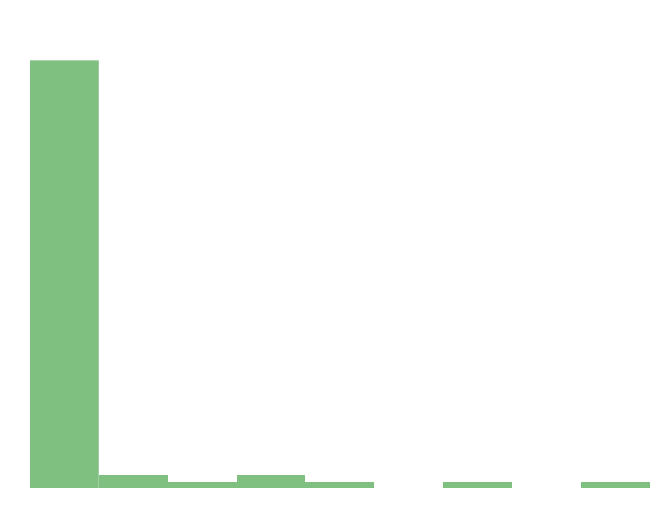} \\ 
Laptops & 10.0 & 10.0 & 15.0 & \includegraphics[width = 2cm, height = 0.5cm]{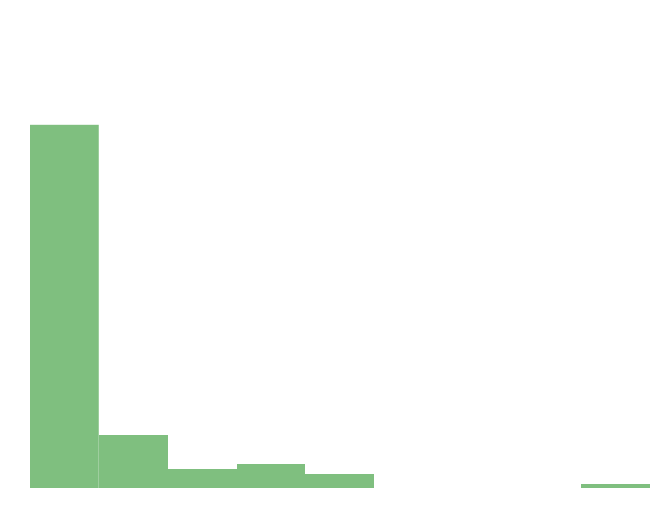} \\ 
Smartphones & 10.0 & 10.0 & 20.0 & \includegraphics[width = 2cm, height = 0.5cm]{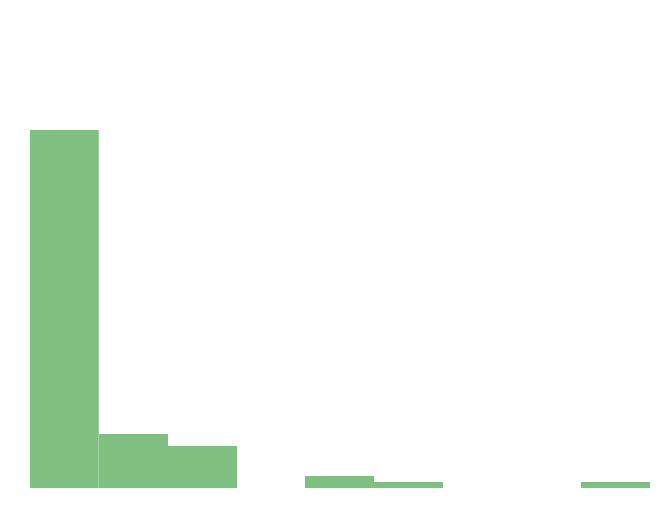} \\ 
Google Glass & 10.0 & 10.0 & 20.0 & \includegraphics[width = 2cm, height = 0.5cm]{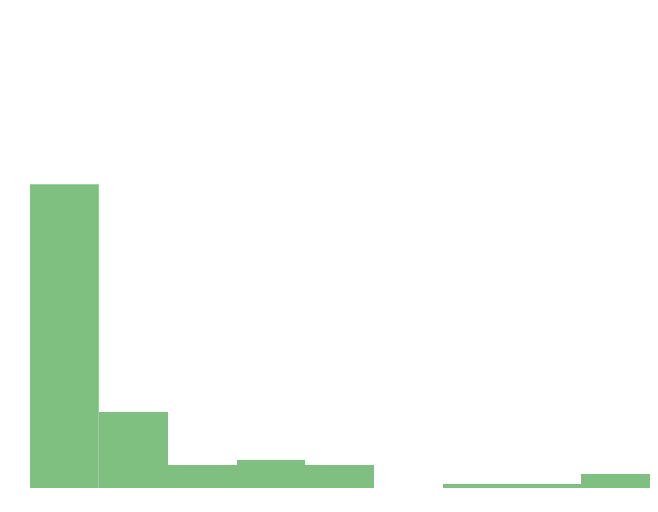} \\ 
Cubetastic & 10.0 & 10.0 & 30.0 & \includegraphics[width = 2cm, height = 0.5cm]{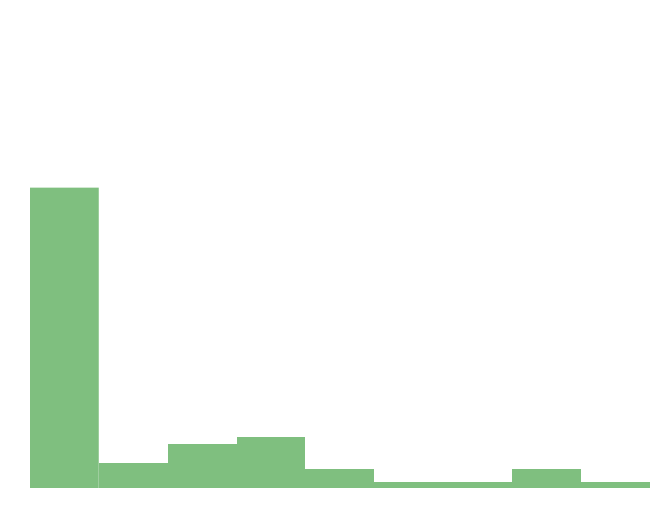} \\ 
Gender Detection & 10.0 & 10.0 & 13.5 & \includegraphics[width = 2cm, height = 0.5cm]{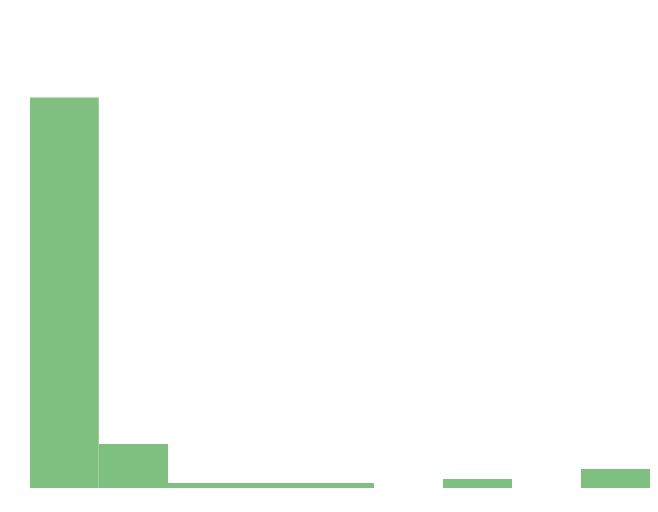} \\ 
Voice Recognition & 10.0 & 10.0 & 15.0 & \includegraphics[width = 2cm, height = 0.5cm]{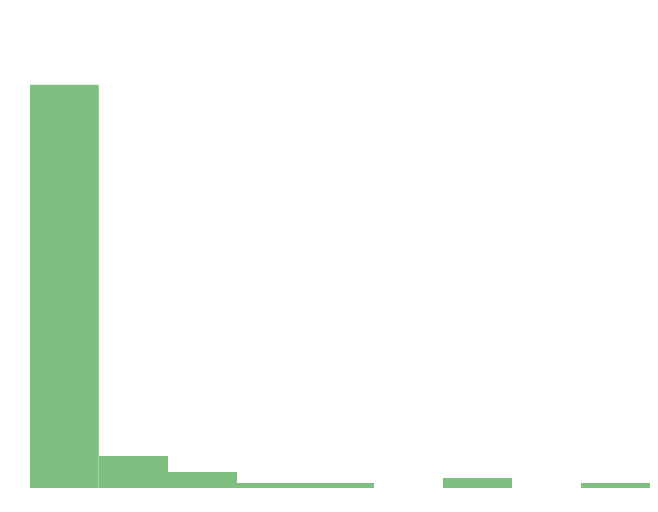} \\ 
Voice Based Emotion Detection & 10.0 & 10.0 & 15.0 & \includegraphics[width = 2cm, height = 0.5cm]{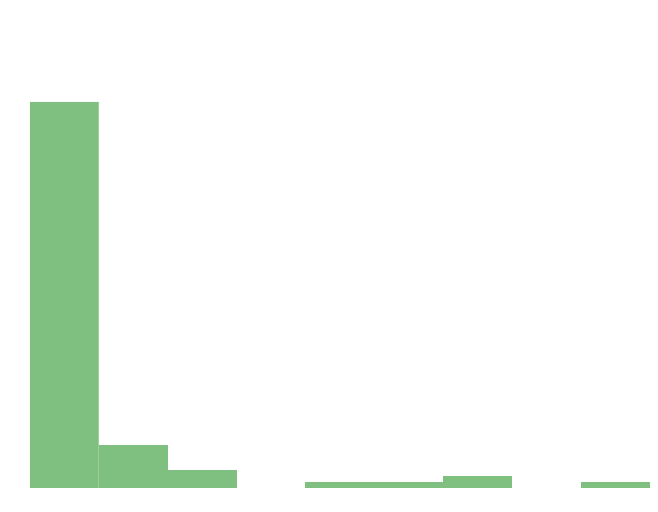} \\ 
Fitness Trackers & 10.0 & 10.0 & 10.0 & \includegraphics[width = 2cm, height = 0.5cm]{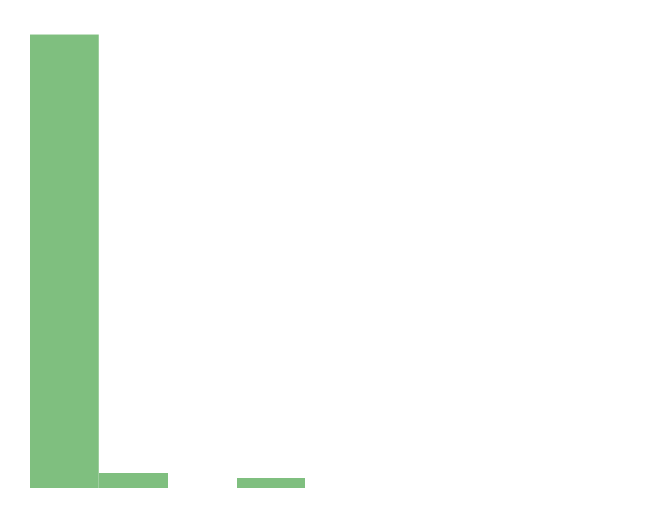} \\ 
Age Detection & 10.0 & 10.0 & 15.0 & \includegraphics[width = 2cm, height = 0.5cm]{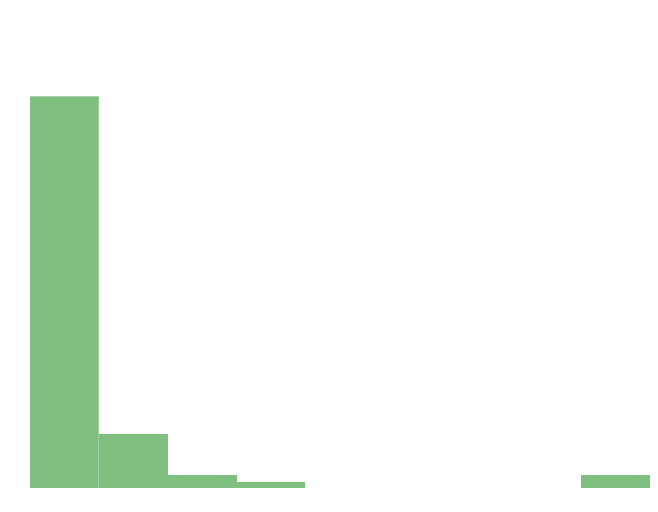} \\ 
Facial Detection & 10.0 & 10.0 & 25.0 & \includegraphics[width = 2cm, height = 0.5cm]{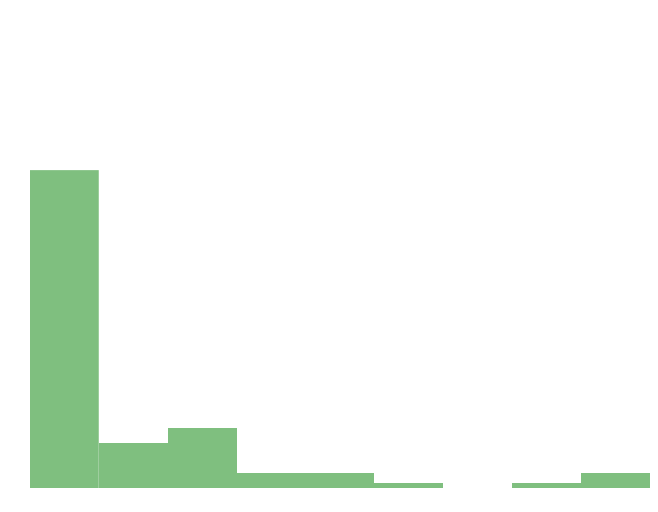} \\ 
Email & 10.0 & 10.0 & 18.0 & \includegraphics[width = 2cm, height = 0.5cm]{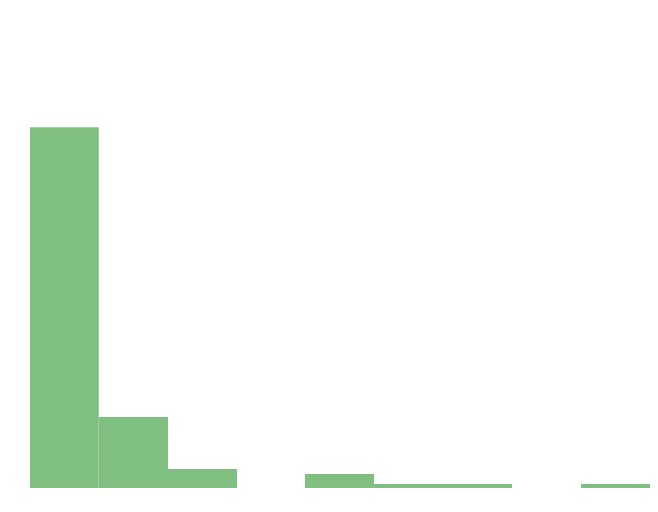} \\ 
Heart Rate Detection & 10.0 & 10.0 & 10.0 & \includegraphics[width = 2cm, height = 0.5cm]{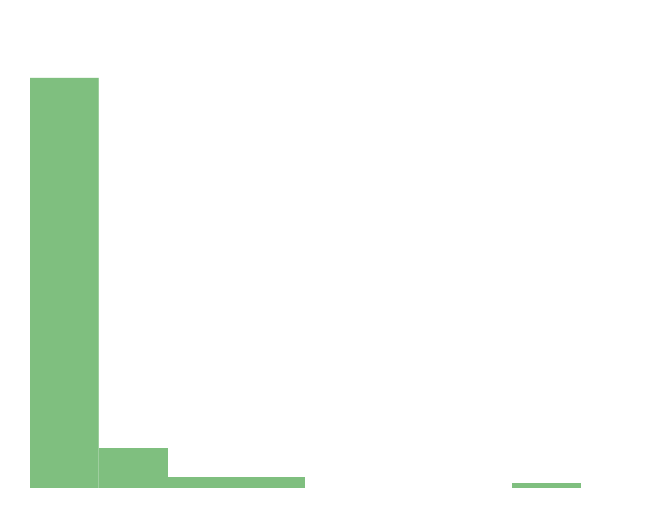} \\ 
Discreet Camera & 12.0 & 10.0 & 30.0 & \includegraphics[width = 2cm, height = 0.5cm]{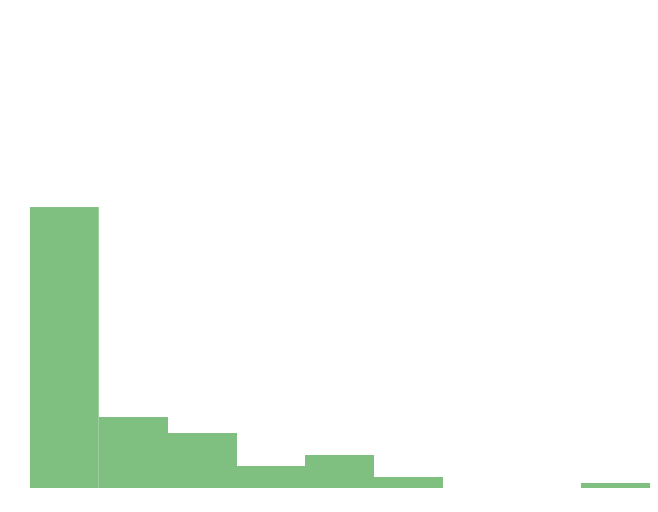} \\ 
Internet & 15.0 & 10.0 & 31.0 & \includegraphics[width = 2cm, height = 0.5cm]{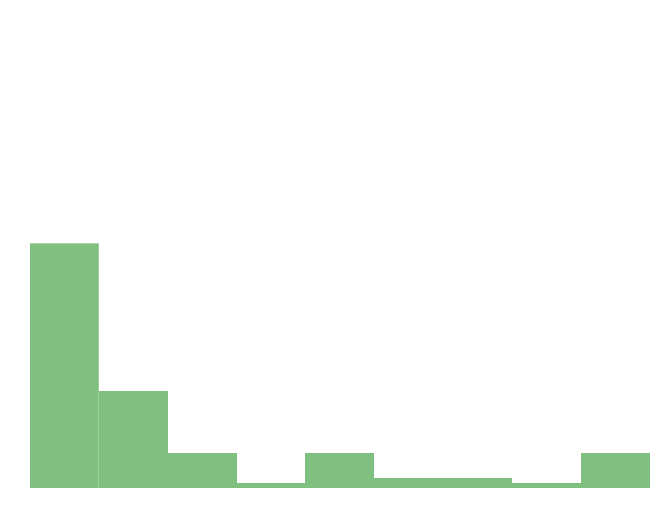} \\ 
Facial Recognition & 17.0 & 10.0 & 30.0 & \includegraphics[width = 2cm, height = 0.5cm]{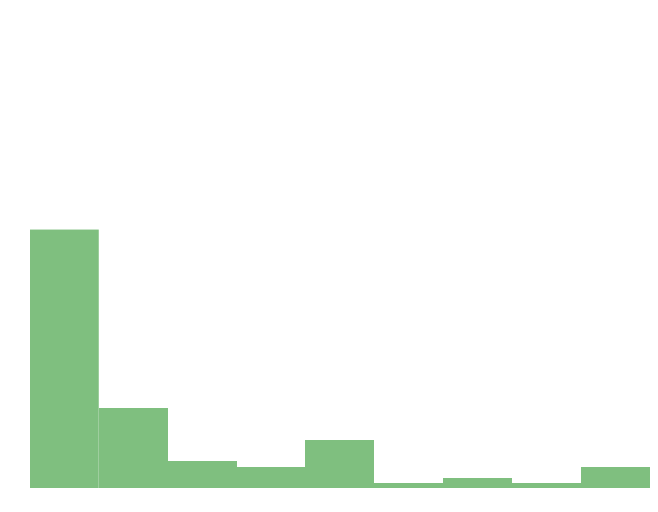} \\ 
Lawnmower & 20.0 & 12.0 & 30.0 & \includegraphics[width = 2cm, height = 0.5cm]{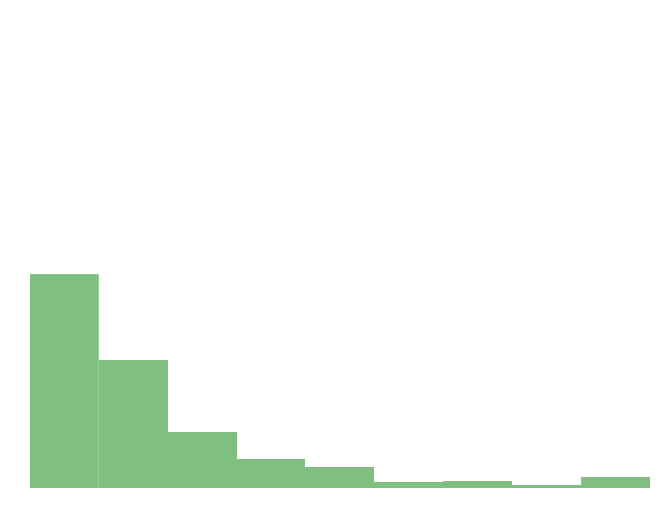} \\ 
Electricity & 25.0 & 15.0 & 40.0 & \includegraphics[width = 2cm, height = 0.5cm]{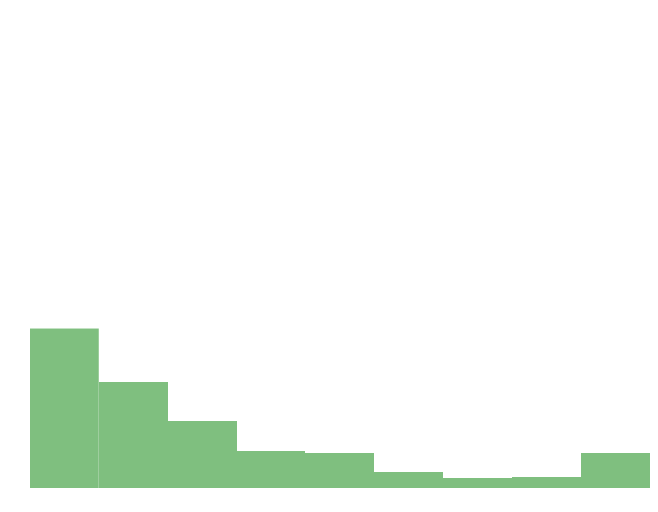} \\ 
Motorcycle & 45.0 & 27.0 & 70.0 & \includegraphics[width = 2cm, height = 0.5cm]{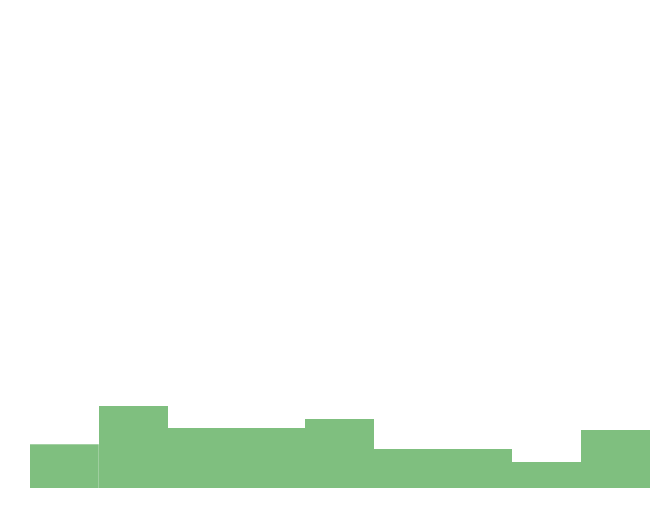} \\ 
Handgun & 60.0 & 40.0 & 100.0 & \includegraphics[width = 2cm, height = 0.5cm]{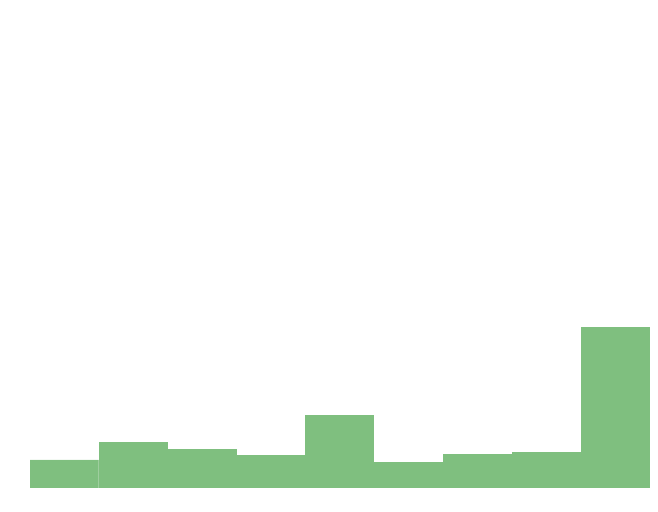} \\ 
\hline
\end{tabular}
\caption{Risk rankings of various technologies in response to the Fischoff-style prompt.}
\label{risk}
\end{center}
\end{table}

\begin{table}[h]
\begin{center}
\small
\begin{tabular}{| p{2.6cm} | p{.8cm} | p{.8cm} | p{.8cm} | c |}
\hline
Technology & Q1 &  Median & Q3 & Distribution  \\ 
\hline
Gender Detection & 10.0 & 10.0 & 15.0 & \includegraphics[width = 2cm, height = 0.5cm]{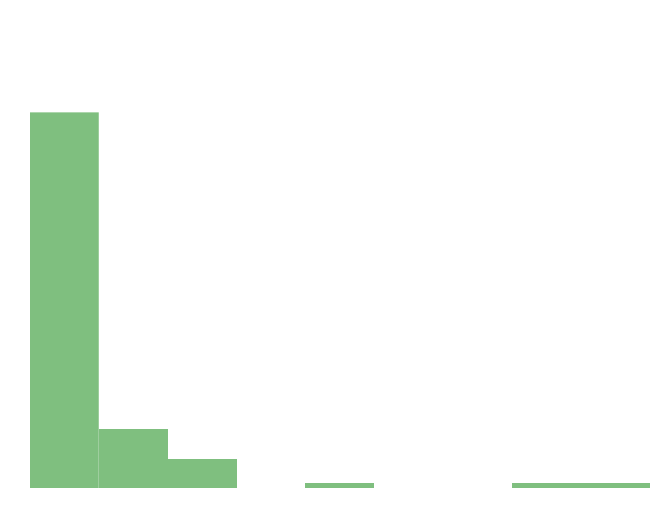} \\ 
Age Detection & 12.0 & 10.0 & 22.0 & \includegraphics[width = 2cm, height = 0.5cm]{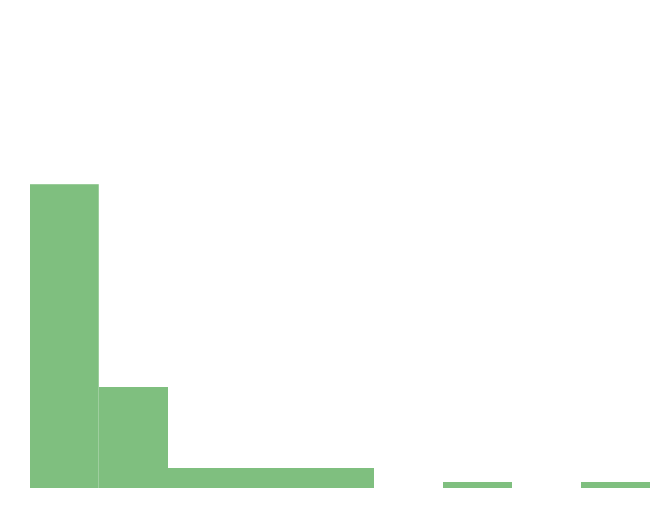} \\ 
Discreet Microphone & 15.0 & 10.0 & 20.0 & \includegraphics[width = 2cm, height = 0.5cm]{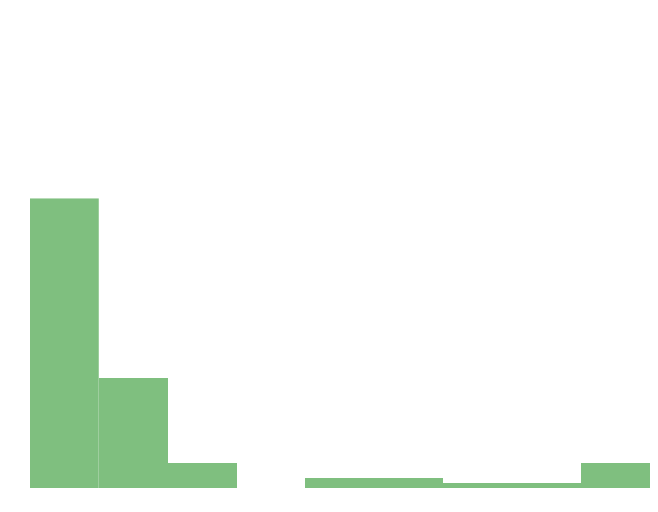} \\ 
Cubetastic & 15.0 & 10.0 & 30.0 & \includegraphics[width = 2cm, height = 0.5cm]{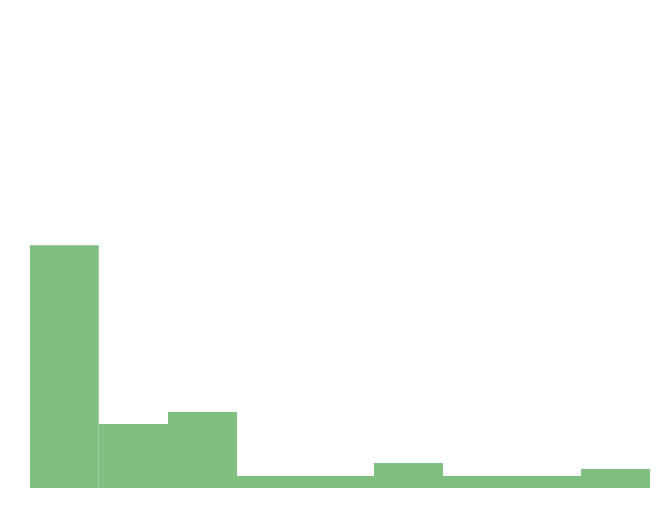} \\ 
Fitness Trackers & 18.5 & 10.0 & 30.0 & \includegraphics[width = 2cm, height = 0.5cm]{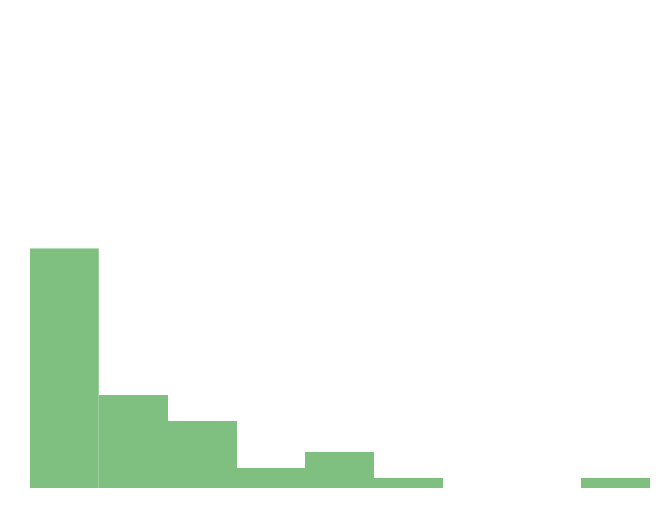} \\ 
Voice Based Emotion Detection & 20.0 & 10.0 & 30.0 & \includegraphics[width = 2cm, height = 0.5cm]{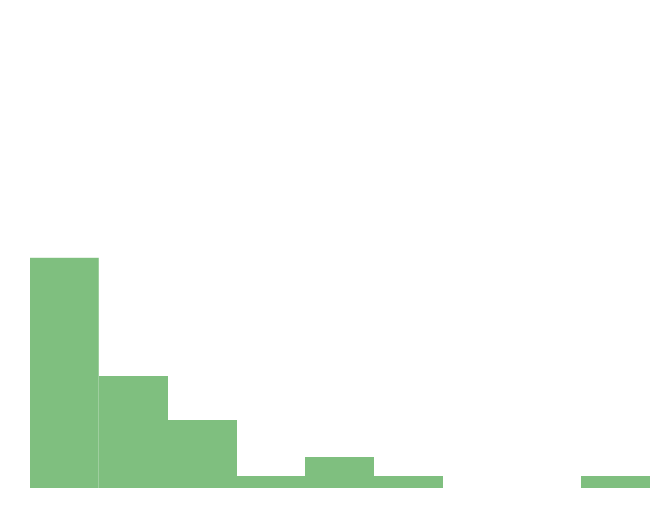} \\ 
Facial Detection & 20.0 & 10.0 & 34.0 & \includegraphics[width = 2cm, height = 0.5cm]{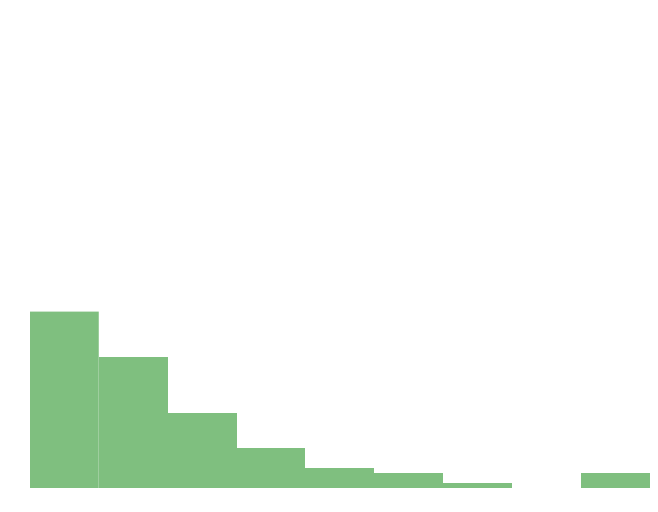} \\ 
Discreet Camera & 20.0 & 15.0 & 30.0 & \includegraphics[width = 2cm, height = 0.5cm]{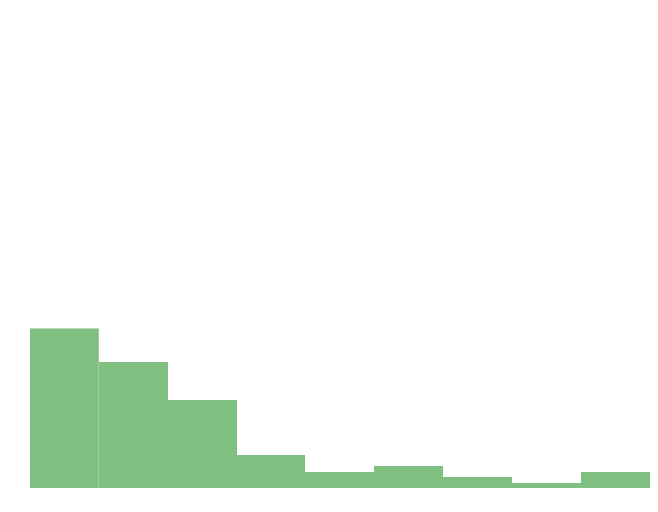} \\ 
Google Glass & 20.0 & 12.0 & 40.0 & \includegraphics[width = 2cm, height = 0.5cm]{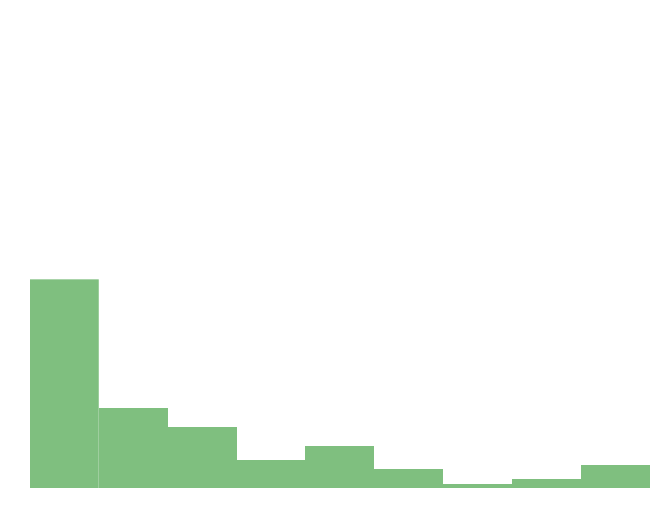} \\ 
Smartwatches & 20.0 & 10.0 & 35.0 & \includegraphics[width = 2cm, height = 0.5cm]{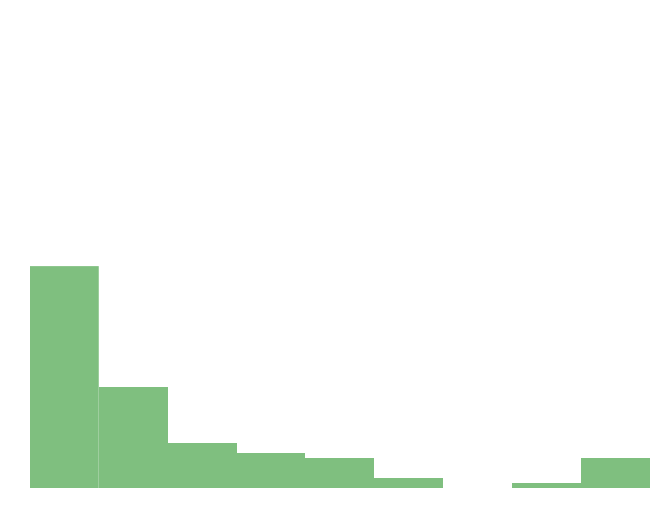} \\ 
Motorcycle & 20.0 & 12.0 & 40.0 & \includegraphics[width = 2cm, height = 0.5cm]{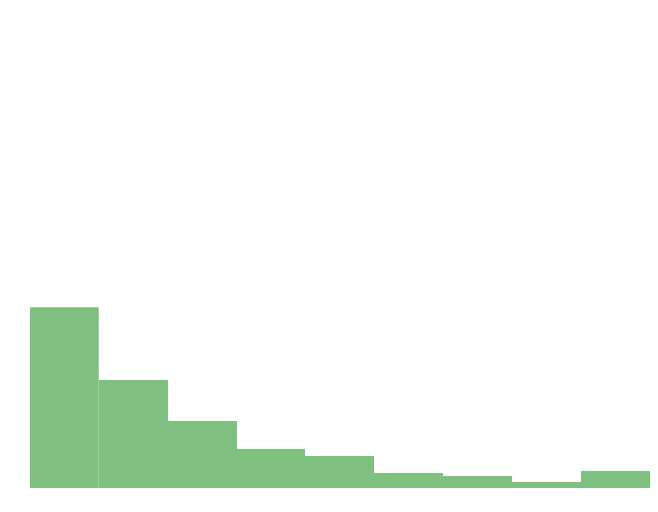} \\ 
Handgun & 20.0 & 10.0 & 30.0 & \includegraphics[width = 2cm, height = 0.5cm]{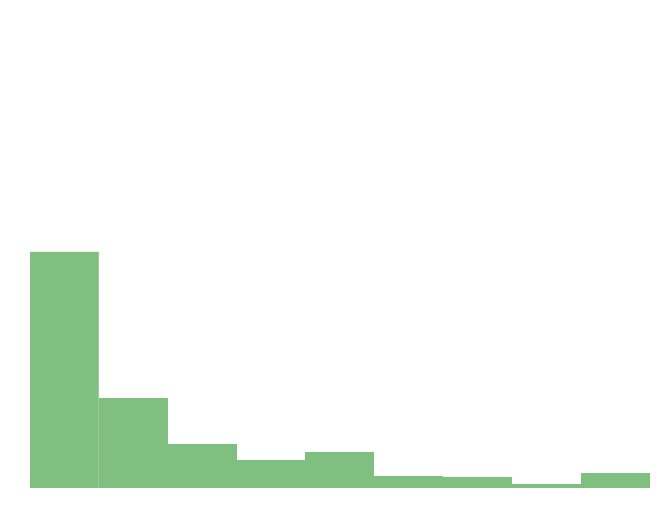} \\ 
Facial Recognition & 22.0 & 12.5 & 42.5 & \includegraphics[width = 2cm, height = 0.5cm]{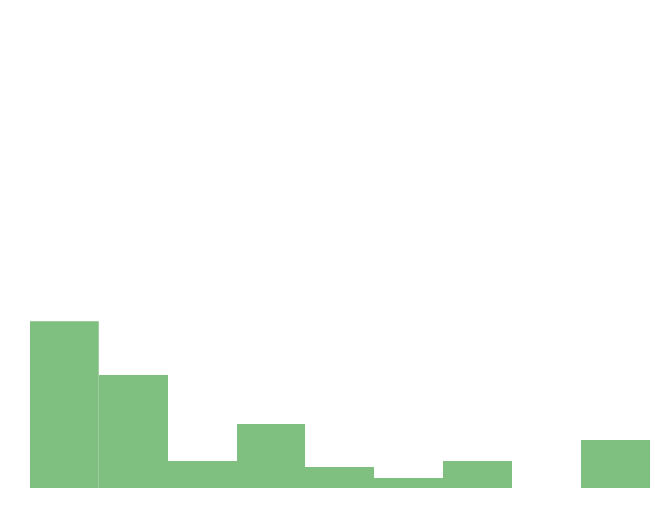} \\ 
Lawnmower & 24.0 & 15.0 & 40.0 & \includegraphics[width = 2cm, height = 0.5cm]{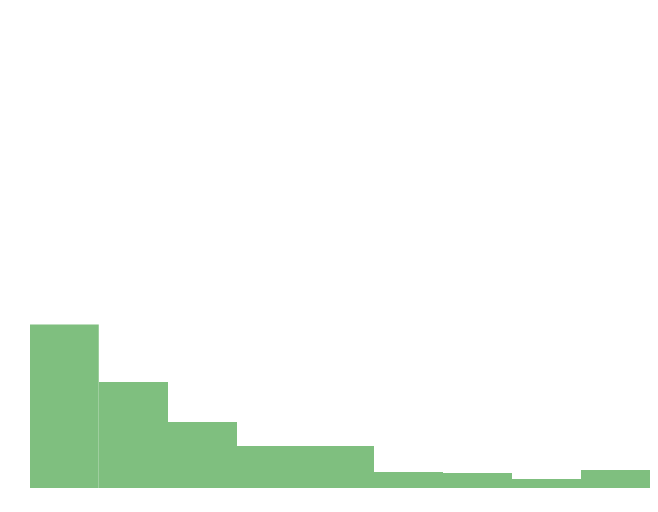} \\ 
Speech To Text & 25.0 & 15.0 & 40.0 & \includegraphics[width = 2cm, height = 0.5cm]{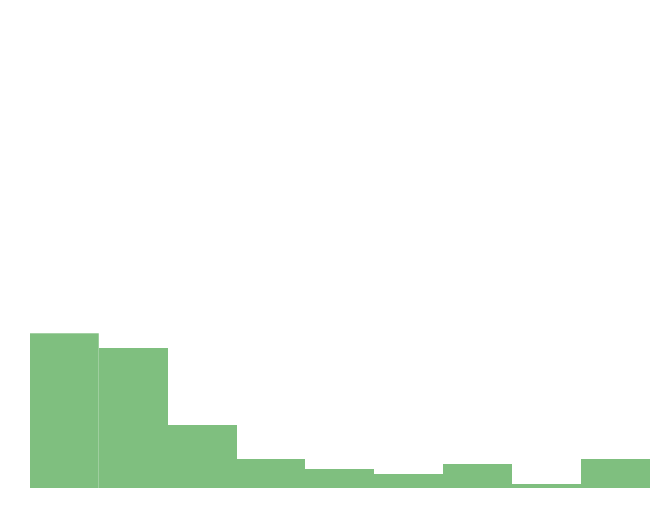} \\ 
Voice Recognition & 25.0 & 15.0 & 40.0 & \includegraphics[width = 2cm, height = 0.5cm]{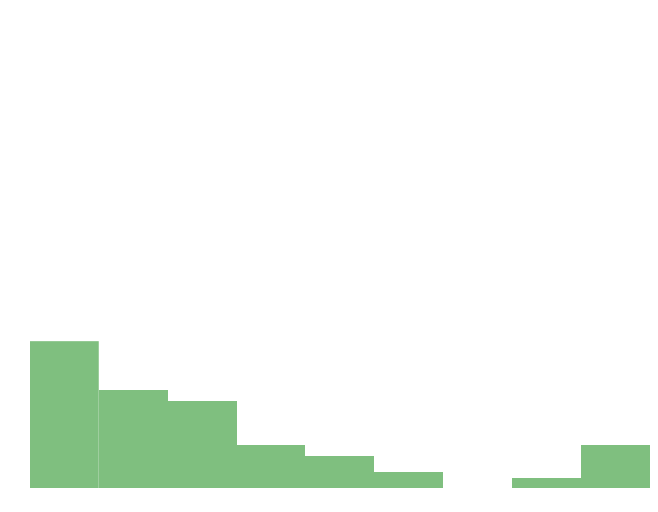} \\ 
Language Detection & 35.0 & 15.0 & 60.0 & \includegraphics[width = 2cm, height = 0.5cm]{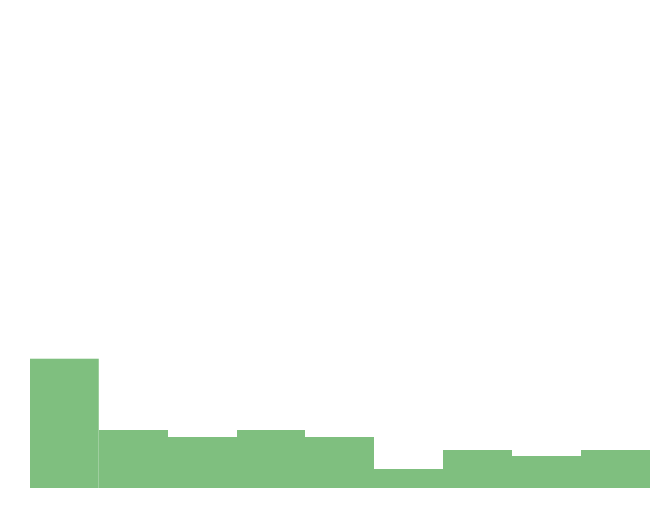} \\ 
Heart Rate Detection & 40.0 & 26.0 & 65.0 & \includegraphics[width = 2cm, height = 0.5cm]{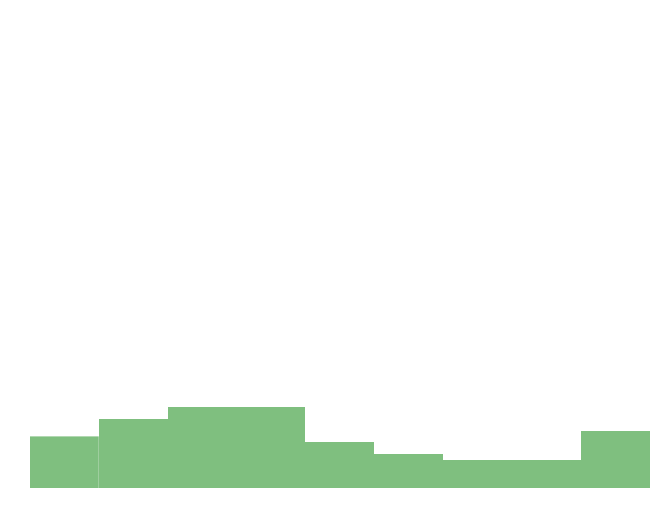} \\ 
Location Tracking & 40.0 & 20.0 & 70.0 & \includegraphics[width = 2cm, height = 0.5cm]{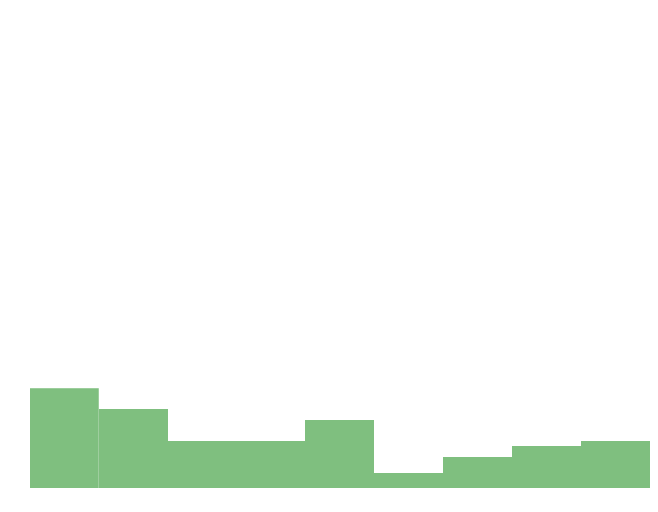} \\ 
Email & 50.0 & 29.0 & 77.5 & \includegraphics[width = 2cm, height = 0.5cm]{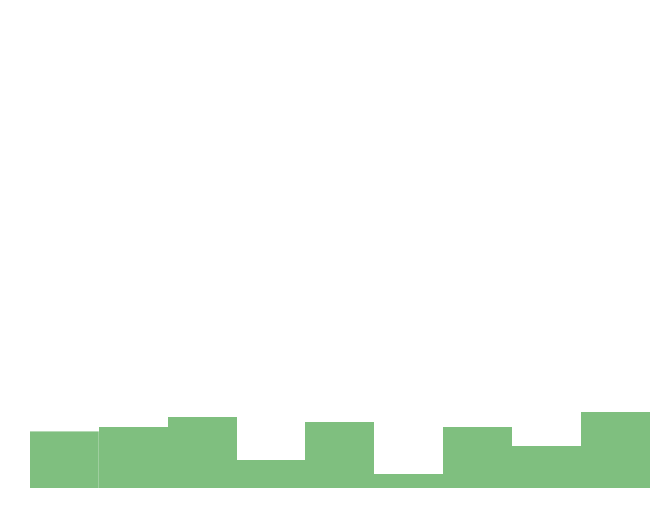} \\ 
Smartphones & 50.0 & 30.0 & 75.0 & \includegraphics[width = 2cm, height = 0.5cm]{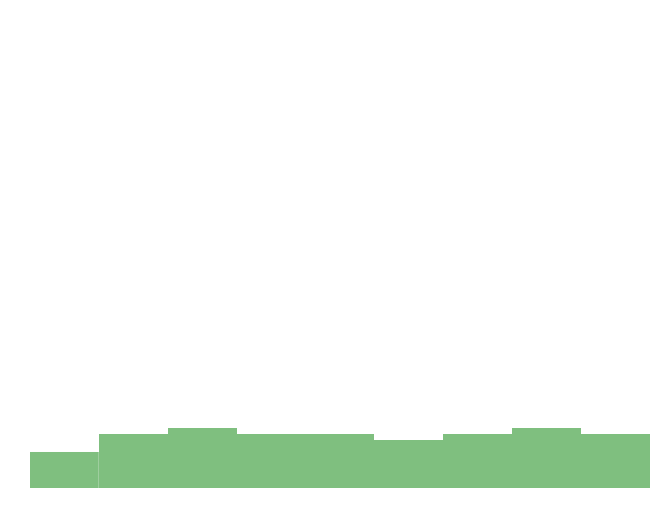} \\ 
Laptops & 60.0 & 40.0 & 80.0 & \includegraphics[width = 2cm, height = 0.5cm]{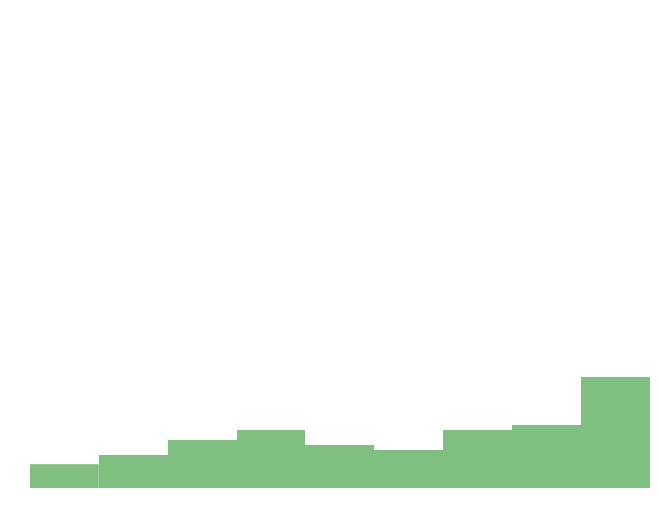} \\ 
Internet & 65.0 & 45.0 & 100.0 & \includegraphics[width = 2cm, height = 0.5cm]{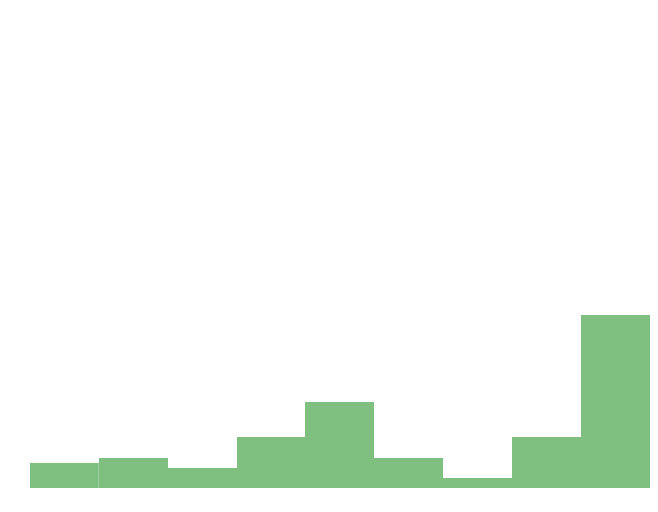} \\ 
Electricity & 88.0 & 50.0 & 100.0 & \includegraphics[width = 2cm, height = 0.5cm]{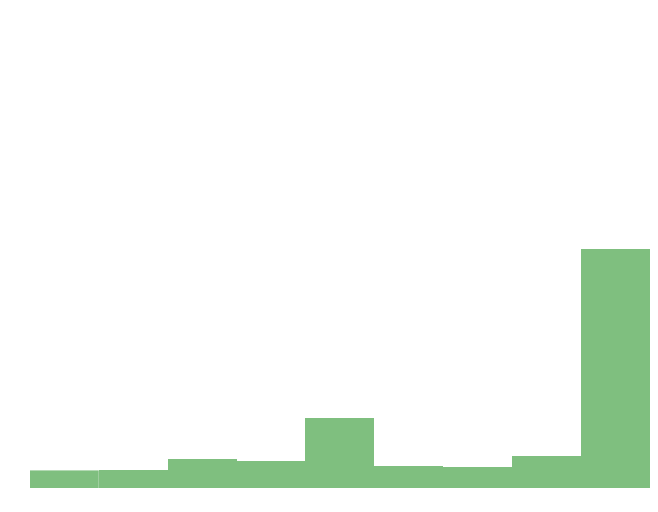} \\ 
\hline
\end{tabular}
\caption{Benefit rankings of various technologies in response to the Fischoff-style prompt. }
\label{benefit}
\end{center}
\end{table}

%% file: tex-inputs/VURcombined.tex


\begin{table*}[t]
\begin{center}
\small
\begin{tabular}{| l | r | r | r | r | r |}
\hline
Question & All & Friends & Public & Work& App \\
\hline
video of you unclothed & 95\% (1) & 97\% (4) & 94\% (10) & 100\% (1) & 90\% (2) \\ 
bank account information & 95\% (2) & 94\% (10) & 95\% (7) & 100\% (1) & 90\% (1) \\ 
social security number & 94\% (3) & 100\% (1) & 100\% (1) & 93\% (9) & 88\% (3) \\ 
video entering in a PIN at an ATM & 92\% (4) & 100\% (1) & 93\% (12) & 87\% (20) & 88\% (4) \\ 
photo of you unclothed & 92\% (5) & 96\% (6) & 91\% (16) & 100\% (1) & 77\% (6) \\ 
photo of you that is very embarrassing & 91\% (6) & 94\% (8) & 100\% (1) & 94\% (6) & 78\% (5) \\ 
username and password for websites & 89\% (7) & 96\% (5) & 95\% (9) & 94\% (7) & 64\% (14) \\ 
credit card information & 88\% (8) & 100\% (1) & 93\% (13) & 95\% (5) & 65\% (13) \\ 
video of you that is very embarrassing & 88\% (9) & 91\% (13) & 94\% (11) & 94\% (7) & 71\% (9) \\ 
photo of you at home & 87\% (10) & 85\% (19) & 96\% (5) & 93\% (10) & 71\% (10) \\ 
audio recording of work conversations & 86\% (11) & 94\% (9) & 96\% (6) & 100\% (1) & 53\% (24) \\ 
video of entering in a passcode to a door & 85\% (12) & 95\% (7) & 89\% (21) & 81\% (35) & 75\% (7) \\ 
audio recording of phone conversations & 85\% (13) & 93\% (11) & 97\% (4) & 90\% (14) & 56\% (20) \\ 
amount of money you have & 84\% (14) & 90\% (14) & 100\% (1) & 93\% (11) & 63\% (15) \\ 
video of you intoxicated & 83\% (15) & 81\% (26) & 91\% (16) & 88\% (17) & 68\% (11) \\ 
when you have sex & 81\% (16) & 78\% (31) & 87\% (23) & 90\% (15) & 73\% (8) \\ 
how much debt you have & 81\%(17) & 85\% (19) & 90\% (20) & 87\% (22) & 59\% (18) \\ 
video of you at home & 81\% (18) & 87\% (16) & 86\% (24) & 89\% (16) & 60\% (17) \\ 
photo of you intoxicated & 78\% (19) & 80\% (27) & 90\% (18) & 87\% (23) & 53\% (25) \\ 
photo of you at random  & 78\% (20) & 82\% (24) & 83\% (29) & 81\% (32) & 66\% (12) \\ 
audio recording of conversations & 78\% (21) & 86\% (18) & 85\% (26) & 87\% (20) & 55\% (21) \\ 
medical conditions & 77\% (22) & 92\% (12) & 85\% (25) & 85\% (27) & 40\% (37) \\ 
video of you at random & 76\% (23) & 73\% (40) & 90\% (19) & 88\% (19) & 48\% (31) \\ 
video of you off-guard & 76\% (24) & 85\% (21) & 79\% (34) & 91\% (13) & 53\% (23) \\ 
photo of your work or workplace & 74\% (25) & 76\% (33) & 82\% (31) & 81\% (32) & 62\% (16) \\ 
username for websites & 73\% (26) & 90\% (15) & 74\% (43) & 84\% (28) & 50\% (29) \\ 
address & 72\% (27) & 62\% (50) & 93\% (14) & 81\% (31) & 51\% (28) \\ 
audio recording you captured & 72\% (28) & 87\% (17) & 75\% (40) & 72\% (46) & 50\% (29) \\ 
photo of you off-guard & 72\% (29) & 83\% (23) & 80\% (32) & 80\% (37) & 45\% (33) \\ 
photo downloaded from internet & 71\% (30) & 79\% (29) & 76\% (38) & 86\% (25) & 32\% (47) \\ 
photo others sent you & 71\% (31) & 85\% (21) & 84\% (27) & 75\% (44) & 41\% (35) \\ 
video others sent you & 70\% (32) & 82\% (24) & 95\% (7) & 80\% (37) & 30\% (49) \\ 
video of your work or workplace & 70\% (33) & 74\% (36) & 83\% (28) & 70\% (49) & 51\% (26) \\ 
fingerprint & 70\% (34) & 77\% (32) & 80\% (32) & 70\% (48) & 55\% (22) \\ 
when you were lying nervous or stressed & 69\% (35) & 74\% (35) & 74\% (42) & 91\% (12) & 41\% (34) \\ 
audio recording of you \% (voice notes) & 69\% (36) & 80\% (28) & 78\% (35) & 88\% (18) & 38\% (39) \\ 
medication taken & 69\% (37) & 79\% (29) & 73\% (44) & 81\% (34) & 37\% (40) \\ 
videos taken on device & 68\% (38) & 58\% (52) & 82\% (30) & 79\% (40) & 51\% (27) \\ 
photo of your signature & 68\% (39) & 63\% (48) & 64\% (51) & 85\% (26) & 59\% (19) \\ 
web history & 66\% (40) & 74\% (36) & 70\% (45) & 86\% (24) & 37\% (40) \\ 
photos already on device & 66\% (41) & 75\% (34) & 77\% (36) & 79\% (39) & 27\% (53) \\ 
home address & 65\% (42) & 61\% (51) & 87\% (22) & 69\% (50) & 40\% (36) \\ 
fine-grained location tracking (+/-  cm) & 63\% (43) & 73\% (39) & 76\% (37) & 78\% (41) & 30\% (50) \\ 
photo of people at random & 61\% (44) & 72\% (41) & 61\% (54) & 82\% (30) & 38\% (38) \\ 
video downloaded from the internet & 61\% (45) & 63\% (47) & 75\% (40) & 82\% (29) & 33\% (45) \\ 
when you are alone & 61\% (46) & 51\% (55) & 69\% (46) & 80\% (36) & 35\% (43) \\ 
location tracking (+/- m) & 61\% (47) & 57\% (53) & 92\% (15) & 63\% (55) & 25\% (56) \\ 
videos of people at random & 61\% (48) & 63\% (49) & 75\% (39) & 71\% (47) & 28\% (52) \\ 
where you are currently going & 60\% (49) & 74\% (36) & 68\% (48) & 65\% (54) & 35\% (44) \\ 
recording of sound around you & 60\% (50) & 71\% (42) & 64\% (50) & 75\% (43) & 35\% (42) \\ 
people you spend time with & 60\% (51) & 71\% (42) & 60\% (55) & 76\% (42) & 31\% (48) \\ 
workplace address & 58\% (52) & 69\% (45) & 64\% (49) & 57\% (61) & 46\% (32) \\ 
sounds on device \% (notifications, etc) & 54\% (53) & 70\% (44) & 59\% (56) & 66\% (52) & 22\% (58) \\ 
phone usage & 51\% (54) & 67\% (46) & 56\% (57) & 68\% (51) & 15\% (64) \\ 
purchased products & 50\% (55) & 57\% (54) & 55\% (58) & 62\% (57) & 26\% (54) \\ 
when you are sick or healthy & 48\% (56) & 40\% (64) & 61\% (52) & 62\% (58) & 26\% (55) \\ 
how close you are to interacting people & 46\% (57) & 50\% (57) & 61\% (53) & 51\% (62) & 13\% (66) \\ 
feelings (based on biometrics) & 46\% (58) & 50\% (57) & 55\% (58) & 63\% (56) & 18\% (61) \\ 
computer usage& 44\% (59) & 51\% (56) & 52\% (60) & 45\% (63) & 28\% (51) \\ 
eating patterns & 42\% (60) & 41\% (62) & 45\% (62) & 75\% (45) & 12\% (67) \\ 
name & 42\% (61) & 50\% (57) & 68\% (47) & 26\% (71) & 32\% (46) \\ 
sleeping patterns & 40\% (62) & 43\% (61) & 41\% (63) & 62\% (59) & 21\% (59) \\ 
eye patterns \% (for eye tracking) & 40\% (63) & 48\% (60) & 50\% (61) & 61\% (60) & 6\% (71) \\ 
exercise patterns & 38\% (64) & 33\% (67) & 34\% (66) & 66\% (52) & 16\% (63) \\ 
when you are happy or having fun & 34\% (65) & 40\% (64) & 32\% (69) & 43\% (65) & 24\% (57) \\ 
television shows watched & 30\% (66) & 38\% (66) & 33\% (67) & 36\% (68) & 11\% (68) \\ 
when you are busy or interruptible & 29\% (67) & 40\% (63) & 28\% (70) & 36\% (68) & 17\% (62) \\ 
music on device & 28\% (68) & 4\% (72) & 37\% (64) & 42\% (66) & 20\% (60) \\ 
heart rate & 27\% (69) & 21\% (68) & 36\% (65) & 44\% (64) & 9\% (70) \\ 
age & 24\% (70) & 17\% (69) & 33\% (67) & 36\% (67) & 14\% (65) \\ 
language spoken & 15\% (71) & 17\% (70) & 18\% (72) & 28\% (70) & 27\% (53) \\ 
gender & 15\% (72) & 15\% (71) & 19\% (71) & 15\% (72) & 9\% (69) \\ 
\hline
\end{tabular}
\caption{The VUR of all questions for all recipients.}
\label{all-vur}
\end{center}
\end{table*}